\PassOptionsToPackage{warn}{textcomp}
\documentclass[acmsmall,authorversion,screen]{acmart}

\usepackage{enumitem}
\usepackage{textcomp}
\usepackage{graphics} 
\usepackage{epsfig}
\usepackage{bbm}
\usepackage{subfig}
\usepackage{color}
\usepackage{multirow}
\usepackage{times} 
\usepackage{amsmath}
\usepackage{mathrsfs}

\usepackage[makeroom]{cancel}
\usepackage{amssymb}
\usepackage{flushend}
\usepackage[n,operators,advantage,sets,adversary,landau,probability,notions,logic,ff,mm,primitives,events,complexity,asymptotics,keys]{cryptocode}
\usepackage{algorithm}
\usepackage{algpseudocode}
\usepackage{listings}
\usepackage{float}
\usepackage[normalem]{ulem}
\usepackage{tcolorbox}
\usepackage{hyperref}

\newlength\mytemplength

\newcommand\parboxc[3]{%
    \settowidth{\mytemplength}{#3}%
    \parbox[#1][#2]{\mytemplength}{\centering #3}%
}

\usepackage{amsthm}
\newtheoremstyle{mydef}
{ }{}%
{}{}%
{\bfseries}{}
{}
{%
  \thmname{#1}~\thmnumber{#2}\thmnote{\ -\ #3}.{ }%\\*[-1.5ex]%
}%

\newtheorem{definition}{Definition}

\newcommand{\stkout}[1]{\ifmmode\text{\sout{\ensuremath{#1}}}\else\sout{#1}\fi}
\newcommand{\add}{\texttt{add}}
\newcommand{\mult}{\texttt{mult}}
\newcommand{\Eval}{\texttt{eval}}

\renewcommand{\mod}{\mathrm{mod~}}

\newcommand\highlight[1]{\textcolor{black}{#1}}

\PassOptionsToPackage{warn}{textcomp}

\AtBeginDocument{%
  \providecommand\BibTeX{{%
    \normalfont B\kern-0.5em{\scshape i\kern-0.25em b}\kern-0.8em\TeX}}}

\newtcolorbox[auto counter]{pabox}[2][]{float,left=0mm,colframe=black!5,coltitle=black,title=\centering Scheme~\thetcbcounter: #2,#1}

\begin{document}

\title{Computing Blindfolded on Data Homomorphically Encrypted under Multiple Keys: An Extended Survey}

\author{Asma Aloufi}
 \authornote{Asma Aloufi is also affiliated with Taif University, Saudi Arabia.}
\email{ama9000@rit.edu}

\author{Peizhao Hu}
\email{Peizhao.Hu@rit.edu}

\affiliation{%
  \institution{Rochester Institute of Technology, New York}
}

\author{Yongsoo Song}
\email{yongsoo.song@microsoft.com}

\author{Kristin Lauter}
\email{klauter@microsoft.com}
\affiliation{%
  \institution{Microsoft Research, Redmond, WA}
}

\renewcommand{\shortauthors}{Aloufi et al.}

\begin{abstract}
    New cryptographic techniques such as homomorphic encryption (HE) allow computations to be outsourced to and evaluated blindfolded in a resourceful cloud. These computations often require private data owned by multiple participants, engaging in joint evaluation of some functions. For example, Genome-Wide Association Study (GWAS) is becoming feasible because of recent proliferation of genome sequencing technology. Due to the sensitivity of genomic data, these data should be encrypted using different keys. However, supporting computation on ciphertexts encrypted under multiple keys is a non-trivial task. In this paper, we present a comprehensive survey on different state-of-the-art cryptographic techniques and schemes that are commonly used. We review techniques and schemes including Attribute-Based Encryption (ABE), Proxy Re-Encryption (PRE), Threshold Homomorphic Encryption (ThHE), and Multi-Key Homomorphic Encryption (MKHE). We analyze them based on different system and security models, and examine their complexities. We share lessons learned and draw observations for designing better schemes with reduced overheads. 
\end{abstract}

%\keywords{Homomorphic Encryption, Privacy, Security, Cloud Computing}

\maketitle

% Paper sections:
\section{Introduction}
\label{sec:intro}
Modern data-driven applications involve highly sensitive data such as genome sequences~\cite{wang2017community}, biometric data including iris scans and fingerprints~\cite{yasuda2013secure}, and location data and information of users' whereabouts~\cite{hu2017geosocial}. These data are hard to change, sometime irreplaceable, once exposed. Therefore, privacy enhancing techniques have to be applied in order to ensure the privacy of user data. Conventional encryption protects data \emph{in-transit} and \emph{in-storage}. However, data has to be decrypted before performing any computation on it. Advanced encryption technique is needed to extend the protection of user data during computation; that is, \emph{in-use}. 

Homomorphic encryption (HE) supports arithmetic operations, such as addition and multiplication, on encrypted data without decrypting it first. Specifically, given two messages $m$ and $m'$ and homomorphic addition and multiplication operations $\oplus$ and $\otimes$, we have $Enc(m) \oplus Enc(m')$ which decrypts to $m+m'$ and $Enc(m) \otimes Enc(m')$ which decrypts to $m \times m'$. For simplicity, we will use normal arithmetic operators to represent homomorphic operations in the rest of the paper. Generally speaking, any algorithm that can be reduced to just these arithmetic operations can be homomorphically evaluated on encrypted data. Logic gates, such as $\mathsf{AND}$, $\mathsf{OR}$, $\mathsf{XOR}$, and $\mathsf{NOT}$, can be translated into arithmetic forms; for example, $\mathsf{XOR}(x,y) = x + y - 2xy$ if $x,y\in \ZZ$, or simply $\mathsf{XOR(x,y) = x + y}$ if $x,y\in \ZZ_2$. Additionally, checking equality of two numbers $x,y \in \ZZ$ can be performed as $\mathsf{EQ}(x,y) = \prod( \mathsf{XNOR}(x_i,y_i))$ where $x_i, y_i$ are bits of the input numbers and $\mathsf{XNOR}(x_i,y_i)=\mathsf{NOT}(\mathsf{OR}(x_i,y_i))$, or simply  $\mathsf{EQ}(x, y) = x + y + 1$ if $x,y\in \mathbb{Z}_2$. The first set of HE schemes~\cite{goldwasser1984probabilistic, elgamal1985public, benaloh1994dense, paillier1999public} only realized \emph{partial} homomorphisim, which supports either addition or multiplication on ciphertexts but not both, such as Paillier~\cite{paillier1999public} and ElGamal~\cite{elgamal1985public} schemes. Boneh-Goh-Nissim scheme~\cite{boneh2005evaluating} supports arbitrary number of additions and a single multiplication; hence, it is \emph{somewhat} homomorphic.
There are \emph{leveled} HE schemes which support a predetermined number of multiplications based on the targeted function. In 2009, Gentry~\cite{gentry2009fully} proposed the first plausible construction to achieve a \emph{fully} HE (FHE) scheme based on ideal lattices that can support arbitrary number of additions and multiplications. Homomorphic addition is performed almost free of cost but multiplication significantly increases the noise elements in the ciphertext. Hence, it is important to reduce the multiplicative depth (i.e., consecutive multiplications) of a homomorphic function. When the noise growth in the ciphertext is too large, it is not possible to retrieve the correct message after decryption. Gentry's FHE scheme, which was built on a leveled HE scheme, can homomorphically evaluate its own decryption circuit in a \emph{bootstrapping} step. Hence, it can refresh the encryption of an evaluated ciphertext, when the noise level reaches a defined threshold, to obtain a fresh ciphertext with small noise. In practice, the bootstrapping step is computationally expensive; therefore, most applications sufficiently use a leveled HE scheme. 
Also, HE schemes can be designed to be either symmetric (uses one key to encrypt and decrypt) or asymmetric (uses one key pair, a public key to encrypt and a secret key to decrypt).

\subsection{System and security models for secure computation}
\label{subsec:systemAndSecurityModel}
An ideal form of secure computation is to send the sensitive data to a trusted party, who performs the computations and returns the result. However, the existence of this trusted party is not always possible especially with frequent security breaches and possibility of participants' corruption. In a practical setting, the security of constructed protocols is emulated to the security of an ideal model, following the Ideal/Real model paradigm~\cite{lindell2009secure}. In other words, we focus on a security model which considers the existence of semi-honest or malicious adversaries.

\begin{figure*}
\centering
  \subfloat[Two-party setting]
 {\includegraphics[width=.26\textwidth]{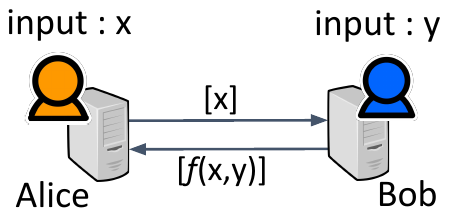} \label{fig:2pc}}\hfill
  \subfloat[Multi-party setting]
 {\includegraphics[width=.28\textwidth]{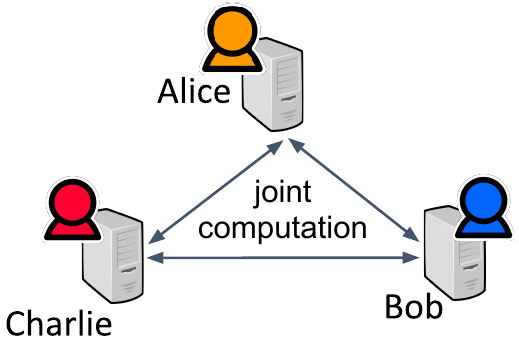} \label{fig:mpc}}\hfill
 \subfloat[Outsourced setting]
 {\includegraphics[width=.37\textwidth]{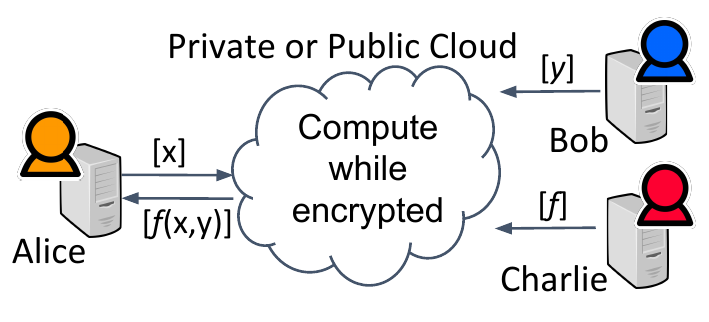} \label{fig:outsourced}}
\caption{System models for secure computation.}
\label{fig:system-setup}
\end{figure*}

The ability to compute while encrypted allows sensitive data to be outsourced for computations without compromising privacy. This ability enables many applications to be designed based on different system models, as illustrated in Fig.~\ref{fig:system-setup}. These system models are often referred to as \emph{coopetitive}~\cite{zheng2019helen,coopetition} models (i.e., cooperation in the presence of competition). The simplest system model for secure outsourced computation is in a two-party setting (Fig.~\ref{fig:2pc}) where Alice wants Bob to perform a computation $f(x, y)$, where $f$ is a function or a trained machine learning model, on sensitive inputs $x$ and $y$ owned by them without revealing the data to each other. In this case, Alice encrypts the input $[x]$ under her public key and sends it to Bob, who homomorphically computes the function $f$ on the inputs $[x],y$ and returns the encrypted result $[f(x,y)]$ to Alice. Alice can decrypt the evaluated results using her private key. In this system model, user data privacy is protected against passive adversary by encryption as long as the function $f$ does not leak information about Bob's input $y$. For some leaky functions such as summing two input numbers, HE primitives do not prevent leakage.   

The two-party setting can be extended to a more general setting, secure outsourced computation among multiple parties. As illustrated in Fig.~\ref{fig:mpc}, multiple parties may want to perform a joint computation on their encrypted data without revealing sensitive data to each others. This computation model is often based on interactive secure multi-party computation (MPC) protocols, which are characterized by the high communication overhead and require participation of all the parties during the computation. 

There is an emerging system model which supports non-interactive computations and better suits the cloud computing model because data owners delegate computations on their data to a private or public cloud, as shown in Fig.~\ref{fig:outsourced}. Although there are existing work~\cite{yasuda2013secure,bos2014private, bost2014machine, chen2018labeled} on outsourcing homomorphic computations to the cloud, these works focus on settings similar to the two-party setting; that is, the computations are performed on data from one data owner. In this paper, we focus on a more challenging computation model in which multiple parties delegate their sensitive data or functions to the cloud. The cloud is an evaluator of some functions and should not learn anything from the encrypted data. Similar to previous system models, data owners do not want to reveal their data to others. To ensure user data privacy, encrypting data using a single key is not very practical in many application scenarios. \emph{Can we realize a secure computation model in which data are encrypted using different keys of the corresponding owners?} This cloud-friendly computation model has been discussed in previous literature~\cite{shan2018practical, naehrig2011can}. However, it has not been realized due to the challenges in supporting homomorphic evaluation on data encrypted under different keys, in addition to issues in key management and lack of efficient decryption protocols. Note, we focus on the system model in Fig.~\ref{fig:outsourced} and assume inputs are encrypted under their owners' keys, but we do not define under which key the result $[f(x,y)]$ is encrypted for now. We will be specific about this key in the subsequent sections when we review different approaches. 

To the best of our knowledge, none of existing surveys provide a comprehensive review on how to support efficient homomorphic computations on encrypted data under multiple keys in the new computation model we discussed earlier. In this survey, our goal is to fill this gap and provide a comprehensive review, analysis, and lesson-learned of the state-of-the-art multi-key homomorphic encryption techniques for secure computation outsourcing.

\subsection{Computing with different keys}
\label{subsec:taxonomy}
\begin{figure}[t]
\centering
\includegraphics[width=.93\textwidth]{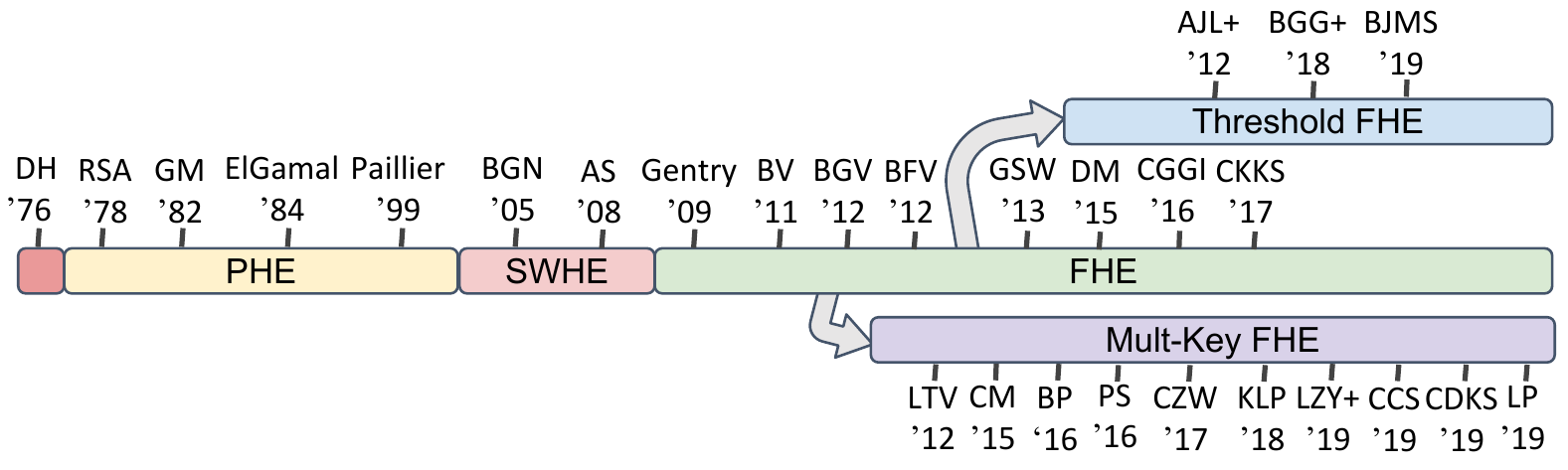}
\caption{Timeline featuring recent developments in HE schemes.}
\label{fig:timeline}
\end{figure}

Spanning over the last 40 years, there are many homomorphic encryption schemes, each provides different supports for performing arithmetic operations on homomorphically encrypted data. Figure~\ref{fig:timeline} shows a timeline of different types of HE schemes (improved based on \cite{acar2018survey, yang2019comprehensive}), including recent schemes that support homomorphic computations on encrypted data under multiple keys.

\begin{figure}[t]
\centering
\includegraphics[width=.93\textwidth]{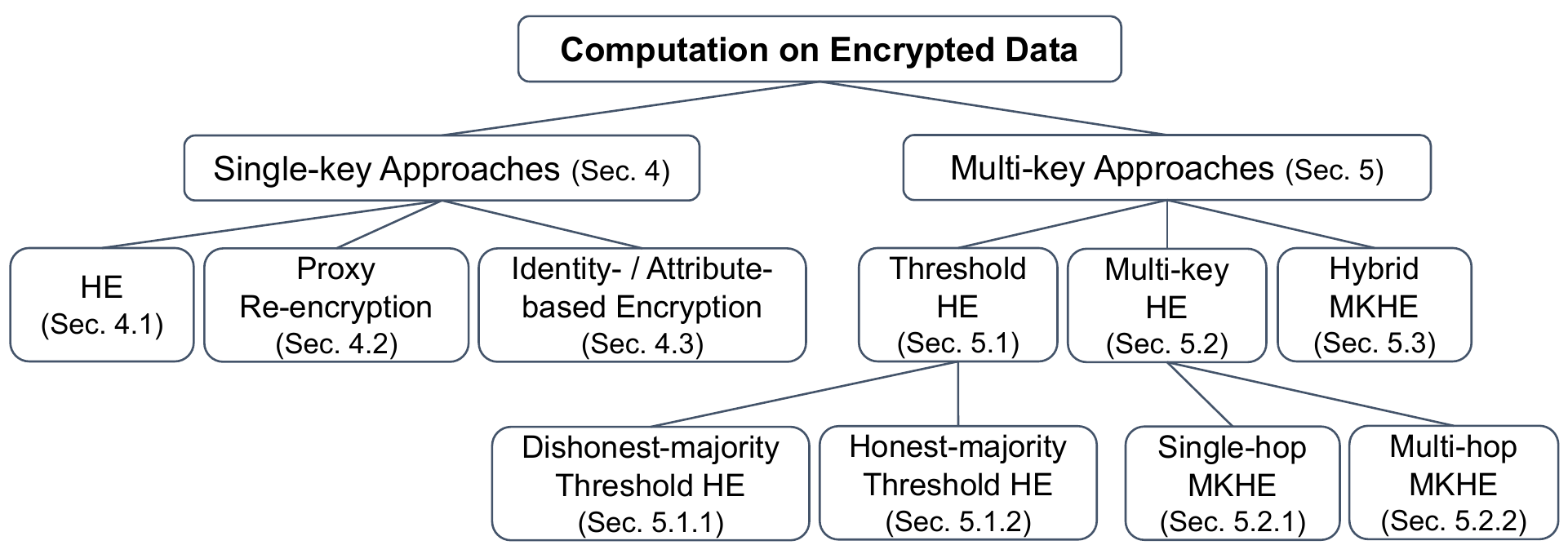}
\caption{Taxonomy of techniques for secure computation on encrypted data.}
\label{fig:taxonomy}
\end{figure}

In this paper, we review and categorize these techniques according to how the multiple keys are used to secure data and how homomorphic computations are supported. Figure~\ref{fig:taxonomy} shows a taxonomy of these surveyed techniques within this paper. Generally speaking, they can be roughly divided into two categories:
\begin{itemize}[leftmargin=*]
\item[-] A subset of these techniques is based on a \emph{single} key, where ciphertexts are encrypted under a key owned by a single party at all times and decrypted using the corresponding secret key or can be made decryptable under another key. Most existing works on HE assumed data encrypted under a single key. However, it is clearly impractical in our \emph{coopetitive} system models illustrated in Fig.~\ref{fig:system-setup}. Another example of single key schemes but offering additional functionalities is Proxy Re-Encryption (PRE) technique~\cite{blaze1998divertible, yang2016cloud} that transforms a stored ciphertext $c_i = \enc(\pk_i, m)$ into a ciphertext $c_j = \reenc(\rk_{\pk_i\rightarrow \pk_j}, c_i)$ that is decrypted by authorized user's secret key $\sk_j$. Also, there are Identity-based encryption (IBE)~\cite{boneh2001identity} and Attribute-based encryption (ABE)~\cite{sahai2005fuzzy} schemes that provide access control mechanisms where data can be decrypted by authorized keys generated based on the user's identity or attributes.
\item[-] On the other hand, \emph{multiple} keys may be involved in the encryption of ciphertexts and required to contribute to its decryption. Threshold HE (ThHE)~\cite{asharov2012multiparty} and Multi-key HE (MKHE) are examples of these techniques. An example of the latter allows the ciphertext $c_i$, encrypted under $\pk_i$, to be extended to an additional key $\pk_j$ such that $\Bar{c} = \enc(\{\pk_i, \pk_j\}, m)$. The message can be retrieved when using the two corresponding secret keys $m = \dec(\{\sk_i, \sk_j\}, \Bar{c})$. Recently, there is a hybrid approach~\cite{aloufi2019collaborative} which combines the advantages of both threshold and multi-key HE with a goal to reduce computation complexity and ciphertext size. 
\end{itemize}

\subsection{Related work}
As discussed before, many surveys focused on secure outsourced computation techniques~\cite{tang2016ensuring, shan2018practical, yang2019comprehensive} and the construction of HE schemes~\cite{fontaine2007survey, vaikuntanathan2011computing, martins2017Survey, acar2018survey}. There is a notable lack in the literature for detailed review on secure outsourced computation with different keys in the \emph{coopetitve} system model. Tang~et~al.~\cite{tang2016ensuring} focused on threat modeling for secure outsourced computation and studied techniques to realize security requirements including confidentiality, integrity, privacy, and access control. Techniques with different keys were studied only for data sharing in the cloud and did not address supporting homomorphic computations on encrypted data. Secure outsourcing of fundamental and application-specific homomorphic computations based on a single key were surveyed in~\cite{shan2018practical, yang2019comprehensive}. 

On the other hand, extensive reviews of the HE schemes construction have been provided both from engineering~\cite{martins2017Survey} and theoretical \cite{fun2016survey,acar2018survey} perspectives. Unfortunately, existing survey articles did not discuss extended HE schemes, such as ThHE and MKHE, which support computing with multiple keys. Particularly, these multi-key techniques are witnessing rapid development in recent years as can be observed in Fig.~\ref{fig:timeline}. Some proposed MKHE work~\cite{chen2019MKSEAL, Lee2019Security, aloufi2019collaborative} give brief analysis of related work. A more-in-depth review~\cite{bongenaar2016multi} studied the first two MKHE constructions (i.e., LaTV12, CM15). Overall, no existing survey provides detailed analysis and categorization of state-of-the-art multi-key approaches, which is a primary aim of our survey.

\subsection{Contributions and organization}
In this paper, we fill the gap in the literature and conduct a comprehensive survey on cryptographic techniques that enable multiple parties to compute on their data encrypted under multiple keys. In particular, we investigate design trends in the state-of-the-art schemes for different system and threat models. We share lesson learned and discuss new directions that is yet to be explored to achieve more practical solutions for secure computation outsourcing. We discuss potential applications that can be benefit from the ability of homomorphically computing on encrypted data under multiple keys.

The rest of this survey is organized as follows. Different system models and security models for secure computation is defined in Section~\ref{sec:models}. After that, preliminaries and HE commonly used techniques are presented in Section~\ref{sec:prelim}. Single-key approaches such as attribute-based encryption and proxy re-encryption are presented in Section~\ref{sec:single}. Following that, the multi-key approaches, threshold HE, multi-key HE, and hybrid approaches are discussed in Section~\ref{sec:multi}. In Section~\ref{sec:lessons}, we share lesson learned and open research directions. In Section~\ref{sec:applications}, we review application scenarios for computation on data with multiple keys. 
Finally, Section~\ref{sec:conclusion} concludes the survey.
\section{Security Considerations} 
\label{sec:models}
To design a secure protocol, we need to determine possible attacks that target a system model. Adversaries who launch the attacks often aim to compromise security requirements such as confidentiality, integrity, and availability. It is essential to understand these threats and their impacts on the security during threat modeling. In often cases, malicious adversaries can launch arbitrary attacks that deviates from the protocol, and it is difficult to protect against every threat. Moreover, some countermeasures may be computationally-intensive and affect the practicality of the protocol. Hence, we make security assumptions as a trade-off for efficiency. For example, we assume system users are semi-honest, i.e., they strictly follow the protocol specifications. Careful design of the security model is critical to achieving secure yet practical protocol for a given system model. In the rest of this section, we discuss the security model of a system in more details, including modeling \emph{potential threats} and specifying the appropriate \emph{security requirements} and \emph{assumptions}. 

\subsection{Threat model} 
Threat modeling is the process of identifying potential vulnerabilities, circumstances, and actions in which a capable adversary can compromise the security of the system~\cite{stallings2006cryptography}. In the system models illustrated in Fig.~\ref{fig:system-setup}, an adversary can be either an \emph{internal} or \emph{external}. An internal adversary is a participant in the protocol, e.g., Bob in the multi-party setting in Fig.~\ref{fig:mpc}, and may want to learn confidential information of other users. On the other hand, an external adversary may intercepts or corrupts system users, such as colluding with the cloud evaluator, with the intention to breach the privacy of other users. Moreover, adversaries are assumed to be either \emph{semi-honest}, or \emph{malicious}. 

A passive adversary in the \emph{semi-honest} setting, say the cloud evaluator, follows the protocol execution and does not attempt to cheat; however, it passively collects the transmitted inputs and tries to infer useful information about the data owners. It is difficult to detect this type of passive attacks because no abnormal behaviour is observed~\cite{schneier2007applied}. Hence, it is fundamental that any secure protocol has to provide defense, rather than detection, against such passive adversary attacks.

Unlike passive adversaries, active or \emph{malicious} adversary can launch arbitrary attacks such as deviating from the protocol's specifications. A malicious cloud evaluator may use homomorphic proprieties to alter ciphertexts without decryption, or simply corrupt them. Active attacks may also take other forms to breach security, such as corrupting and impersonating authorized system users and feeding new or changed inputs to the protocol, or interfering with the communication channel by delaying and replaying messages. A system model is proven to be secure if it applies countermeasures against any possible attack defined in the threat model and launched by any corrupted user. But, it is often computationally expensive to account for all possible attacks.

\subsection{Security requirements}
Security requirements include the confidentiality and privacy of system inputs, the integrity of data and correctness of the evaluation, and the availability of data for authorized users. In threat modeling, attacks are often categorized based on which requirement they target. Subsequently, cryptographic primitives are applied to meet those requirements. Data confidentiality can be ensured by encryption, especially with those that offer semantic security, i.e., the ciphertext does not leak information about its plaintext. Data integrity can be checked by obliviously verifying data inputs. Suppose a classification protocol that requires users' inputs to be in the range $\{0,1\}$, and maliciously providing inputs not in this range may leak information about the evaluation model~\cite{wu2016privately}. To ensure integrity, each user must send additional information, in the form of Zero-Knowledge proofs, with their encrypted inputs to prove that the inputs are indeed valid without revealing them. Similarly, correctness of evaluation can be checked to prevent corrupted evaluator from applying unwanted functions. For example, the evaluation of a function can be represented as a graph with each computation as a node associated with a hash value. The correctness can be checked by running a proof, using zkSNARKs techniques~\cite{cryptoeprint:2013:507}, with the hash values generated during the evaluation. 

Another important requirement for security is key management. The life cycle of cryptographic keys includes \textit{generation}, \textit{distribution}, \textit{storage}, \textit{use}, and \textit{revocation}. It is vital to carefully manage keys through each of these phases because poor key management can easily defeats the purpose of using cryptography~\cite{ferguson2010cryptography}. In key generation, cryptographic parameters such as key bit-length must be chosen appropriately based on a security parameter $\lambda$ to protect against known attacks such as brute-force attack. Generated secret keys must be kept private by their rightful owners. Owners may provide information about their secret keys in a form of evaluation keys to perform homomorphic techniques like key switching or bootstrapping (discussed in Sec.~\ref{subsec:HEtech}). Those evaluation keys must be securely distributed and stored under encryption. The keys, or key pairs, should also be properly revoked when expired or changed across the system. Throughout this survey, we will focus on the case of asymmetric cryptosystems that generates key pairs, but the key management also applies to symmetric cryptosystems.

\subsection{Security assumptions}
\label{subsec:assumptions}
Secure protocols must take countermeasures against possible attacks. As mentioned, modeling threats and defining security requirements are two essential steps. However, designing for a practical setting often faces a trade-off between security and efficiency. Therefore, it is common to make security assumptions to achieve a more efficient design. For example, many HE schemes base their security on the Learning with Errors (LWE) problem (formally defined in Sec.~\ref{subsec:notation}). But studies show it is more efficient to operate in a ring of polynomials instead of matrices, which enables Single Instruction Multiple Data (SIMD) operations. Many recent HE schemes that show practical performance have their ring variant constructed under the ring-LWE assumption. Another example is in the distributed decryption process of a Threshold HE scheme~\cite{asharov2012multiparty}. In this scheme, $N$ participants create a joint public key, such that  $\pk^*=(\pk_1+\dots+\pk_N)$, and use it for encryption. The corresponding secret key is secretly shared among the $N$ participants. Thus, users must jointly construct the secret key in order to decrypt. There are two assumptions that are considered for this case. First, we may assume the presence of at most $N-1$ corrupted users. Similar to $(N,N)$ secret sharing schemes~\cite{shamir1979share}, this means \emph{all} $N$ users are required to participate in the decryption; otherwise, the ciphertext will not be correctly decrypted. This design is based on the dishonest-majority assumption. The second assumption is the honest-majority assumption, which relaxes the security to achieve efficiency. It allows a subset $T < N$ of the participants to reconstruct the corresponding secret key and collaboratively decrypt the ciphertext, i.e., $(T,N)$ scheme. 
\section{Preliminaries} 
\label{sec:prelim}
In this section, we define notations and definitions that will be used throughout this survey and provide background on general homomorphic primitives and techniques used in HE schemes.

\subsection{Notations and definitions}
\label{subsec:notation}
We denote vectors as bold lowercase, such as $\boldsymbol{a}$, and matrices as bold uppercase $\boldsymbol{B}$. Specifically, given a vector $\boldsymbol{a} = (a_1, \dots, a_n)$, we define $\boldsymbol{a}[i] = a_i$ as the $i$-th element. Let $n\times n'$ be the matrix dimensions and $\boldsymbol{B}[i,j]$ be the $j$-th element of the $i$-th row. The dot product of the two vectors $\boldsymbol{a}, \boldsymbol{b}$ is denoted by $\langle \boldsymbol{a},\boldsymbol{b}\rangle = \sum_{i=1}^{n}\boldsymbol{a}[i]\cdot \boldsymbol{b}[i]$, and the tensor product is denoted by $\boldsymbol{a}\otimes \boldsymbol{b}$. The multiplication of two elements $a,b$ is denoted by $ab$ or $a\cdot b$. For an element $a \in \RR$, $\near{a}$ denotes the rounding to its nearest integer, and $\ceil{a}, \floor{a}$ denote the rounding up and down, respectively. For a symmetric (asymmetric) encryption scheme, we denote the secret key and the public and private key pair as $\sk$ and $(\pk, \sk)$, respectively. For scheme-specific notations, we will introduce them when needed in the subsequent sections.

\begin{table}[t]
\centering
\caption{Description of notations used throughout the survey.}
\label{tab:notations}
\resizebox{.9\textwidth}{!}{
\begin{tabular}{|c|c|l|}
\hline
Category   &   Notation   &   \multicolumn{1}{c|}{Description}  \\ \hline
\multirow{5}{*}{General}   &   $\boldsymbol{a}$ &   A vector of $n$ elements where $a_i = \boldsymbol{a}[i]$ is the $i$-th element.
\\ \cline{2-3}    
&   $\boldsymbol{B}$ &   A matrix of elements where $b_{i,j}$ is the $j$-th element of the $i$-th row.
\\ \cline{2-3} 
% &   $|\boldsymbol{a}|$   &   \highlight{Asma: Norm is to be removed since it was never used}
%The $l_1$-norm of vector, $|\boldsymbol{a}| = \sum_{i=0}^{n}{a_i}$.\YS{why don't we use the standard notation $\|\cdot\|_1$?} 
% \\ \cline{2-3}
&   $\langle \boldsymbol{a},\boldsymbol{b}\rangle$    &   The dot-product of the two vectors $\boldsymbol{a},\boldsymbol{b}$ as $\langle \boldsymbol{a}, \boldsymbol{b}\rangle = \sum_{i=1}^{n}(a_ib_i)$.
\\ \cline{2-3}
&   $N$ &   Total number of users in the system.
\\ \hline
\multirow{5}{*}{(R)LWE-specific} &   $\Phi(x)$   & A cyclotomic polynomial $\Phi(X)=X^d+1$, where $d$ is a power of 2.
\\ \cline{2-3}
&   $R$ &   A ring over polynomials with integer coefficients $R=\ZZ[X]/(\Phi(X))$.  
\\ \cline{2-3}
&   $R_t$  &  The ring of polynomials with coefficients in $\ZZ_t$, where $t$ is plaintext modulus.
\\ \cline{2-3}
&   $R_q$  &   The ring of polynomials with coefficients in $\ZZ_q$, where $q$ is ciphertext modulus.
\\ \cline{2-3}
&   $\psi$  &   The key distribution over $R$. %\YS{needed to define RLWE}
\\ \cline{2-3}
&   $\chi$  &   The noise distribution over $R$ or $\mathbb{Z}$ with small standard deviation.
\\ \hline
\multirow{4}{*}{HE Scheme-specific} &   $\pk_i$ &   HE public key of the $i$-th user.
\\ \cline{2-3}
&   $\sk_i$ &   HE secret key of the $i$-th user.
\\ \cline{2-3}
&   $\ek_i$ &   HE evaluation key of the $i$-th user.
\\ \cline{2-3}
&   $[m]_i$ &   HE encryption of a message $m$ under $\pk_i$. Also denoted as $c$
\\ \hline
\end{tabular}
}
\end{table}

\begin{definition}[Learning with Errors (LWE)~\cite{regev2009lattices}]
For a security parameter $\lambda$, let $n=n(\lambda)$ be a dimension, $q=q(\lambda)\geq 2$ be an integer, and $\chi = \chi(\lambda)$ be an error distribution over $\ZZ$. For a secret $\boldsymbol{s} \in \ZZ_q^n$, let $\mathcal{D}$ be a distribution obtained by uniformly sampling $\boldsymbol{a}\sample \ZZ_q^n$ and $e\sample \chi$ and setting the pair as $(\boldsymbol{a},b = \langle \boldsymbol{a},\boldsymbol{s} \rangle + e)$.
Let $\mathcal{U}$ be the uniform distribution over $\ZZ_q^{n+1}$. The (decisional) LWE problem is to distinguish between the two distributions $\mathcal{D}$ and $\mathcal{U}$ for a fixed secret $\boldsymbol{s}$ sampled according to a key distribution $\psi=\psi(\lambda)$ over $\ZZ_q^n$.
% Sample a secret $\boldsymbol{s} \sample \ZZ_q^n$, and error $\e_i \sample \chi$.
% $\boldsymbol{a}_i \sample \ZZ_q^n$,\
%Sample a vector $\boldsymbol{a}_i \sample \ZZ_q^n$, $\boldsymbol{s} \sample \ZZ_q^n$, and $e_i \sample \chi$. The LWE problem is to distinguish the pair $(\boldsymbol{a}_i, b_i = \langle \boldsymbol{a}_i,\boldsymbol{s} \rangle + e_i) \in \ZZ_q^{n+1}$ from a uniformly sampled $(\boldsymbol{a}_i, b'_i)$ from $\ZZ_q^{n+1}$. The LWE assumption is that LWE problem is \emph{computationally} infeasible.
\end{definition}

The LWE problem operates in the $n$-dimensional integer space $\ZZ^n_q$ and consists of vectors and matrices. To achieve more efficient computations, the problem can be extended to rings of polynomials with integer coefficient. The aim is significantly reduce the dimension, such that $n = 1$, and correspondingly the size of keys and ciphertexts. 

\begin{definition}[Ring Learning with Errors (RLWE)~\cite{lyubashevsky2013ideal}] For a security parameter $\lambda$, let $\Phi(x)=x^d+1$ be a cyclotomic polynomial where $d=d(\lambda)$ is a power of $2$, and $q=q(\lambda)\geq 2$ be an integer. Define the ring $R$ over polynomials with integer coefficients $R=\ZZ[x]/(\Phi(x))$.
Let $\chi=\chi(\lambda)$ be an error distribution over $R$. For a secret $s\in R$, let $\mathcal{D}$ be a distribution over $R_q^2$ which samples $a \sample R_q$ and $e\sample \chi$ and returns the pair $(a,b = as+ e)$. Let $\mathcal{U}$ be the uniform distribution over $R_q^2$. The (decisional) RLWE problem is to distinguish between the two distributions $\mathcal{D}$ and $\mathcal{U}$ for a fixed secret $s$ sampled according to a key distribution $\psi=\psi(\lambda)$ over $R$.
%Let $a$ and $s$ be uniformly sampled elements from $R_q$, and let $e\sample\chi$ be a sampled error term. The RLWE problem is to distinguish the pair of $(a_i, b_i=a_is + e)$ from any uniformly sampled pair $(a_i, b'_i) \sample R^2_q$.
The RLWE assumption is that the RLWE problem is \emph{computationally} infeasible.
\end{definition}

%and bounded by $\mathcal{B}=\mathcal{B}(\lambda)$ such that $\mathcal{B} \ll q$

An amortized version of the RLWE problem~\cite{lyubashevsky2013ideal,applebaum2009fast} shows that it is equivalent to sampling $s$ from a small distribution $\psi$ instead of uniformly from the ring $R_q$. This yields a smaller secret key in an RLWE-based cryptosystem, e.g.
the BGV scheme~\cite{brakerski2012leveled}
% \fix{Pei: should we define the CRS model here because it is about the shared lattice element?}

\begin{definition}[Common Reference String (CRS)] For some distribution $D$, the CRS model starts in a trusted setup with sampling a value such that $d\sample D$. This value is a common reference string~\cite{damgaard2000efficient,canetti2001universally} (or a public parameter~\cite{fischlin2000efficient, pass2005unconditional}) that is made available to all participants before any computation starts.  
\end{definition}

In the RLWE-based schemes described in this paper, the CRS is a vector of elements uniformly sampled from $R_q$ which is provided as a public parameter. When computing with multiple keys, it is a requirement for those keys to be related in a manner to ensure correct computation and decryption. Hence, schemes are often designed in the CRS model where participants are given access to this public parameter to generate their individual public keys. More details on this model is discussed in Sec.~\ref{sec:multi}. 

\subsection{General homomorphic primitives} 
\label{subsec:primitives}
Homomorphic encryption (HE) is a class of encryption schemes that support computations such as addition and multiplication on encrypted data. Existing HE schemes can be divided into three main types based on the homomorphic operations supported by the evaluation function. In \emph{Partial HE} (PHE) schemes, the evaluation function supports either addition (i.e. additive homomorphism), such as Goldwasser-Micali~\cite{goldwasser1984probabilistic} and Paillier cryptosystem~\cite{paillier1999public}, or multiplication (i.e. multiplicative homomorphism), such as ElGamal cryptosystem \cite{elgamal1985public}, but \emph{not both}. In contrast, \emph{Fully HE} (FHE) allows arbitrary number of additions and multiplications. \emph{Somewhat HE} (SWHE) schemes support \emph{both} addition and multiplication on the ciphertexts. Yet, the number of multiplications allowed is limited due to the inherited construction of the scheme where ciphertexts contain noise that exponentially scales with multiplications. In general, HE scheme is a tuple of probabilistic polynomial-time (\textsf{PPT}) algorithms $\mathsf{HE} =(\kgen,\enc,\eval,\dec)$. We define each algorithm as follow. 
\begin{itemize}
\item[-] $\mathsf{HE}.\kgen(1^\lambda)\rightarrow (\pk, \sk)$: Given a security parameter $\lambda$ determining the security level, the key generation algorithm outputs a public key $\pk$, a private key $\sk$.
\item[-] $\mathsf{HE}.\enc(\pk, m)\rightarrow c$: Given a public key $\pk$ and a message $m$, the encryption algorithm outputs a ciphertext $c$. 
\item[-] $\mathsf{HE}.\eval(\pk, f, c, c')\rightarrow c_{\Eval}$: Given a public key $\pk$, two ciphertexts $c, c'$, and a homomorphic function $f$, the evaluation algorithm outputs the evaluated ciphertext $c_{\Eval} = f(c,c')$.
\item[-] $\mathsf{HE}.\dec(\sk,c)\rightarrow m$: Given a ciphertext $c$ encrypted under $\pk$ and the corresponding secret key $\sk$, the decryption algorithm outputs the message $m$. 
\end{itemize}

Note, the $\mathsf{HE}.\eval$ algorithm homomorphically performs a defined function $f$ on the ciphertexts. This function is constructed using $\mathsf{HE}.\evaladd$ and $\mathsf{HE}.\evalmult$ which are homomorphic addition and multiplication respectively. In this paper, we will focus on HE schemes that are based on the LWE problem and its ring variant, RLWE.

\subsection{Common techniques in HE} 
\label{subsec:HEtech}

% \subsection{Gadget and Bit Decomposition}
\subsubsection{Gadget Toolkits / Key Switching}
\label{subsec:keyswitching}
%  (rewritten) \YS{Please take a look at this subsection: all bitdecomp in the paper can be replaced by general gadget decomposition}  Asma: It looks great, thanks. 
In many of the RLWE-based HE schemes, the initial output of homomorphically multiplying two ciphertexts is a longer ciphertext encrypted under a new secret key element $s^2$. For instance, observe in the BGV scheme~\cite{brakerski2012leveled} presented in Scheme~\ref{scheme:BGV}, the product between two ciphertexts $\bc=(c_0, c_1)$, $\bc'=(c_0',c_1')$ is defined by
\begin{align*}
\tilde{\bc}_{\mult} = (c_0\cdot c'_0, c_0\cdot c'_1+c'_0\cdot c_1, c_1\cdot c'_1) = (\tilde c_0, \tilde c_1, \tilde c_2) \in R^3_{q}.
\end{align*}

The additional component $c_2$ corresponds to the quadratic element $s^2$ resulted from the multiplication $(c_0+c_1\cdot s)(c_0'+c_1'\cdot s)=\tilde c_0+\tilde c_1\cdot s+\tilde c_2\cdot s^2$. This new ciphertext $\tilde{c}_{\mult}$ is no longer decryptable by the secret key $s$. Therefore, HE schemes employ \emph{key switching}~\cite{brakerski2012leveled} as a transformation technique to reduce the dimension after each homomorphic multiplication. This transformation is accomplished with the aid of auxiliary information provided as evaluation key $\ek$ which encrypts $s^2$ under $s$.
The following \emph{gadget toolkit} is needed to perform the key switching operation:
\begin{itemize}
    \item[-] Gadget vector: $\bg=(g_0,\dots,g_{\ell-1})\in R^\ell$ for some integer $\ell\ge 1$.
    \item[-]$\decomp(x)$: Given an element $x\in R_q$, decompose it into a \emph{short} vector $\bu=(u_0, \dots, u_{\ell-1}) \in R^\ell$ such that $\langle \bu, \bg \rangle = x \pmod q$.
\end{itemize}
The decomposition function is often denoted by $\decomp(\cdot)=\bg^{-1}(\cdot)$ since $\langle \bg^{-1}(x), \bg \rangle = x \pmod q$ for all $x$.
There have been proposed several gadget toolkits in the literature. A typical example of the gadget vector is $\bg=(1,2,\dots,2^{\lceil \log q\rceil-1})$ which corresponds to the bit decomposition $\bitdecomp:x=\sum_{0\le i< \lceil  \log q\rceil}2^i\cdot u_i\mapsto \bu=(u_i)_{0\le i<\lceil \log q\rceil}$.

As discussed above, key switching is a commonly used building block of RLWE-based HE schemes %\ref{scheme:BFV},\ref{scheme:CKKS}) , BFV, and CKKS
which reduces the dimension of a long ciphertext after homomorphic multiplication.
Here for example, we review how to generate an evaluation key $\ek$ and perform the key switching operation in the BGV scheme on a three-dimensional ciphertext $\tilde{c}$ encrypted under $(1,s,s^2)$.

\begin{itemize}
    \item[-]$\evalkgen(s)$: The evaluation key is a special encryption of $s^2$ under $s$. The evaluation key $\ek = (\hat \bb, \hat \ba)\in R_q^{\ell\times 2}$ is generated during the scheme. It is generated by $\hat \ba\sample R_q^\ell$, $\hat \be\sample \chi^\ell$, and $\hat \bb = -s\cdot \hat \ba+t\cdot \hat \be +s^2\cdot \bg\pmod q$.
    \item[-]$\keyswitch(\ek, \tilde{c})$: Given an evaluated ciphertext $\tilde{c} = (\tilde{c}_0, \tilde{c}_1, \tilde{c}_2)$ encrypted under the secret element $s^2$, and the evaluation key $\ek=(\hat \bb, \hat \ba)$, compute $\bg^{-1}(\tilde c_2)=(u_0, \dots, u_{d-1})$. Then, return the ciphertext $(c_0,c_1)=(\tilde c_0, \tilde c_1)+ \sum_{0\le i<\ell} u_i\cdot (\hat \bb[i], \hat \ba[i]) \pmod q$.
\end{itemize}

Other RLWE-based HE schemes, such as BFV (Scheme~\ref{scheme:BFV}) and CKKS (Scheme~\ref{scheme:CKKS}), perform key switching similar to the BGV but with a few technical differences which will be described in Sec.~\ref{sec:single}.
This technique can be generalized and used for other purposes beside dimension reduction. For example, it can be employed in proxy re-encryption (see Sec.~\ref{subsec:PRE}) to transform a ciphertext from one encrypting a message $m$ under one key $s_1$ to one encrypting the same message $m$ under a different key $s_2$. It can also be used to facilitate the \emph{bootstrapping} step~\cite{gentry2009fully, brakerski2014efficient}, which accomplishes fully HE scheme from a leveled SWHE scheme.

\subsubsection{Bootstrapping}
\label{subsec:bootstrapping}
Following the blueprint proposed by Gentry~\cite{gentry2009fully}, one can construct a fully HE scheme from a somewhat HE scheme. Mainly, when a ciphertext reaches the maximum defined level where the noise in the ciphertext is large, a \emph{bootstrapping} technique is applied. The technique \emph{recrypts} the ciphertext by homomorphically evaluating the decryption circuit and outputting a fresh ciphertext with small noise. This new ciphertext can be further evaluated homomorphically. More technically, Let $c = \enc(\pk, m)$ be a ciphertext at the maximum level encrypted under $\pk$ and the corresponding secret key is $\sk$. To bootstrap, the user sends the evaluator a bootstrapping key, which is essentially an encryption of the secret key $\enc(\pk', \sk)$. Note, the secret key may be encrypted with the same key $\pk' = \pk$, but this requires using circular security assumption. A bootstrapped ciphertext can be obtained then by performing $\Check{c} = \bootstrap(\enc(\pk', c), \enc(\pk', \sk)) = \enc(\pk', \dec(\sk,c))$. Generally, this function has to be of a limited depth to be performed with the somewhat HE scheme. This mean the degree of the decryption polynomial must be low for the ciphertext to be bootstrapable~\cite{gentry2009fully}. Integer-based schemes, such as Gentry's scheme~\cite{gentry2009fully}, require a squashing technique which additionally makes a sparse subset-sum assumption to decrease the degree of the decryption polynomial. Recent LWE-based HE schemes, such as the BV scheme~\cite{brakerski2011efficient}, uses relinearization to reduce the ciphertext dimension after each multiplication. As a result, the decryption polynomial has a low degree and does not need squashing.

\subsubsection{Distributed decryption}
\label{subsec:distdec}
In many HE protocols based on the multi-party computation (MPC) protocol, the involved parties may be required to help decrypting the final evaluation result. For example, in threshold encryption (discussed in Sec.~\ref{subsec:THE}), the final result is encrypted under the combination of all the parties' keys; hence, each party must participate to partially decrypt the result with their own secret key. This process becomes a \emph{distributed decryption protocol} since the decryption now is executed as an MPC protocol. Two algorithms are preformed within this protocol. First, a partial decryption that is performed by each party who locally decrypts the result with their own secret key and shares the output. Then, a final decryption is performed by the designated party who aggregates the shared partial decryptions to obtain the final decrypted result. 

Two main security issues has to be addressed for this decryption protocol. First, the shared partially decrypted ciphertext may leak information on the party's secret key; hence, additional security measures, such as \emph{noise smudging} (or noise flooding), are considered to prevent this leakage. To make sure that no secret shares can be learned, we need to add larger errors following the Smudging Lemma~\cite{asharov2012multiparty, mukherjee2016two}, which states that adding a large noise ``smudges out'' the small values in the ciphertext. Hence, adding a large noise to the decryption component prevents leaking information about the secret share.

\begin{lemma}[Smudging Noise~\cite{asharov2012multiparty}]
Let $B_1=B_1(\lambda)$ and $B_2=B_2(\lambda)$ be two positive integers and let $e_0\in[-B_1,B_1]$ be a fixed integer. Let $e_1\sample [-B_2,B_2]$ be chosen uniformly at random. Then the distribution of $e_1$ is \emph{statistically} indistinguishable from that of $e_1 + e_0$ if $B_1/B_2 = \epsilon$, where $\epsilon = \epsilon(\lambda)$ is a negligible function. 
\end{lemma}

The second security issues is related to the final decryption algorithm, which may leak the encrypted result. Specifically, an attacker who falsely acquired the shared partial decryptions can perform the final decryption by combining the shares and retrieve the final result. 

\subsubsection{Noise reduction techniques}
\label{subsubsec:noisereduction}
% $c_i\cdot (1,s) = q\cdot I_i+\Delta m_i +e_i$

% $I_i=\lceil q^{-1}\cdot c_i\cdot(1,s)\rfloor$ its size depends on the key distribution.

% $(t/q)\cdot (c_1\cdot c_2) ((1,s)\cdot (1,s))=(t/q)\cdot (q\cdot I_1+\Delta m_1+e_1)(q\cdot I_2+\Delta m_2+e_2) = \Delta m_1m_2+e_{mult}$

% $e_{mult}=t(I_1e_2+I_2e_1)+...$

As a result of homomorphic computations, the embedded noise in ciphertexts increases in magnitude. For a noise magnitude $\mathcal{B}$, each homomorphic addition doubles the noise as $2B$, and each homomorphic multiplication squares it as $\mathcal{B}^2$. However, in BFV scheme (Sec.~\ref{subsubsec:BFV}) multiplication of two ciphertexts obtains $\langle\boldsymbol{c}_{\mult}, \sk \rangle = qI_\mult + \Delta m_1m_2 + e_{\mult}$ where the noise growth is dominated by $e_{\mult}\approx t(I_1e_2+I_2e_1)$ for $I_i \approx \lceil q^{-1}\langle\boldsymbol{c}_i, \sk\rangle\rfloor$. 
% However, in BFV scheme (Sec.~\ref{subsubsec:BFV}) the noise growth is dominated by $I_1\mathcal{B}+I_2\mathcal{B}$ where \red{$I_i=\Delta m_i - m_i$ is a factor introduced when we scale up the messages $m_i$ with $\Delta$}.
%\YS{BFV-like schemes have different noise growth.} 
To ensure correct decryption of the evaluated result, the noise must not exceed the $\frac{q}{2}$ range. We briefly discuss some of the noise reduction techniques used in HE schemes and refer the readers to~\cite{brakerski2012leveled,brakerski2012fully,cheon2017homomorphic} for more details. 

\paragraph{Modulus switching} Modulus switching is a technique proposed by Brakerski~\emph{et~al.}~\cite{brakerski2012leveled} to scale down the noise after each multiplication. For a $L$-depth circuit (i.e., requires at most $L$ multiplications), define $L+1$ moduli $\{q_L, q_{L-1},\dots, q_0\}$. After the $i$-th multiplication, we can transform a ciphertext $c\bmod q_i$ into the ciphertext $c'\bmod q_{i-1}$, where $q_{i-1}<q_i$, such that the noise scales down approximately by a factor of $\frac{q_{i-1}}{q_{i}}$. This brings the noise magnitude back to $\mathcal{B}$ and changes to a smaller modulus. This is proven to increase the number of supported homomorphic multiplications before the need for bootstrapping.

\paragraph{Scale-invariant} A following work of Brakerski~\cite{brakerski2012fully} proposed the scale-invariant technique which uses \emph{one} modulus $q$ to homomorphically evaluate $L$-depth circuit. This technique completely removes the need for modulus switching, which requires choosing $L+1$ decreasing moduli. In a nutshell, a message is scaled at encryption by a factor of $\Delta = \near{q/t}$, where $q$ is significantly larger than $t$; hence, the message is scaled up. After each homomorphic multiplication and at decryption, the ciphertext is scaled down by a factor of $(q/t)$, resulting in a significant decrease in the noise magnitude. The origin technique was proposed for LWE-based schemes with $t=2$, and was later ported~\cite{Fan:2012aa} to the RLWE assumption for efficiency. 

% \paragraph{scaling}
These noise reduction techniques are applied in the base HE schemes discussed in Sec.~\ref{sec:single} and can be readily extended to the multi-key (Sec.~\ref{sec:multi}) setting.
\section{Single-key approaches}
\label{sec:single}
Many existing work on homomorphic encryption assume data is always encrypted under one key in symmetric schemes or one key pair in asymmetric schemes. This key or key pair is typically owned by a user or an organization. However, in some scenarios such as in the multi-party and outsourced system models (in Figs.~\ref{fig:mpc} and~\ref{fig:outsourced}), data owners may want to reveal the homomorphically computed result to data consumers who should not be able to decrypt any input data. To accomplish this, additional step may be performed to transform the encrypted result to an encryption under a different key through proxy re-encryption (PRE). In other scenarios, we may want to generate decryption keys for a specific group of authorized users. In this case, we can construct protocols based on identity-based encryption (IBE) or attribute-based encryption (ABE). In this section, we review these single-key approaches and discuss their robustness in supporting secure computation. For simplicity, we list descriptions of the notations used throughout this section in Table~\ref{tab:SKnotations}.

\begin{table}[t]
\centering
\caption{Notations specific to single-key approaches.}
\label{tab:SKnotations}
\resizebox{\textwidth}{!}{
\begin{tabular}{|c|c|l|}
\hline
Category & Notation & \multicolumn{1}{c|}{Description}  \\ \hline
\multirow{6}{*}{HE-specific} & $\param$ & HE scheme's public parameters including security parameter $\lambda$ and circuit depth $L$. \\ \cline{2-3} 
& $\Delta$ & A scaling factor used in scale-invariant or rescaling techniques, $\Delta=\near{q/t}$ \\ \cline{2-3} 
% & $l$ & The $l$-th level of evaluated circuit, such that $l \in L$. \\ \cline{2-3} 
& $\ell$ & The gadget dimension.  \\ \cline{2-3} 
& $\boldsymbol{G}$ & The gadget matrix constructed as $\boldsymbol{G} = \boldsymbol{I}_n\otimes \boldsymbol{g}$ for a gadget vector $\boldsymbol{g}$.  \\ \cline{2-3} 
& $\boldsymbol{G}^{-1}(\cdot)$ & The bit-decomposition function (inverse gadget), such that $\boldsymbol{GG}^{-1}(a) = a$.\\ \hline 
PRE-specific & $\rk_{A \rightarrow B}$ & The re-encryption key switching an encryption from under Alice's key to Bob's key. \\ \hline
\multirow{3}{*}{IBE/ABE-specific} & $\mathcal{U}$ & The universe set of attributes/identities. \\ \cline{2-3} 
& $\omega'$ & A subset of attributes $\omega' \subset \mathcal{U}$ chosen for encrypting a ciphertext. \\ \cline{2-3} 
& $\omega$ & The set of attributes provided for decrypting a ciphertext.\\ \hline 
\end{tabular}
}
\end{table}

\subsection{Homomorphic encryption}
\label{subsec:HEscheme}
Many cloud-based applications can leverage homomorphic encryption to support secure computation outsourcing without compromising the privacy of user data. In majority of proposed HE protocols for these applications~\cite{naehrig2011can, yasuda2013secure, shan2018practical, yang2019comprehensive}, data used in computation is encrypted under the same key. The primary focus in these works is on the construction of efficient homomorphic algorithms to evaluate a targeted function in the encrypted domain. Applying these protocols in the outsourced system model (Fig.~\ref{fig:outsourced}), where privacy of individual data owners is important, requires making security assumptions such as semi-honest cloud and non-collusion between the cloud evaluator and participating parties. These assumptions obviously weaken the security of the protocol, especially in malicious settings where corruption or collusion compromises data privacy of other data owners. Ideally, users should keep private data protected under their own keys. However, supporting multi-key HE is complex and inefficient today. In this survey, we start with reviewing the construction of some well-known base HE schemes. Then, we review the techniques applied to support homomorphic computations on data encrypted under multiple keys in later sections. 

\subsubsection{The Brakerski-Gentry-Vaikuntanathan (BGV) scheme} \label{subsubsec:BGV}
Brakerski~et~al.~\cite{brakerski2012leveled}, proposed an efficient \emph{leveled} HE scheme to allow arbitrary number of additions but \emph{limited} consecutive multiplications determined according to the depth of the evaluated circuit. 

The BGV scheme is based on the original Brakerski and Vaikuntanathan's scheme~\cite{brakerski2011efficient}, which is the base of the second generation of HE schemes that improve the efficiency after Gentry's breakthrough~\cite{gentry2009fully}. The BV scheme was the first scheme basing its security solely on the hardness of standard LWE assumption, which has proven to be as hard as solving the shortest vector problem in lattices~\cite{lyubashevsky2010ideal}. With this reduction, the scheme removes the need for making additional strong assumptions, such as the secret subset sum assumption, and avoids the squashing technique in bootstrapping. The main construction starts with a SHWE scheme of depth $L$, which also considers the depth of its own decryption circuit, then converts to a fully HE scheme through bootstrapping. The original LWE-based BV scheme encrypts a message bit $m \in \{0,1\}$ with the secret $\boldsymbol{s} \in \ZZ^n_q$ as
$\boldsymbol{c} = (b=-\langle\boldsymbol{a},\boldsymbol{s}\rangle + m + 2e, \boldsymbol{a})\in \ZZ^{n+1}_q$, where $\boldsymbol{a}\in \ZZ^n_q$ is a random vector and $e$ is a small noise. Essentially, we observe the pattern of masking the message with a large element and some noise to hold the LWE assumption. To decrypt, we use the secret $\boldsymbol{s}$ and the provided vector $\boldsymbol{a}$ in the ciphertext to compute $b+\langle\boldsymbol{a},\boldsymbol{s}\rangle \pmod q$, which effectively removes the mask and obtains $m + 2e$.
The message is then retrieved by removing the even noise by computing $m=m + 2e \pmod 2$. In subsequent BV variants, such as the BGV scheme~\cite{brakerski2012leveled}, the message can be an integer in $\ZZ_t$, where $t$ is a chosen plaintext module that is significantly smaller than $q$. The noise is scaled at-most by $t$. 

The BGV scheme can be instantiated based on LWE or its ring variant RLWE, which operates on a ring or polynomials and is proven to be more efficient. It also applies optimizations such as the Smart-Vercauteren~\cite{smart2014fully} batching technique, which packs multiple plaintext messages into one ciphertext so that computations can be performed in a SIMD manner. For a circuit depth $L$ and public parameters $\param = (\lambda, d, q, t)$, we summarize the ring variant of the BGV scheme in Scheme~\ref{scheme:BGV}.

% \item[-] \YS{Revised some algorithms. It might be better to have setup algorithm which generates pp and decides key/error distributions}
% Given a security parameter $\lambda$ and a circuit depth $L$, choose  an error distribution $\chi$ bounded by $\mathcal{B}_\chi$, and a modulus $q$ such that the LWE problem holds. Set $n' = n\log(q)$ and choose a random matrix $\boldsymbol{B} \sample \ZZ^{n-1\times n'}_q$. Output the scheme public parameters as $\param = (q,n,n',\chi,\mathcal{B}_\chi,\boldsymbol{B})$. 

% \red{\chi \rightarrow \psi?}
% \red{R_t \rightarrow \psi?}

\begin{pabox}[label={scheme:BGV}]{The Brakerski-Gentry-Vaikuntanathan (BGV) scheme~\cite{brakerski2012leveled}}
\begin{itemize}[leftmargin=*]
\item[-] {$\mathsf{BGV.}\setup(1^\lambda,1^L)$}: Given the security parameter $\lambda$ and a multiplicative depth $L$, choose a cyclotomic polynomial $\Phi(x)=x^d+1$, where $d$ is a power of 2. Define $R=\mathbb{Z}[x]/(\Phi(x))$ as a polynomial ring of degree $d$ with integer coefficients. Generate the error and key distributions $\chi$ and $\psi$ over $R$, respectively. Choose the ciphertext modulus $q$ and the plaintext modulus $t$. Finally, output $\param = (d, q, t, \chi, \psi)$ as the public parameter. We assume all following algorithms implicitly take $\param$ as an input.

\item[-]{$\mathsf{BGV.}\kgen(\param)$}: Given the scheme's public parameters $\param$, sample a small element $s\sample \psi$ and a small noise $e \sample \chi$. Also uniformly sample $a \sample R_q$. Set the secret key as $\sk = s$ and the public key as $\pk= (b,a) \in R^2_q$ where $b=-as+te \pmod q$. 

\item[-] $\mathsf{BGV.}\evalkgen(s)$: Given the secret key $s$, generate the evaluation key $\ek$ by sampling $\tilde{\boldsymbol{a}} \sample R_q^\ell$, $\tilde{\boldsymbol{e}} \sample \chi^\ell$ and setting 
$\tilde{\boldsymbol{b}} = \tilde{\boldsymbol{a}}\cdot s+\tilde{\boldsymbol{e}} + \bg\cdot s^2 \pmod q$. Output $\ek = (\tilde{\boldsymbol{b}}, \tilde{\boldsymbol{a}})$.

\item[-]{$\mathsf{BGV.}\enc(\pk, m)$}: Given a plaintext message $m \in R_t$, a public key $\pk = (b, a)$, uniformly sample a random $r \sample \psi$ and errors $e_0,e_1\sample \chi$, and encrypt the message $m$ as $\boldsymbol{c}= (c_0, c_1) \in R^2_{q}$, where $c_0 = rb + m+te_0$ and $c_1 = ra+te_1$.
%For clarity, we omit the $r$ associated with the noise $rte$ and write $c_0 = ras + te + m$.

\item[-]{$\mathsf{BGV.}\dec(\sk, \boldsymbol{c})$}: Given a ciphertext $\boldsymbol{c} = (c_0, c_1) \in R^2_{q}$ and the secret key $\sk = s$, set $\boldsymbol{s} = (1, s) \in R^2_{q}$ and decrypt by computing $m=(\langle\boldsymbol{c},\boldsymbol{s}\rangle \pmod q) \pmod t \in R_t$.
%The decryption of a ciphertext in the BGV scheme is correct if and only if $(\tilde{m} \bmod t = m)$.

\item[-]{$\mathsf{BGV.}\evaladd(\boldsymbol{c}, \boldsymbol{c}')$}:
Adding two ciphertexts $\boldsymbol{c} = (c_0, c_1)$, $\boldsymbol{c}'= (c'_0, c'_1) $ results in $\boldsymbol{c}_{\add} = (c_0 + c'_0, c_1 + c'_1) \in R^2_q$.
%Decryption is still correct because $-(r+r')a$ can be canceled when using the secret key $\sk$.

\item[-]{$\mathsf{BGV.}\evalmult(\boldsymbol{c}, \boldsymbol{c}')$}:
Given two ciphertexts $\boldsymbol{c}, \boldsymbol{c}' \in R^2_q$, their homomorphic multiplication yields first an extended ciphertext $\tilde{\boldsymbol{c}}_{\mult} = (\tilde{c}_0, \tilde{c}_1, \tilde{c}_2)$ that is encrypted under the element $s^2$. 

\item[-]$\mathsf{BGV.}\relin(\ek, \tilde{\boldsymbol{c}}_{\mult})$: Given the evaluation key $\ek = (\tilde{\boldsymbol{b}}, \tilde{\boldsymbol{a}})$ and a long ciphertext $\tilde{\boldsymbol{c}}_{\mult}=(\tilde c_0, \tilde c_1, \tilde c_2)$, perform relinearization as follows:
%$\ek = (\boldsymbol{\hat{b}}, \boldsymbol{\hat{b}}) = (b_i = a_is + te_i + \powersoftwo(s^2, q), a_i \sample R_q)$ to perform relinearization (using key switching) as following.  special evaluation key must be provided by $\ek=(\tilde \bb, \tilde\ba)\leftarrow\evalkgen(s^2, s)\in R_q^{\ell\times 2}$ to 
\begin{itemize}
    \item[-] Apply gadget decomposition $\bg^{-1}(c_2)=(u_0,\dots,u_{\ell-1})$ as described in Sec.~\ref{subsec:keyswitching}.
%    \item[-] Compute $u = \langle\hat{c}_2, \boldsymbol{\hat{a}}\rangle$ and $v = \langle\hat{c}_2, \boldsymbol{\hat{b}}\rangle$. 
    \item[-] Output the new relinearized ciphertext as 
    $\bc_\mult = (\tilde c_0, \tilde c_1) + \sum_{i} u_i \cdot (\tilde \bb[i], \tilde \ba[i]) \in R^2_q$.
    %$\boldsymbol{c}_\mult = (\hat{c}_0 + v, \hat{c}_1 + u) \in R^2_q$.
\end{itemize}
\end{itemize}
\end{pabox}
% A special \emph{key switching} technique is used to transform it to a normal ciphertext $\boldsymbol{c}_{\mult}$ that is the product of two ciphertexts decryptable under $s$. We denote this process by $\boldsymbol{c}_{\mult} =\mathsf{KeySwitch}(\tilde{\boldsymbol{c}}_{\mult},\ek)$, where $\ek$ is a set of auxiliary keys used to perform the technique as discussed in Sec.~\ref{subsec:keyswitching}.

Additive homomorphisim is straightforward in BV-type schemes, but multiplicative homomorphisim is more complicated. Multiplying two ciphertexts $\boldsymbol{c}, \boldsymbol{c}'$ results in a higher dimensional ciphertext that is now encrypted under $s^2$ rather than $s$. Additionally, the noise within the ciphertext grows significantly (i.e., $e^2$). Without control, the noise can grow exponentially with respect to the number of multiplications. Hence, two new techniques were proposed, namely \emph{relinearization} and \emph{modulus switching}, to address these two issues. We briefly explain here how relinearization works and defer the modulus switching to the next subsection. The core technique used in relinearization is key switching (described in Sec.~\ref{subsec:keyswitching}), which will transform the encryption from $s^2$ back into $s$ without decryption.
Given the ciphertext product $\boldsymbol{\hat{c}}_\mult = (\hat{c}_0, \hat{c}_1, \hat{c}_2) \in R^3_q$, where $\hat{c}_0 = c_0c'_0$, $\hat{c}_1 = c_0c'_1 + c'_0c_1$, and $\hat{c}_2 = c_1c'_1$. The latter contains the product $mm'$ encrypted under $s^2$. Hence, we need an evaluation key $\ek = (\boldsymbol{\hat{b}}, \boldsymbol{\hat{a}})$, where $\boldsymbol{\hat{a}} \sample R_q^\ell$, $\be\leftarrow \chi^\ell$ and $\boldsymbol{\hat{b}} = -\hat \ba\cdot s + t \cdot\be + s^2\cdot \bg \pmod q$.
The evaluation key encrypts information about $s^2$ encrypted under $s$. For stronger security, the secret $s^2$ should be encrypted under a different secret $s'$ to avoid making the circular security assumption, which states that the security of the secret key is ensured under the protection of its own public key. The transformation of the ciphertext is done by computing the new relinearized ciphertext as following:
(1) apply the gadget decomposition algorithm to output the vector $\bg^{-1}(\hat c_2)$;
(2) compute the two dot products $u = \langle \bg^{-1}(\hat c_2), \hat \ba\rangle$ and $v = \langle \bg^{-1}(\hat c_2), \hat \bb\rangle$;
(3) finally, add the results $u,v$ to the first two elements and output the new ciphertext as $\boldsymbol{c}_\mult = (\hat{c}_0 + v, \hat{c}_1 + u)$. This is essentially a way of homomorphically computing the decryption $\hat c_2\cdot s^2$ with secret $s^2$ and embedding it in the ciphertext result, so it becomes decryptable with $s$.

\subsubsection{The Brakerski/Fan-Vercauteren (BFV) scheme} \label{subsubsec:BFV}

\begin{pabox}[label={scheme:BFV}]{The Brakerski/Fan-Vercauteren (BFV) scheme~\cite{brakerski2012fully,Fan:2012aa}}
\begin{itemize}[leftmargin=*]

\item[-] {$\mathsf{BFV.}\setup(1^\lambda,1^L)$}: Given the security parameter $\lambda$ and a multiplicative depth $L$, choose a cyclotomic polynomial $\Phi(x)=x^d+1$, where $d$ is a power of 2. Define $R=\mathbb{Z}[x]/(\Phi(x))$ as a polynomial ring of degree $d$ with integer coefficients. Generate the error distribution $\chi$ over $R$. Generate the key distribution $\psi$ over $R$.
Choose two integers $q$ and $t$ as the ciphertext and plaintext moduli, respectively.
Finally, output $\param = (d, q, t, \chi, \psi)$ as the public parameters of scheme. We assume all following algorithms implicitly take $\param$ as an input.

\item[-]{$\mathsf{BFV.}\kgen(\pp)$}: Given the public parameters $\pp$, sample a secret $s \sample \psi$, an element $a \sample R_q$, and a small noise $e \sample \chi$.
Set the secret key as $\sk = s$ and the public key as $\pk = (b, a) \in R_q^2$ where $b=-as+e \pmod q$.

\item[-]$\mathsf{BFV.}\evalkgen(s)$: Given the secret $s$, generate the evaluation key as $\ek = (\tilde{\boldsymbol{b}}, \tilde{\boldsymbol{a}})$, where $\tilde{\boldsymbol{a}} \sample R_q^\ell$, $\tilde{\boldsymbol{e}}\sample \chi^\ell$ and $\tilde{\boldsymbol{b}} = \tilde{\boldsymbol{a}}\cdot s+\tilde{\boldsymbol{e}} + \bg\cdot s^2 \pmod q$.

\item[-]{$\mathsf{BFV.}\enc(\pk, m)$}: Given a public key $\pk = (b,a)$ and a message $m \in R_t$, sample a random $r \sample \psi$ and noise elements $e_0, e_1 \sample \chi$.
Scaled up the message by a factor $\Delta$ such that $\Delta m = \floor{\frac{q}{t}} m$.
Encrypt the scaled message as $\boldsymbol{c}= (c_0, c_1) \in R^2_{q}$ where $c_0 = rb + e_0 + \Delta m$ and $c_1 = ra + e_1$.

\item[-]{$\mathsf{BFV.}\dec(\sk,\boldsymbol{c})$}: Given a ciphertext $\boldsymbol{c} = (c_0, c_1) \in R^2_{q}$ and the corresponding secret key $\sk$, set $\boldsymbol{s} = (1, s)$ and decrypt by computing $\near{\frac{t}{q}\langle\boldsymbol{c}, \boldsymbol{s}\rangle} \in R_t$. 
% compute $\langle\boldsymbol{c}, \boldsymbol{s}\rangle \in R_q$ as following. 
% \begin{align*}
% \langle\boldsymbol{c}, \boldsymbol{s}\rangle & = c_0 + c_1s = (r(as+e) + e_1 + \Delta m) + (-ra + e_2)s
% \\ &= (ras - ras) + (re + e_1 + e_2s) + \Delta m
% \\ &= \tilde{e} + \Delta m
% \end{align*}
% Then for a factor $\frac{t}{q}$, decrypt by computing $\near{\frac{t}{q}\langle\boldsymbol{c}, \boldsymbol{s}\rangle} \in R_t$. The proof of decryption correctness is as follows.
% \begin{align*}
% \near{\frac{t}{q} \langle\boldsymbol{c}, \boldsymbol{s}\rangle} &=  \near{\frac{t}{q}(\tilde{e} + \Delta m)} = \near{\frac{t}{q}\tilde{e} + \frac{t}{q}\Delta m} \\
% &= \highlight{\near{\frac{t}{q}\tilde{e}} + m'\approx m}~(\bmod~t)
% \end{align*}

\item[-]{$\mathsf{BFV.}\evaladd(\boldsymbol{c}, \boldsymbol{c}')$}:
To add two ciphertexts $\boldsymbol{c} = (c_0, c_1)$ and $\boldsymbol{c}'= (c'_0, c'_1)$, compute $\boldsymbol{c}_{\add} = (c_0 + c'_0, c_1 + c'_1) \pmod q$.

% \item[-] \red{evalkeygen and keyswitch}
\item[-]{$\mathsf{BFV.}\evalmult(\boldsymbol{c}, \boldsymbol{c}')$}: To multiply two ciphertexts $\boldsymbol{c}, \boldsymbol{c}' \in R^2_q$, compute first the ciphertext $\tilde{\boldsymbol{c}}_{\mult} = (\tilde{c}_0, \tilde{c}_1, \tilde{c}_2) \in R^3_q$ by $\tilde{c}_0 = \near{(t/q)\cdot (c_0c'_0)}$, $\tilde{c}_1 = \near{(t/q)\cdot (c_0c'_1 + c_1c'_0)}$, and $\tilde{c}_2 = \near{(t/q)\cdot (c_1c'_1)}$. 

\item[-]$\mathsf{BFV.}\relin(\ek, \tilde{\boldsymbol{c}}_{\mult})$: Given the evaluation key $\ek = (\tilde{\boldsymbol{b}}, \tilde{\boldsymbol{a}})$ and a long ciphertext $\tilde{\boldsymbol{c}}_{\mult}=(\tilde c_0, \tilde c_1, \tilde c_2)$, perform relinearization as follows:
%$\ek = (\boldsymbol{\hat{b}}, \boldsymbol{\hat{b}}) = (b_i = a_is + te_i + \powersoftwo(s^2, q), a_i \sample R_q)$ to perform relinearization (using key switching) as following.  special evaluation key must be provided by $\ek=(\tilde \bb, \tilde\ba)\leftarrow\evalkgen(s^2, s)\in R_q^{\ell\times 2}$ to 
\begin{itemize}
    \item[-] Apply gadget decomposition $\bg^{-1}(c_2)=(u_0,\dots,u_{\ell-1})$.
    \item[-] Output the ciphertext 
    $\bc_\mult = (\tilde c_0, \tilde c_1) + \sum_{i} u_i \cdot (\tilde \bb[i], \tilde \ba[i]) \in R^2_q$.
    %$\boldsymbol{c}_\mult = (\hat{c}_0 + v, \hat{c}_1 + u) \in R^2_q$.
\end{itemize}
% The ciphertext is encrypted under the element $s^2$ and need to be transformed to one under element $s$ through the \emph{key switching} technique. We denote this process by $\boldsymbol{c}_{\mult} =\mathsf{KeySwitch}(\tilde{\boldsymbol{c}}_{\mult},\ek)$.  
\end{itemize}
\end{pabox}

Similar to the BGV scheme, the BFV scheme~\cite{brakerski2012fully, Fan:2012aa} is also designed based on the original BV scheme~\cite{brakerski2011efficient}. Yet, instead of the modulus switching, it introduces an approach, called \emph{scale-invariant}, to reduce noise accumulated through homomorphic multiplications. The BFV scheme also adopts the original BV's key switching technique for relinearization, which brings back the initial ciphertext product to be an encryption under $s$ as explained in the previous subsection. We present the RLWE-based BFV scheme in Scheme~\ref{scheme:BFV} for a circuit depth $L$ and a set of public parameters $\param = \{\lambda, d, q, t\}$.

The intuition behind scale-invariance is that elements' features should not change when scaled with a common factor. Given two messages $m$ and $m'$ scaled by a shared factor $\Delta$. If we aggregate $\Delta m + \Delta m'$ and scaled back by $\frac{1}{\Delta}$, then the result is roughly $m + m'$, i.e., the relation is preserved regardless to the scale. Now, let $\Delta = \near{q/t}$ be the factor used to scale up the message, since $q$ is significantly larger than $t$, at encryption then add an initial noise term as described in $\mathsf{BFV}.\enc$. If we decrypt, we need to scale back by the factor $\frac{t}{q}$ to retrieve the message $m$. As a result, the noise is significantly reduced by this scaling down operation. Observe the following proof of decryption correctness for the decryption $\near{\frac{t}{q}\langle\boldsymbol{c}, \boldsymbol{s}\rangle} \in R_t$, where we compute $\langle\boldsymbol{c}, \boldsymbol{s}\rangle$ as follows.
\begin{align*}
\langle\boldsymbol{c}, \boldsymbol{s}\rangle & = c_0 + c_1s = (rb+e_0 + \Delta m) + (ra + e_1)s
\\ &= r(b+as) + (e_0 + e_1s) + \Delta m = \tilde{e} + \Delta m \pmod q
\end{align*}
for a small error $\tilde e=re+e_0+e_1s$.
Then, we re-scale the result by a factor $\frac{t}{q}$ as follows to obtain the message $m$. 
\begin{align*}
\near{\frac{t}{q} \langle\boldsymbol{c}, \boldsymbol{s}\rangle} &=  \near{\frac{t}{q}(\tilde{e} + \Delta m)} = m \pmod t
\end{align*}

We can use the scaling aspect to effectively reduce the noise after homomorphic evaluations -- namely multiplication, which causes the embedded noise in the ciphertext to grow. In many HE schemes, the noise growth from multiplication is dominated by $e_1e_2$ from the two original ciphertexts $\boldsymbol{c_1}$ and $\boldsymbol{c_2}$, but in BFV the noise growth is dominated by $I_1e_2+I_2e_1$ where $I_i \approx \lceil q^{-1}\langle\boldsymbol{c}_i, \sk\rangle\rfloor$.% is a factor introduced when we scale up the messages with $\Delta$.
%Because $\Delta=\floor{\frac{q}{t}}$ and $q>>t$, then $I_1e_2+I_2e_1$ is larger than $e_1e_2$.
Moreover, multiplying two ciphertexts results in squaring the scaling factor, such that we get $\Delta^2(mm')$. To address this issue, the ciphertext must be scaled down by a factor of $\frac{t}{q}$ after each multiplication to reduce the noise and brings the result scale back to $\Delta(mm')$. Note, we could have divided the ciphertext by the factor of $\Delta$. However, this will incur twice the rounding error. We may not effectively reduce the large noise in the ciphertext scaled during multiplication; hence, we directly scale by $\frac{t}{q}$ instead. Homomorphic addition does not require a scaling step because the sum of two ciphertexts directly obtains $\Delta(m+m')$ and does not change the shared factor.

Unlike modulus switching, we observe right away that scale-invariance requires \textit{one} modulus $q$ instead of a sequence of decreasing $L+1$  moduli $q_L > q_{L-1} > \dots, q_0$. The modulus switching used in the BGV scheme performs a gradual scaling after each multiplication to map a ciphertext from $R_{q_i}$ to $R_{q_{i-1}}$. This causes the ciphertext to be scaled by a factor of $\frac{q_i}{q_{i-1}}$. This means the noise magnitude after each multiplication remains constant, but at the cost of changing to a smaller modulus $q$. 

\subsubsection{The Gentry-Sahai-Waters (GSW) scheme} \label{subsubsec:GSW} % \YS{Notation has not been unified througout the section}

Gentry~et al.~\cite{gentry2013homomorphic} proposed another LWE-based HE scheme based on eigenvectors and eigenvalues of a transformation matrix. This new scheme is simpler and asymptotically faster than other BV variants because it does not require the complex relinearnization for dimension reduction and modulus switching or scale-invariant for noise reduction. The intuition can be illustrated as follow. Given a transformation matrix (ciphertext), $\boldsymbol{C}$, and an eigenvector (secret key), $\boldsymbol{s}$, we can retrieve an eigenvalue (message), $m$, such that $\boldsymbol{s}\boldsymbol{C} = m\boldsymbol{s} \pmod q$. From this toy construction, homomorphic addition and multiplication of two ciphertexts $\boldsymbol{C}$ and $\boldsymbol{C}'$, encrypted under the same secret $\boldsymbol{s}$, can be realized easily by computing $\boldsymbol{s}(\boldsymbol{C}+ \boldsymbol{C}') = (m + m') \boldsymbol{s} \pmod q$ and $\boldsymbol{s}(\boldsymbol{C}\cdot\boldsymbol{C}') = (mm')\cdot \boldsymbol{s} \pmod q$, respectively.

However, this toy construction is insecure as it can be easily broken because finding eigenvectors for a given transformation matrix is as easy as solving a system of linear equations. To address this problem, the actual GSW scheme is built on the \emph{approximated eigenvectors} method, where a small noise $e$ is added to satisfy the LWE assumption, such that $\boldsymbol{s}\boldsymbol{C} = m\boldsymbol{s} + e ~(\mod q)$. Note, we can view $\boldsymbol{s}\boldsymbol{C}$ as the decryption of some ciphertext $\boldsymbol{C}$ containing a message $m$ with a secret key $\boldsymbol{s}$ associated to a public key $\boldsymbol{A}$. 

\begin{pabox}[label={scheme:GSW}]{\highlight{A variant of the Gentry-Sahai-Waters} (GSW) scheme~\cite{alperin2014faster}}
\begin{itemize}[leftmargin=*]
\item[-]{$\mathsf{GSW.}\setup(1^\lambda,1^L)$}: Given a security parameter $\lambda$ and a circuit depth $L$, choose a lattice dimension $n$, an error distribution $\chi$, and a modulus $q$ such that the LWE problem holds. Set $\ell = \ceil{\log q}$. Output the scheme public parameters as $\param = (q,n,\chi)$. % bounded by $\mathcal{B}_\chi$
% \YS{Does it sample B in the setup phase? I think it should be done in KeyGen}

\item[-]{$\mathsf{GSW.}\kgen(\param)$}: Sample a secret vector $\boldsymbol{\grave{s}}\sample \ZZ^{n-1}_q$ and the noise vector $\boldsymbol{e} \sample \chi^{n\ell}$. \highlight{Choose a random matrix $\boldsymbol{B} \sample \ZZ^{n-1\times n\ell}_q$ and set $\boldsymbol{b} = \boldsymbol{\grave{s}}\boldsymbol{B} + \boldsymbol{e} \in \ZZ^{n\ell}_q$}. Output the secret key $\sk$ as $\boldsymbol{s} = (1,-\boldsymbol{\grave{s}}) \in \ZZ^n_q$ and the public key $\pk$ as $ \boldsymbol{A} = (\boldsymbol{b}, \boldsymbol{B}) \in \ZZ^{n\times n\ell}_q$, observe that $\boldsymbol{s}\boldsymbol{A} \approx 0$ with respect to some small noise. 

\item[-]{$\mathsf{GSW.}\enc(\pk, m)$}: Let $m \sample \{0,1\}$ be a message bit and $\pk$ be a given public key. To encrypt the message, choose a random matrix $\boldsymbol{R}\sample \{0,1\}^{n\ell\times n\ell}$ compute the ciphertext as $\boldsymbol{C} = \boldsymbol{AR} + m\boldsymbol{G} \in\ZZ^{n\times n\ell}_q$, such that the property $\boldsymbol{s}\boldsymbol{C} \approx m \boldsymbol{s}\boldsymbol{G}$ holds.

\item[-]{$\mathsf{GSW}.\dec(\sk, \boldsymbol{C})$}: To decrypt a given ciphertext $\boldsymbol{C}$ with the associated secret key $\sk$, define a vector $\boldsymbol{w} = (0,\dots,0,\ceil{q/2}) \in \ZZ^n_q$. Then compute $\boldsymbol{x} = \boldsymbol{s}\boldsymbol{C}\boldsymbol{G}^{-1}(\boldsymbol{w}^T)\in \ZZ^{n\ell}_q$. Output the decrypted message as $\tilde{m} = \abs{\near{\frac{\boldsymbol{x}}{q/2}}}$. 

\item[-]{$\mathsf{GSW.}\evaladd(\boldsymbol{C}, \boldsymbol{C}')$}: To add two ciphertexts $\boldsymbol{C}=(\boldsymbol{A}\boldsymbol{R} + m \boldsymbol{G}), \boldsymbol{C}'=(\boldsymbol{A}\boldsymbol{R}' + m' \boldsymbol{G})$, directly output $\boldsymbol{C}_{\add} = \boldsymbol{C} + \boldsymbol{C}' = \boldsymbol{A}(\boldsymbol{R}+\boldsymbol{R}') + (m+m')\boldsymbol{G} \in\ZZ^{n\times n\ell}_q$, such that $\boldsymbol{s}\boldsymbol{C}_{\add} \approx (m+m')\boldsymbol{s}\boldsymbol{G}$.

\item[-]{$\mathsf{GSW.}\evalmult(\boldsymbol{C}, \boldsymbol{C}')$}: To multiply two ciphertexts $\boldsymbol{C}, \boldsymbol{C}'$, utilize the gadget matrix and output $\boldsymbol{C}_{\mult} = \boldsymbol{C} \cdot \boldsymbol{C}' = \boldsymbol{C}\boldsymbol{G}^{-1}(\boldsymbol{C}') \in\ZZ^{n\times n\ell}_q$.The product matrix satisfies $\boldsymbol{s}\boldsymbol{C}_{\mult} \approx \boldsymbol{s}\boldsymbol{C}\boldsymbol{G}^{-1}(\boldsymbol{C}') \approx m \boldsymbol{s}\boldsymbol{G}\boldsymbol{C}' \approx (m\cdot m')\boldsymbol{s}\boldsymbol{G}$. 
\end{itemize}
\end{pabox}

Like other LWE-based schemes, the public key in GSW is generated from an LWE instance over $\ZZ_q$. For a given secret vector $\boldsymbol{\grave{s}}$ and a uniformly chosen matrix $\boldsymbol{B}$, we set the public key as $\boldsymbol{A} = (\boldsymbol{B}\boldsymbol{\grave{s}} + \boldsymbol{e}, \boldsymbol{B})$ and the secret key as $\boldsymbol{s} = (1, -\boldsymbol{\grave{s}})$. When the secret key $\boldsymbol{s}$ is provided at decryption, we have $\boldsymbol{s}\boldsymbol{A} \approx 0$ because  $\boldsymbol{s}\boldsymbol{A} = (1, -\boldsymbol{\grave{s}})(\boldsymbol{B}\boldsymbol{\grave{s}} + \boldsymbol{e}, \boldsymbol{B}) =  \boldsymbol{B}\boldsymbol{\grave{s}} + \boldsymbol{e} - \boldsymbol{B}\boldsymbol{\grave{s}} = \boldsymbol{e} \approx 0 $. This operation removes the masking matrix $\boldsymbol{B}$ in the ciphertext. To encrypt a message $m$ under the public key $\boldsymbol{A}$, we sample a random matrix $\boldsymbol{R}$, to ensure semantic security of the scheme, and compute $\boldsymbol{C} = \boldsymbol{A}\boldsymbol{R} + m$ such that the decryption $\boldsymbol{s}\boldsymbol{C} = \boldsymbol{s}(\boldsymbol{A}\boldsymbol{R} + m) = \boldsymbol{s}\boldsymbol{A}\boldsymbol{R} + m\boldsymbol{s} = m\boldsymbol{s} + e\boldsymbol{R}$, where $e\boldsymbol{R}$ is small when $\boldsymbol{R}$ has entries in $\ZZ_2$. If $\boldsymbol{s}$ is an integer vector and $m$ is restricted to the message space $\{0,1\}$, we can determine the original message from $(m\boldsymbol{s} + e)$ with high probability. Specifically, if $m = 0$, then $\boldsymbol{s}\boldsymbol{C}$ is close to $0$; or an integer vector otherwise. 

Homomorphic evaluations are performed as natural matrix operations. The product of multiplying two GSW ciphertexts does not contain a long secret $s^2$. Therefore, it does not require the complex relinearization operation used in the previous BV variant schemes. Yet, homomorphic multiplication still increases the noise. The product of multiplying two GSW ciphertexts $\boldsymbol{C}$ and $\boldsymbol{C}'$ is
\begin{align*}
\boldsymbol{s}\boldsymbol{C}_\mult &= \boldsymbol{s}\boldsymbol{C}G^{-1}(\boldsymbol{C}') = (m\boldsymbol{s} + \boldsymbol{e})\boldsymbol{C}' \\
& = m(m'\boldsymbol{s} + \boldsymbol{e}') + \boldsymbol{C}'\boldsymbol{e} \\
& = mm'\boldsymbol{s} + (m \boldsymbol{e}' + \boldsymbol{C}'\boldsymbol{e})
\end{align*}

The new error term is $(m \boldsymbol{e}' + \boldsymbol{e}\boldsymbol{C}')$. Multiplication works if this term is small. While $\boldsymbol{e}$ and $\boldsymbol{e}'$ are small noise, we need to make $m$ and $\boldsymbol{C}'$ small. The former can be achieved by restricting input message, $m \in \{0, 1\}$, but $\boldsymbol{C}' \pmod q$ is large making $\boldsymbol{e}\boldsymbol{C}'$ a large noise. Without optimizations, the noise growth for multiplying two ciphertexts is exponential. % and bounded around \red{$(n\ell+1)\mathcal{B}^2$} for some bound $\mathcal{B}$. 
This shows a significant noise increase compared to homomorphic addition that is bounded by $2\mathcal{B}$. To achieve a better noise management, the scheme applies a \emph{flattening} technique to transform $\boldsymbol{C}'$ with integer coefficient module $q$ into bits through $\bitdecomp$ and $\powersoftwo$, as discussed in Section~\ref{subsec:HEtech}. Hence, the noise growth becomes linear and bounded by $(n\ell+1)\mathcal{B}$ as a result of multiplying the noise term with small values in $\ZZ_2$. 

A variant of GSW~\cite{alperin2014faster} later suggested the use of special \emph{gadget matrix} for a simpler design with a tighter bound on the noise growth. The gadget matrix $\boldsymbol{G}$ is a diagonal matrix containing powers of two $\boldsymbol{g}= (1,2,4,\dots,2^{\ell-1})$, where $\ell = \ceil{\log q}$ and associated with a function $\boldsymbol{G}^{-1}(\cdot)$ such that $\boldsymbol{G}\boldsymbol{G}^{-1}(a) = a$ for any given $a\in \ZZ_q$. It also proposed a bootstrapping technique for GSW to support arbitrary homomorphic evaluations. We describe the main functions of this GSW variant in Scheme~\ref{scheme:GSW}.

% \highlight{
Due to the construction of the GSW, there is another way to decrypt and retrieve the message bit $m\in\{0,1\}$. More specifically, if we view a GSW ciphertext $\boldsymbol{C} = \boldsymbol{AR} + m\boldsymbol{G}$ as two aggregated parts, a randomized public key $\boldsymbol{AR}$ and a message $m$ masked by a gadget matrix $\boldsymbol{G}$ represented in the powers-of-2 form.
Then the $i$-th column $\boldsymbol{c}_i$ of this ciphertext contains the encryption of $m2^{i}$ and can be decrypted by computing $\langle\boldsymbol{c}_i, \boldsymbol{s}\rangle \pmod 2$. The collection of entries are required for homomorphic evaluations, but decryption can be done using a single column. The penultimate (i.e., second-to-last)  column corresponds to encryption of $m\cdot 2^{\ell-2}$, which decrypts to plaintext in the interval $(q/4, q/2]$ appropriate for decoding the message bit. Hence, extracting the penultimate column is sufficient to use for decryption. To illustrate, we provide a proof of correctness for the GSW decryption as follows. Let $\boldsymbol{w} = (0,0,\dots, \ceil{q/2})$ be the penultimate column of the gadget matrix $\boldsymbol{G}$ corresponding to the power $2^{\ell-2}$ and compute $\boldsymbol{x}$ as following.
% }
\begin{align*}
    \boldsymbol{x} &= \boldsymbol{s}\boldsymbol{C}\boldsymbol{G}^{-1}(\boldsymbol{w}^T)= \boldsymbol{s}(\boldsymbol{A}\boldsymbol{R} + m \boldsymbol{G})\boldsymbol{G}^{-1}(\boldsymbol{w}^T) \\
    &= m \boldsymbol{s}\boldsymbol{G}\boldsymbol{G}^{-1}(\boldsymbol{w}^T) \\ 
    &= m\boldsymbol{s}(\boldsymbol{w}^T)
\end{align*}
% \highlight{
This process extracts the penultimate column of the GSW ciphertexts, which is sufficient to for decryption. Compute $m = \abs{\near{\frac{\boldsymbol{x}}{q/2}}}$ and retrieve the message $\tilde{m}\in\{0,1\}$ after checking whether it is closer to $0$ or $q/2$. Using a single column from the ciphertext to decrypt certainly has its benefits. One of them is reducing the size of the ciphertext results transmitted over the network, especially in cases we discuss later in Sec.~\ref{subsec:MKHE}, where ciphertexts are extended to multiple keys and may suffer exponential increase in their size. More importantly, this property provides a mechanism to decrypt a ciphertext without fully revealing information about the embedded secret key. We review in later sections how this property is exploited to build a secure proxy re-encryption that is based on the GSW scheme.
% }

% #todo
% \subsubsection{\red{The TFHE scheme}}

% fixed-point arithmetic, significand and scaling factors, encoding and decoding, noise and approximation error, modulus switching and rescaling
\subsubsection{The Cheon-Kim-Kim-Song (CKKS) scheme} \label{subsubsec:CKKS}
Previous HE schemes support linear transformations of encrypted integers or bits through homomorphic addition and multiplication. However, many machine learning algorithms, such as logistic regression and neural networks, often operate on real numbers (e.g., $12.345$). In integer-based HE schemes, a common way to deal with these real numbers is to scale them up by a relatively large factor (e.g., $10^{6}$) prior to encryption so that we only need to work with integers (e.g., $12.345 \times 10^{6}$). The scaling dramatically increases the input values and subsequently requires a much larger plaintext modulus and even larger ciphertext modulus. As a result, the computation time will be significantly impacted.

Cheon~et~al.~\cite{cheon2017homomorphic} proposed a HE scheme that supports fixed-point arithmetic over encrypted real numbers. The scheme bases its security on the RLWE problem and is based on a construction that is similar to the rest of BV-type schemes discussed in Sections~\ref{subsubsec:BGV} and \ref{subsubsec:BFV}. The main intuition is treating the noise embedded at encryption, to realize the LWE assumption, as the rounding error in fixed-point arithmetic since real numbers inherently carry error through rounding. We give an overview of the RLWE-based CKKS scheme in Scheme~\ref{scheme:CKKS}.

\begin{pabox}[label={scheme:CKKS}]{\highlight{The Cheon-Kim-Kim-Song (CKKS) scheme}~\cite{cheon2017homomorphic}}
\begin{itemize}[leftmargin=*]

\item[-] {$\mathsf{CKKS.}\setup(1^\lambda,1^L)$}: Given the security parameter $\lambda$ and a multiplicative depth $L$, choose a cyclotomic polynomial $\Phi(x)=x^d+1$, where $d$ is a power of 2. Define $R=\mathbb{Z}[x]/(\Phi(x))$ as a polynomial ring of degree $d$ with integer coefficients. Generate the error distribution $\chi$ over $R$. Generate the key distribution $\psi$ over $R$. Choose an integer $q$ as the ciphertext modulus. Finally, output $\param = (d, q, \chi, \psi)$ as the public parameters of scheme. We assume all following algorithms implicitly take $\param$ as an input. 
%Uniformly sample a vector $\highlight{\boldsymbol{a \sample R_{q}$.

\item[-]{$\mathsf{CKKS.}\kgen(\param)$}: Given the public parameters $\pp$, sample a small element $s \sample \chi$ and a small noise $e \sample \psi$.%, where $\chi$ is a Gaussian distribution over $R_q$.
Uniformly sample $a \sample R_q$. Output the secret key as $\sk = s$ and the public key as $\pk= (b,a) \in R^2_q$ where $b=-as+e \pmod q$.
Sample $\tilde{\boldsymbol{a}} \sample R_q^\ell$, $\tilde{\boldsymbol{e}}\sample \chi^\ell$ and set
$\tilde{\boldsymbol{b}} = \tilde{\boldsymbol{a}}\cdot s+\tilde{\boldsymbol{e}} + \bg\cdot s^2 \pmod q$.
Output $\ek = (\tilde{\boldsymbol{b}}, \tilde{\boldsymbol{a}})$ as the evaluation key.

% \{Encoding/Decoding}
\item[-]$\mathsf{CKKS.}\encode(\boldsymbol{z}, \Delta)$: 
Given a vector of complex numbers $\boldsymbol{z}=(z_1,\dots,z_{d/2})$, the encoding algorithm applies canonical embedding mapping into the polynomial message such that $m(X) = \sigma^{-1}(\lfloor \Delta\cdot\pi^{-1}(\boldsymbol{z})\rceil)$, using the canonical embedding map $\sigma$ and the natural projection $\pi: \mathbb{H} \rightarrow \mathbb{C}^{d/2}$ and $\mathbb{H} = \{(z_j)_{j\in\mathbb{Z}^*_M} : z_{-j}=\overline{z_j}, \forall j\in \mathbb{Z}^*_{M}\}$.

\item[-]$\mathsf{CKKS.}\decode(m,\Delta)$: Given an encoded polynomial message $m(X) \in R$, apply the decoding algorithm to transform $m$ into a complex vector $\boldsymbol{z}$ such that $\boldsymbol{z}=\pi \circ \sigma(\Delta^{-1} \cdot m)$.

\item[-]{$\mathsf{CKKS.}\enc(\pk, m)$}: Given a public key $\pk$ and a plaintext message $m$, uniformly sample a random $r \sample \psi$ and noise elements $e_0, e_1 \in \chi$. Encrypt the message $m$ as $\boldsymbol{c} = (c_0,c_1) \in R^2_{q}$ where $c_0 = rb + e_0 + m$ and $c_1 = ra + e_1$.

\item[-]{$\mathsf{CKKS}.\dec(\sk, \boldsymbol{c})$}: Given a ciphertext $\boldsymbol{c} = (c_0, c_1) \in R^2_{q}$ and the secret key $\sk = s$, set $\boldsymbol{s} = (1, s)$ and decrypt by computing $m=\langle\boldsymbol{c}, \boldsymbol{s}\rangle \pmod q$.

\item[-]{$\mathsf{CKKS.}\evaladd(\boldsymbol{c}, \boldsymbol{c}')$}:
For two ciphertexts $\boldsymbol{c} = (c_0, c_1)$ and $\boldsymbol{c}'= (c'_0, c'_1)$, output $\boldsymbol{c}_{\add} = (c_0+c'_0, c_1 + c'_1) \in R^2_q$.
%where, $(c_0+c'_0) = (r+r')(as+e)+(e_1+e'_1) + (m+m')$ and $(c_1 + c'_1) = -(r+r')a + (e_2+e'_2)$.

\item[-]{$\mathsf{CKKS.}\evalmult(\boldsymbol{c}, \boldsymbol{c}')$}: For two ciphertexts $\boldsymbol{c}, \boldsymbol{c}' \in R^2_q$, their initial homomorphic multiplication yields a long ciphertext $\tilde{\boldsymbol{c}}_\mult = (\tilde c_0, \tilde c_1, \tilde c_2) \in R^3_q$ encrypted under the secret $(1,s,s^2)$.
%The ciphertext can be decryption as $mm' \approx (c_0 + c_1s + c_2s^2)$.

\item[-]$\mathsf{CKKS.}\relin(\ek, \tilde{\boldsymbol{c}}_\mult)$: Given the evaluation key $\ek = (\tilde{\boldsymbol{b}},\tilde{\boldsymbol{a}})$ and a long ciphertext $\tilde{\boldsymbol{c}}_\mult = (\tilde c_0, \tilde c_1, \tilde c_2)$, apply key switching and output $\boldsymbol{c}_{\mult} =(c_0, c_1)+(\langle \bg^{-1}(c_2), \tilde{\boldsymbol{b}}\rangle, \langle\bg^{-1}(c_2), \tilde{\boldsymbol{a}}\rangle) \in R^2_q$.

\item[-]{$\mathsf{CKKS.}\rescale(\boldsymbol{c})$}: For an evaluated ciphertext $\boldsymbol{c} = (c_0, c_1) \in R^2_{q}$, compute $c'_i = \near{\Delta^{-1}\cdot c_i}$ for $i \in \{0,1\}$ and a scaling factor $\Delta$ and return the re-scaled ciphertext $\boldsymbol{c}' = (c'_0, c'_1) \in R^2_{q'}$ where $q'=\Delta^{-1}\cdot q$.
\end{itemize}
\end{pabox}

% The key pair are generated similar to other BV variants, where a small secret $s$, a random ring element $a$, and a small noise are sampled and used to generate the public key $\pk = (as+e, -a)$, such that the secret $s$ remains protected under the LWE assumption. 

\begin{figure}[t]
\centering
\includegraphics[width=.5\textwidth]{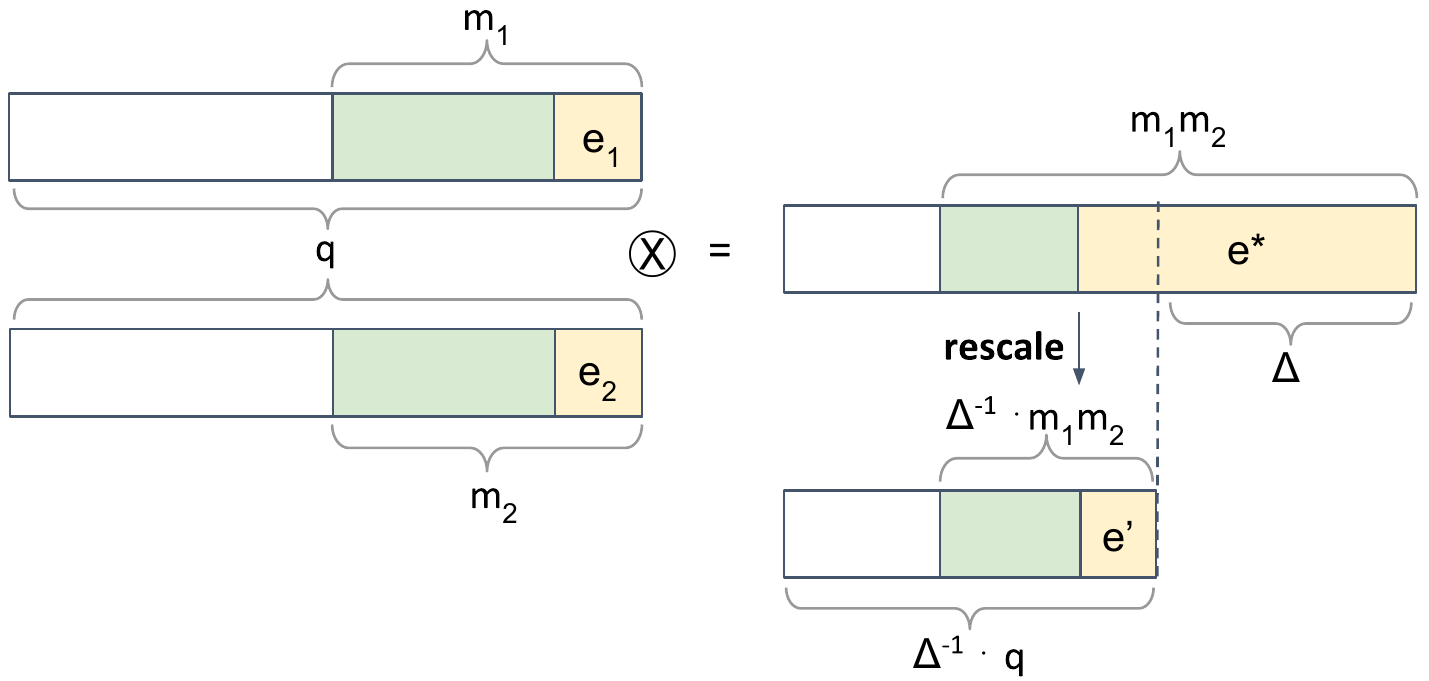}
\caption{Illustration of ciphertext re-scaling in CKKS}
\label{fig:ckks}
\end{figure}

%  $\sigma: \mathcal{S} \rightarrow \CC^N$ (in the form of polynomials with real coefficients in the ring $\mathcal{S}$).
What is unique in the CKKS scheme is the proposal of new message encoding and decoding methods and a rescaling operation that reduces the accumulating noise in a similar way as rounding in plaintext fixed-point arithmetic.
Before encryption, the encoding method applies a canonical embedding to map a vector of complex or real numbers into a polynomial in $R$. Let $d$ being the degree of the cyclotomic polynomial $\Phi(x)=x^d+1$ and determines the number of slots for ciphertext packing, the encoding procedure allows $d/2$ messages $(z_1, z_2, \dots, z_{d/2})$ to be packed into a polynomial $m'\in R$ so that efficient evaluation can be carried out in SIMD manner. If the number of messages in the vector is less than $d/2$, the remaining slots are set to zero. The other half of slots in $m'$ will be filled with conjugates of the corresponding message $z_i$ during the canonical embedding. The polynomial $m'$ is then 
scaled by a large factor $\Delta$, say $\Delta = 2^{40}$, to place the significand far away from the LSBs (least significant bits). Note, placing significand far away from the LSBs is to accommodate the noise growth and to preserve precision during the rescaling operation, as illustrated in Fig.~\ref{fig:ckks}. The output of the encoding method is a plaintext $m\in R$ containing the packed messages scaled by a factor $\Delta$.  At encryption, the plaintext $m$ is masked with a randomized public key. A small noise $e$ is added to ensure the security, such that $\boldsymbol{c} = r(b,a)+(e_0,+m, e_1)$. If we decrypt the ciphertext, we obtain the message $\dec(\sk, \boldsymbol{c}) = m+e$, where $e$ is a small noise which can be considered as small rounding error inherited from fixed-point arithmetic. The decrypted plaintext is then decoded from a polynomial in $R$ to a real or complex message vector $(z_1,z_2,\dots,z_{d/2})$ using a reverse procedure of the encoding.

Homomorphic multiplication causes the scaling factor to square and increases embedded noise. In order to maintain the same precision and prevents the noise from blowing up, the result must be ``\emph{rounded}'' homomorphically. To do so, the scheme rescales the ciphertext after each multiplication. This method can be viewed as similar to the rounding performed on the plaintexts in approximate computations. In particular, the ciphertext $\boldsymbol{c}$ can is multiplied by the scaling factor, such that $\Delta^{-1}\boldsymbol{c}$, which discards $\log \Delta$ least significant bits as illustrated in Fig.~\ref{fig:ckks}.
As a result, we obtain a ciphertext encrypting the message $\frac{m_\mult}{\Delta}$ with a reduced noise $\Delta^{-1}\cdot{e_\mult}$, and the modulus is switched to $q'=\Delta^{-1}\cdot q$ corresponding to the new scaled ciphertext. CKKS uses a technique similar to modulus switching to perform rescaling. For a circuit depth $L$, choose a large modulus $q_0$ and define a decreasing chain of moduli $q_L,q_{L-1},\dots,q_0$, such that the modulus for a level $l \in L$ is defined as $q_l = q_0\cdot\Delta^{l}$. A ciphertext $\boldsymbol{c}~(\mod q_l)$ is rescaled as $\near{\Delta^{-1}\cdot\boldsymbol{c}} \pmod {q_{l-1}}$.
%Based on the definition of the two moduli, the scaling factor $\frac{q_{l-1}}{q_l}$ essentially divides the ciphertext by $\Delta$ because $\frac{q_{l-1}}{q_l} = \frac{q_0\cdot\Delta^{l-1}}{q_0\cdot\Delta^{l}} = \frac{1}{\Delta}$.
The modulus switching technique used here is for a different purpose compared to its use in the BGV. While it preserves the underlying plaintext and manages the noise growth in BGV, it is used in CKKS to remove the LSBs of the ciphertext to maintain the precision. Because of this recurring discard of the LSBs, we need to choose appropriate parameters for the modulus $q_0$ and the scaling factor $\Delta$ according to the circuit depth $L$ to prevent precision loss. Specifically, we ensure that $(\log q_L \geq L \cdot \log \Delta)$ to construct a leveled HE scheme sufficient to evaluate an $L$-depth circuit. After reaching the level $q_0$, bootstrapping~\cite{cheon2018bootstrapping,chen2019improved} may be applied to output a refreshed ciphertext at a higher level that supports further homomorphic operations. % decryption, mention decoding
% refine the reasoning for scaling

% RNS technique improves the efficiency because it decomposes the large modulus with long bits into a product of smaller moduli so they fit into current systems (double-CRT technique). 
A subsequent construction of the CKKS~\cite{cheon2018full} proposed a variant based on the Residue Number System (RNS) to optimize the performance. Most HE scheme implementations requires that the modulus $q$ is at least $128$ bits (often significantly larger), which may not fit in the native $64$-bit modern hardware. As an optimization, they apply Chinese Remainder Theorem (CRT) and similar techniques to represent the modulus as a set of moduli less than 64-bit. The RNS-variant decomposes the large modulus $q$ into a sequence of smaller distinct prime moduli $q'_0, q'_1,\dots, q'_L$, such that $q = \prod^{L}_{i=0}q'_i$. By choosing a chain of moduli that are pairwise coprimes, we can no longer perform the original rescaling technique since the moduli do not correspond to the scaling factor $\Delta$. To address this issue, the RNS-variant proposes an \textit{approximate} rescaling technique, which chooses the primes such that their ratio is as close as the scaling factor.
For example, the prime moduli $q'_l$ is chosen such that $q_l'\approx \Delta$ so that $\frac{q_l}{q_{l-1}} = q'_l \approx \Delta$. After homomorphic multiplication, the ciphertext product
with a squared scale $\Delta^2$ is rescaled by the factor $\frac{q_l}{q_{l-1}}$ to a ciphertext encrypting of $m_1m_2$ with an approximate scale $\Delta$. The new ciphertext contains an additional small approximation error, but the most significant bits of the plaintext remain extractable.

% TFHE

\subsubsection{Discussion}
\label{subsubsec:HEDiscussion}
From reviewing these base HE schemes, we observe similarities in their design pattern and methods to decrease in ciphertext dimension and to deal with noise. We share these similarities and discuss some initial ideas to support homomorphic computation on ciphertexts encrypted under multiple keys.

In term of similarity, the most recent HE schemes are constructed based on the LWE problem, or its ring variant, RLWE. The first unique characteristic is in the way that some small noise is added when masking a secret. Another characteristic is that the public key or the ciphertext contains a counter-part that can cancel the large masking element sampled from some mathematical space. These two characteristics are clearly visible in the $\kgen$ and $\enc$ algorithms of the presented HE schemes. Homomorphic multiplication is highly inefficient because it often raises the ciphertext dimension from an encryption of $s$ to $s^2$ (except GSW due to its construction) and dramatically increases the embedded noise. Hence, focuses are on designing techniques to reduce the ciphertext dimension and noise. Among these techniques, key switching used in dimension reduction is useful to support MKHE. Whereas common noise reduction techniques, such as modulus switching in BGV, scale invariant in BFV, rescaling in CKKS, follow a typical pattern of factorizing the noise by a scaling factor after each multiplication. Often, these reduction techniques have impact on the embedded messages, hence plaintext messages are typically scaled up before encryption. 

There are two common tricks we can apply to extend the base HE schemes to support homomorphic computation on ciphertexts encrypted using multiple keys. 

The first trick is to follow a concept called \emph{layered encryption}~\cite{lopez2012fly}. Let $c$ be a ciphertext encrypting a message $m$ under Alice's key $\pk_A$. We can add second layer of encryption under Bob's key $\pk_B$, such that $\tilde{c} = \enc(\pk_B, \enc(\pk_A, m))$. In theory, the ciphertext can be further extended indefinitely. Obviously, the problem is at decryption. Bob requires Alice, and other involved parties if any, to remove the corresponding layer of encryption. Homomorphic evaluations can be applied, but all ciphertexts must be encrypted in the same order. The number of evaluation is very restricted due to exponential increase of ciphertext size and noise magnitude as we further discuss in Sec.~\ref{sec:multi}. 

On the other hand, one may leverage \emph{key switching} to change the encryption of a message from one key $\pk$ to another $\pk'$. This method follows the notion of proxy re-encryption technique we discuss in the following section. Recall that a special evaluation key $\ek_{\pk\rightarrow \pk'}$ which embeds information about both keys, we can perform $\keyswitch(\ek_{\pk\rightarrow \pk'}, \enc(\pk, m)) = \enc(\pk', m)$ to achieve this purpose. The major issue in this method is to ensure the security of the secret key $\sk$ because the generation of the evaluation key requires encrypting $\sk$ under the new key $\pk'$. If the secret key $\sk'$ is owned by another party, then he or she can obtain $\sk$ violating the privacy requirement of data encrypted under $\pk$.

These simple tricks are neither secure nor efficient in practice. Hence, we review additional techniques which offer more than just simple extension of the base HE schemes.

\subsection{Proxy Re-encryption}
\label{subsec:PRE}
Changing a ciphertext from the encryption of one key to another \emph{without} exposing the plaintext is a useful feature in many applications. Especially, in cloud computing applications where a user Alice stores her encrypted data in the cloud as illustrated in Fig.~\ref{fig:outsourced} and another user Bob wants to perform some evaluations and retrieve the result from the cloud. It is reasonable to assume that Alice does not want to share her private key. 

Without Alice's private key, Bob will not be able to decrypt the result. To address this problem, we can design the system using an interactive model. The cloud can blind the homomorphically evaluated ciphertext with a randomness and sends it to Alice, who decrypts it with $\sk_A$ and sends back another ciphertext re-encrypted under $\pk_B$. Upon receiving the new ciphertext, the cloud evaluator homomorphically removes the blinding element and continues remaining evaluations or forwards the result to Bob, who can decrypt with $\sk_B$. Although this method is also applicable for changing the encryption under different schemes as needed, e.g., to convert from a SWHE scheme to a PHE scheme for efficiency~\cite{bost2014machine}, it is cumbersome due to the interactiveness.  

However, changing the encryption under different keys within the same scheme can be done non-interactively. This gives the advantage of fully outsourcing computations to the cloud, i.e., Alice does not need to participate in evaluations.

\begin{figure}
\centering
\includegraphics[width=.9\textwidth]{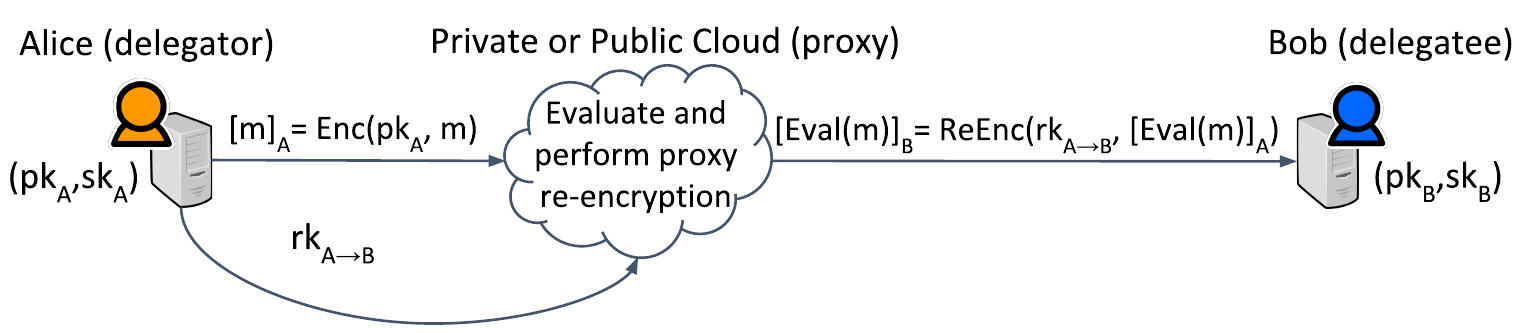}
\caption{Illustration of proxy re-encryption.}
\label{fig:proxy}
\end{figure}

Proxy re-encryption (PRE) is the process of converting a ciphertext from an encryption under a public key $\pk$ to one under another $\pk'$ without decryption. A third party (the \textit{proxy}) is delegated to perform this re-encryption without disclosing the underlying message or the secret key $\sk$. Consider the previous cloud application addressed by the PRE scheme as shown in Fig.~\ref{fig:proxy}. With a PRE scheme, Alice (the \textit{delegator}) provides to the proxy cloud a special re-encryption key $\rk_{A\rightarrow B}$ used to re-encrypt the ciphertext, so it becomes decryptable by Bob (the \textit{delegatee}). The key is generated based on her secret key and Bob's public key, such as $\rk_{A\rightarrow B}=\rekeygen(\sk_A, \pk_B)$. Variations of proposed PRE schemes can be unidirectional $\rk_{A\rightarrow B}$~\cite{libert2008unidirectional, shao2009cca}, where Alice's ciphertext can be only converted into Bob's, but not the other way around, or bidirectional $\rk_{A\leftrightarrow B}$~\cite{blaze1998divertible, canetti1999efficient}, where the ciphertext can be converted from Alice's to Bob's, or vice versa. We focus on the former in this review. 

The formal construction of PRE consists of five \textsf{PPT} algorithms, $\mathsf{PRE} =$ ($\kgen$, $\rekeygen$, $\enc$, $\reenc$, $\dec$). Existing HE schemes can be extended to PRE by adding the two PRE-specific algorithms $(\rekeygen, \reenc)$. We briefly describe these algorithms with the PRE variant of ElGamal scheme in Scheme~\ref{scheme:ElGamal} as an example. The generation of the re-encryption key in $\rekeygen$ is scheme-dependent, i.e., it depends on the mathematical construction of the base scheme. For example, the key $\rk$ in ElGamal scheme is generated as $\rk_{A\rightarrow B} = g^{s_B/s_A}$ from Bob's public key $\pk_B = g^{s_B}$ and Alice's secret key $\sk_A = s_A$. The re-encryption process leverages basic rules for exponentiation to remove $s_A$ from the encryption. Regarding security, recall that ElGamal bases its security on the discrete logarithm problem\footnote{For a carefully chosen cyclic group $\mathbb{G} = \ZZ_p^\times$ with a prime modulus $p$ and a generator $g$, the problem states that given an element $h = g^x \in \mathbb{G}$, finding the discrete logarithm $x$ is difficult.}. This mean it is difficult for the proxy, or Bob, to learn the secret key $s_A$ from the re-encryption key.

\begin{pabox}[label={scheme:ElGamal}]{A PRE-variant of ElGamal scheme~\cite{blaze1998divertible}}
\begin{itemize}[leftmargin=*]
\item[-]{$\mathsf{ElGamalPRE.}\kgen(\mathbb{G},p,g)$}: Let $\mathbb{G}$ be a cyclic group of order $p$ and a generator $g$. Sample two random values $s_A, s_B\sample \ZZ_p$ and compute $h_A = g^{s_A}, h_B = g^{s_B}$. Set the key pairs $\{\pk_A=(\mathbb{G},p,g,h_A), \sk_A = s_A\}$ for Alice, and $\{\pk_B=(\mathbb{G},p,g,h_B), \sk_B = s_B\}$ for Bob. 

\item[-]{$\mathsf{ElGamalPRE.}\rekeygen(\sk_A, \pk_B)$}: To allow Bob to transform a ciphertext under Alice's public key to one under his, Alice generates a re-encryption key $\rk_{A\rightarrow B}$. Given Alice's own secret key $\sk_A$ and Bob's public key $\pk_B$, compute $\rk_{A\rightarrow B} = \pk_B^{1/\sk_A} = (g^{s_B})^{1/s_A} = g^{s_B/s_A}$  

\item[-]{$\mathsf{ElGamalPRE.}\enc(\pk_A, m)$}: To encrypt a given a message $m$ under Alice's public key, choose a random value $r\in\{1,\dots,p-1\}$ and compute the ciphertext $\boldsymbol{c} = (c_1, c_2) = (g^r, m h_A^r)$. 

\item[-]{$\mathsf{ElGamalPRE.}\reenc(\rk_{A\rightarrow B}, \boldsymbol{c})$}: Given a ciphertext $\boldsymbol{c}=(c_1, c_2)$ and a re-encryption key $\rk_{A\rightarrow B}$, a proxy can re-encrypt the ciphertext as $c_2 \cdot \rk_{A\rightarrow B} = (m g^{rs_A} \cdot g^{s_B/s_A}) = m g^{r(s_As_B/s_A)} = m g^{rs_B}$ to obtain $\boldsymbol{\hat{C}} = (c_1, c'_2)$ under Bob's public key.

\item[-]{$\mathsf{ElGamalPRE.}\dec(\sk_B,\boldsymbol{\hat{C}})$}: Given a re-encrypted ciphertext $\boldsymbol{\hat{C}} = (c_1, c'_2)$ and a corresponding secret key $\sk_B$, Bob decrypts by performing $c'_2/c_1^{s_B} = (m g^{rs_B})/(g^r)^{s_B} = m$. 
\end{itemize}
\end{pabox}

In LWE-based HE schemes, performing key switching obtains a re-encrypted message under a different key without revealing it. The process requires an evaluation key, which is conceptually the same as re-encryption key in PRE. In this context, we will refer to the evaluation key as the re-encryption key. As discussed in Sec.~\ref{subsec:HEscheme}, the technique is often utilized for relinearization to convert initial results of homomorphic multiplication from one under $s^2$ back to one under $s$. The re-encryption key is generated as a special encryption of $s^2$ under $s$. For example in the BGV scheme, the re-encryption key is generated by the party who has access to both keys $\sk = s^2$ and $\pk' = (a's'+te', -a')$. To avoid making circular security assumption, BGV chooses $\pk'$ to be new public key containing a different secret from $s$ for each evaluation level. The security of the secret $s^2$ is ensured by the LWE problem. 

In theory, any HE scheme can be extended to PRE by following the key switching technique. The PALISADE software library~\cite{PALISADE} implements the PRE primitive for its HE schemes, including BGV, BFV, and CKKS, where the re-encryption key is generated based on two provided secret keys. However in practice, due to the scheme construction based on the LWE problem, the secret key is retrievable if one can decrypt the re-encryption key. Specifically, Alice generates $\rk_{A\rightarrow B}$ as an encryption of her secret key $\sk_A$ under Bob's public key $\pk_B$. This design is insecure since Bob may collude with the proxy and obtain $\sk_A$ after decrypting $\rk_{A\rightarrow B}$ with his secret key $\sk_B$. The re-encrypted message under $\pk_B$ may also leak information about $\sk_A$ when decrypted. Therefore, additional measures must be taken when extending LWE-based schemes to PRE to prevent leaking the delegator's secret key.

% generate rk, rencrypt, decrypt, discuss issue, solution, impacts, conclude/link to MK.
% \highlight{
Recently, Yasuda~et~al.~\cite{yasuda2018multi} extended a GSW variant~\cite{gentry2013homomorphic, peikert2016multi} to PRE using key switching. We describe this PRE variant of GSW in Scheme~\ref{scheme:GSWPRE}. Alice generates $\rk_{A\rightarrow B}$ as a GSW encryption of $\sk_A$ under Bob's public key $\pk_B$ and sends it to the proxy. Note, decrypting the re-encryption key with Bob's secret key $\sk_B = \boldsymbol{s}_B$ yields Alice's secret key according to this proof.
% }
\begin{align*}
  \boldsymbol{s}_B\rk_{A\rightarrow B} &= \boldsymbol{s}_B( \boldsymbol{A}_B\boldsymbol{X} + \begin{pmatrix} \boldsymbol{s}_A\boldsymbol{G} \\ \boldsymbol{0}_{(n-1)\times n\ell})  \end{pmatrix} 
  = \boldsymbol{s}_B\boldsymbol{A}_B\boldsymbol{X} + (1, -\boldsymbol{\grave{s}}_B)\begin{pmatrix}\boldsymbol{s}_A\boldsymbol{G} \\ \boldsymbol{0}_{(n-1)\times n\ell})\end{pmatrix} 
  \approx \boldsymbol{s}_A\boldsymbol{G}
\end{align*}
% \highlight{ \fix{illustrate this penultimate column} 
Given a ciphertext $\boldsymbol{C}$ under Alice's public key, the proxy cloud re-encrypts it as $\boldsymbol{\hat{C}} = \rk_{A\rightarrow B}\cdot\boldsymbol{G}^{-1}(\boldsymbol{C})$. However, if Bob receives and decrypts $\boldsymbol{\hat{C}}$, he will be able to learn information about Alice's secret key from the plaintext $\boldsymbol{s}_B\boldsymbol{\hat{C}} \approx m\boldsymbol{s}_A\boldsymbol{G}$. Instead, the PRE scheme leverages the special GSW property where a ciphertext can be decrypted using only its penultimate column, which corresponds the power-of-two $2^{\ell-2}$. The last element of this column contains \textit{partial} elements of the re-encrypted ciphertext, $\boldsymbol{\hat{C}} = \rk_{A\rightarrow B}\cdot\boldsymbol{G}^{-1}(\boldsymbol{C})$, which contains in its last element the message $m2^{\ell-2}$. The proxy sends the penultimate column $\boldsymbol{\hat{c}}$ of the re-encrypted ciphertext to Bob. With releasing this column only, we not only reduce the size of the ciphertext, but also withhold possible information about $s_A$ to prevent its leakage. This design renders a single-hop PRE scheme, which means $\boldsymbol{\hat{c}}$ cannot be re-encrypted or homomorphically evaluated after its re-encryption because it loses the GSW ciphertext structure. Still, it is useful to use the PRE primitive with other extended GSW schemes, such as the multi-key variant~\cite{peikert2016multi} (discussed later in Sec.~\ref{subsec:MKHE}) where the final result is re-encrypted under the receiver's key to avoid distributed decryption.
% }

\begin{pabox}[label={scheme:GSWPRE}]{A PRE-variant of GSW scheme~\cite{yasuda2018multi}}
\begin{itemize}[leftmargin=*]
\item[-]{$\mathsf{GSWPRE.}\setup(1^\lambda,1^L)$}: Given a security parameter $\lambda$ and a circuit depth $L$, choose a lattice dimension $n$, an error distribution $\chi$, and a modulus $q$ such that the LWE problem holds. Set $\ell = \ceil{\log q}$ and choose a random matrix $\boldsymbol{B} \sample \ZZ^{n-1\times n\ell}_q$. Output the scheme public parameters as $\param = (q,n,\chi,\boldsymbol{B})$.

\item[-]{$\mathsf{GSWPRE.}\kgen(\param)$}: Given a security parameter $\lambda$, circuit depth $L$, and a GSW public parameters $\param$, sample a secret vector $\boldsymbol{\grave{s}}\sample \ZZ^{n-1}_q$ and the noise vector $\boldsymbol{e} \sample \chi^{n\ell}$. Set $\boldsymbol{b} = \boldsymbol{\grave{s}}\boldsymbol{B} + \boldsymbol{e} \in \ZZ^{n\ell}_q$. Output the secret key $\sk$ as $\boldsymbol{s} = (1,-\boldsymbol{\grave{s}}) \in \ZZ^n_q$ and the public key $\pk$ as $ \boldsymbol{A} = (\boldsymbol{b}, \boldsymbol{B}) \in \ZZ^{n\times n\ell}_q$, observe that $\boldsymbol{s}\boldsymbol{A} \approx 0$ with respect to some small noise. 

\item[-]{$\mathsf{GSWPRE.}\rekeygen(\sk_A, \pk_B)$}: Given Alice's own secret key $\sk_A = \boldsymbol{s}_A$ and Bob's public key $\pk_B = \boldsymbol{A}_B$, Alice generates a re-encryption key $\rk_{A\rightarrow B}$ as $\rk_{A\rightarrow B} = \boldsymbol{A}_B\boldsymbol{X} + \begin{pmatrix} \boldsymbol{s}_A\boldsymbol{G} \\ \boldsymbol{0}_{(n-1)\times n\ell}\end{pmatrix}$, where $\boldsymbol{X}$  is a random binary matrix. Observe that the property $\boldsymbol{s}_B\cdot\rk_{A\rightarrow B} \approx \boldsymbol{s}_A\boldsymbol{G}$ holds. 

\item[-]{$\mathsf{GSWPRE.}\enc(\pk_A, m)$}: To encrypt a given a message $m$ under Alice's public key $\pk_A$, choose a random matrix $\boldsymbol{R}\sample \{0,1\}^{n\ell\times n\ell}$ and compute the ciphertext as $\boldsymbol{C} = \boldsymbol{A}_A\boldsymbol{R} + m\boldsymbol{G} \in\ZZ^{n\times n\ell}_q$, such that the propriety $\boldsymbol{s}_A\boldsymbol{C} \approx m \boldsymbol{s}_A\boldsymbol{G}$ holds.

\item[-]{$\mathsf{GSWPRE.}\reenc(\rk_{A\rightarrow B}, \boldsymbol{C})$}: Given a ciphertext $\boldsymbol{C}$ and a re-encryption key $\rk_{A\rightarrow B}$, a proxy can re-encrypt the ciphertext by computing $\boldsymbol{\hat{C}} = \rk_{A\rightarrow B}\cdot\boldsymbol{G}^{-1}(\boldsymbol{C})$. Output the re-encrypted ciphertext as the penultimate column $\boldsymbol{\hat{c}}$ in $\boldsymbol{\hat{C}}$. % expand 

\item[-]{$\mathsf{GSWPRE.}\dec(\sk_B,\boldsymbol{\hat{C}})$}: Given a re-encrypted ciphertext $\boldsymbol{\hat{c}}$ and the corresponding secret key $\sk_B$, Bob decrypts as $\langle \boldsymbol{\hat{c}}, \boldsymbol{s}_B\rangle (\mod 2)$.
\end{itemize}
\end{pabox}
 
% link PRE to IBE/ABE as a closing statement
% \highlight{
Proxy re-encryption is a powerful method to support secure outsourced evaluations. Ciphertexts encrypted under different keys can be individually re-encrypted under a receiver's (delegatee's) key so they can be homomorphically evaluated under the same key. Other applications, such as secure file sharing, may use this technique for basic access control to make ciphertexts decryptable with authorized user's secret keys, but it requires providing re-encryption keys to do so. We discuss an alternative approach to support this access control functionality without re-encryption keys in the following section.
% }

\subsection{Identity-/Attribute-based Encryption}
\label{subsec:ABE}
Identity-based encryption (IBE)~\cite{shamir1984identity, boneh2001identity}, and its generalization attribute-based encryption (ABE)~\cite{sahai2005fuzzy}, is a type of encryption scheme that provides a fine-grained access control to the encrypted message based on users' identities or attributes. 

Compared to the PRE scheme, IBE/ABE schemes do not require data owners to provide re-encryption keys for changing the ciphertexts encrypted under their public key to ones encrypted under the intended user's key. A data owner can encrypt their data under a derived public key based on predefined identities or attributes for authorized users. In IBE schemes, the data owner encrypts the data under a key derived from a user's identity $\mathsf{ID}$, e.g., their email address. The ciphertext can be only decrypted with an authorized secret key $\sk_{\mathsf{ID}}$ issued corresponding to the user's identity $\mathsf{ID}$. ABE schemes encrypt data according to a set of attributes predefined by the data owner and often described as an \textit{access policy structure}, such that the ciphertext can only be decrypted by a user whose attributes satisfy this policy. The IBE scheme can viewed as a special type of ABE scheme where the sole defined attribute is the identity of the user. Henceforth, we focus our descriptions on the ABE scheme.  

\begin{figure}[t]
\centering
\includegraphics[width=.5\textwidth]{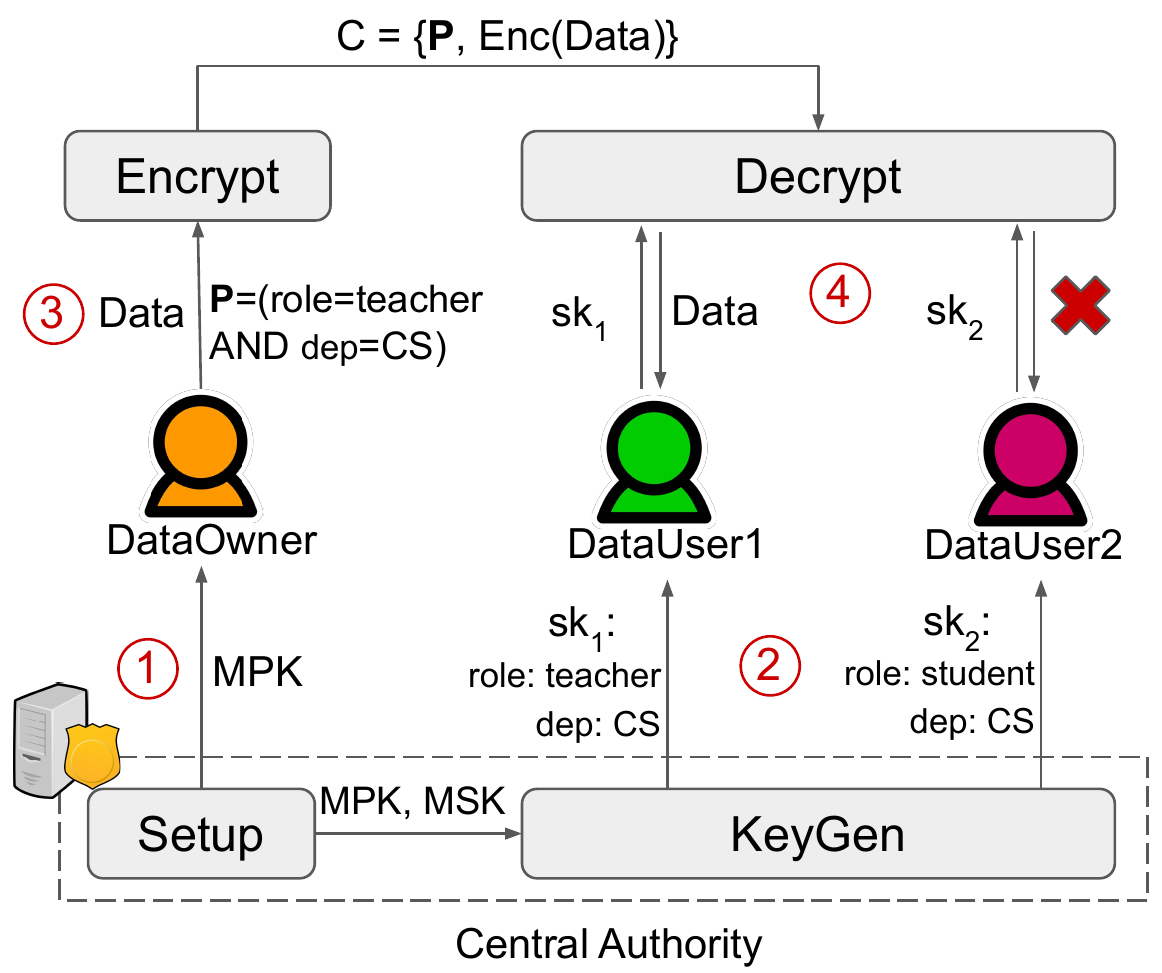}
\caption{Illustration of CP-ABE framework where the access policy is embedded in the ciphertext}
\label{fig:abeTypes}
\end{figure}

An ABE scheme has four core cryptographic functions, $\mathsf{ABE} = (\setup, \kgen, \enc, \dec)$. A trusted central authority, acting as an attribute authority, generates a master key pair $(\mpk, \msk)$ in the $\setup$ function and uses the master secret key $\msk$ to derive decryption keys in the $\kgen$ for authorized users after checking their attributes. Data owners and users interact with this attribute authority to obtain authorized key based on the attributes to perform encryption and decryption. Fig.~\ref{fig:abeTypes} shows an example ABE scheme where a data owner encrypts their data and defines the access policy stating that ``\emph{the ciphertext can be only decrypted by a user who is both a teacher and in the computer science department}''. In general, there are two types of ABE schemes based on where the access structure is defined, Key-policy (KP-ABE)~\cite{goyal2006attribute} and Ciphertext-policy (CP-ABE)~\cite{bethencourt2007ciphertext}. In CP-ABE, the ciphertext is encrypted based on the access policy, and the secret key is generated based on the attributes as illustrated in Fig.~\ref{fig:abeTypes}. In KP-ABE, the ciphertext is encrypted based on a set of attributes, and the user's secret key is generated based on a defined access policy. 

The first construction of ABE was proposed by Sahai and Waters~\cite{sahai2005fuzzy} as a fuzzy form of identity-based encryption (IBE), in which data are encrypted under a key derived from user's identity~\cite{shamir1984identity, boneh2001identity}. The fuzzy IBE scheme (FIBE) realizes access control through combined techniques of Shamir's linear secret sharing~\cite{shamir1979share} and Lagrange polynomial interpolation. The construction is based on Bilinear maps, where $\mathbb{G}_1$ and $\mathbb{G}_2$ are two groups of prime order $p$. The group $\mathbb{G}_1$ has a generator $g$ and a defined bilinear map $e: \mathbb{G}_1 \times \mathbb{G}_1 \rightarrow \mathbb{G}_2$, such that for all $a, b \in \ZZ_p$, we have $e(g^a, g^b) = e(g, g)^{ab}$. The security of the scheme is based on the Decisional Bilinear Diffie-Hellman assumption (DBDH)~\cite{boneh1998decision}, which states that given a group generator $g$ and uniformly chosen values $a, b, c \in \ZZ_p$, then no polynomial-time adversary is able to distinguish the tuple $(g^a,g^b,e( g,g)^{ab})$ and $(g^a,g^b, e(g,g)^c)$ with non-negligible advantage. We briefly describe each function in the FIBE scheme in Scheme~\ref{scheme:FIBE}.

\begin{pabox}[label={scheme:FIBE}]{The Fuzzy Identity-based Encryption (FIBE) scheme~\cite{sahai2005fuzzy}}
\begin{itemize}[leftmargin=*]
\item[-]$\mathsf{FIBE.}\setup(\lambda)$: This function is performed by a trusted third party (TTP), acting as the Attribute Authority, and takes as an input the security parameter $\lambda$ which determines the size of inputs. It defines a universe of attributes $\mathcal{U}$, randomly chooses a unique value $t_i$ where $i=1,..,|\mathcal{U}|$ to corresponds to each defined attribute, and chooses a secret value $y$. The outputs of this function are the public key $\mpk = \{ \{T_i = g^{t_i}\}_{i\in 1,\dots,|\mathcal{U}|}, Y = e(g,g)^y \}$ and the master key $\msk = \{t_1, \dots, t_{|\mathcal{U}|}, y\}$. Both the universe of attributes $\mathcal{U}$ and the public key $\mpk$ are published, but the master key $\msk$ is kept secret with the trusted Attribute Authority. 

\item[-]$\mathsf{FIBE.}\enc(\pk,\omega', m)$: This function takes as input a message $m$, a set of attributes $\omega' \subset \mathcal{U}$, and the public key $\mpk$. It chooses a random $r\in\ZZ_p$ for semantic security and outputs the ciphertext $\boldsymbol{c} = (\omega', m Y^r, \{T_i^r\}_{i\in 1,\dots,|\omega'|})$.

\item[-]{$\mathsf{FIBE.}\kgen(\mk,\omega)$}: This function is performed by the TTP and generates a secret key $\sk$ associated with a given set of user's attributes $\omega$. Specifically for each user, it defines an independent polynomial $q(x)$ of degree $d-1$ and $q(0) = y$, where $y$ is a secret defined in $\mk$. The function outputs $\sk = \{D_i = g^{\frac{q(i)}{t_i}}; i=1,..,|\omega|\}$. 

\item[-]{$\mathsf{FIBE.}\dec(\sk,\omega, \boldsymbol{c})$}: This function takes as input a secret key $\sk$ and a ciphertext $\boldsymbol{c}$ to decrypt based on Lagrange interpolation and secret sharing techniques. The decryption is successful if and only if $|\omega\cap\omega'|\geq d$, i.e., the user has at least $d$ attributes in $\omega$ matching the attributes $\omega'$ embedded in the ciphertext. Let a set $\boldsymbol{s} = \omega\cap\omega'; |\boldsymbol{s}| = d$ be arbitrary elements which can reconstruct the secret element $y$, and $\Delta_{i,\boldsymbol{s}[0]}$ are defined Lagrange coefficients, then the ciphertext is decrypted as:
\begin{align*}
    % m &= E'/\prod_{i= 1,\dots, |\boldsymbol{s}|}(e(D_i, E_i))^{\Delta_{i,\boldsymbol{s}[0]}} \\
    m &= m Y^r/\prod_{i= 1,\dots, |\boldsymbol{s}|}(e(g^{\frac{q(i)}{t_i}}, g^{rt_i}))^{\Delta_{i,\boldsymbol{s}[0]}} \\ 
    &= m e(g,g)^{ry}/\prod_{i= 1,\dots, |\boldsymbol{s}|}(e(g, g)^{rq(i)})^{\Delta_{i,\boldsymbol{s}[0]}} \\ 
    &= m e(g,g)^{ry}/e(g, g)^{ry} = m
\end{align*}
\end{itemize}
\end{pabox}

This scheme enables a generalized controlled access to encrypted data when there are at least $d$ overlapping attributes in the ciphertext and the secret key. Although the attributes are public, the access control is still cryptographically enforced by the underlaying security assumption. Note that each attribute in $\mathcal{U}$ has a corresponding unique secret value $t_i$ in the master secret key. The attribute authority issues a secret key for a user depending on their provided attributes, such that it computes the value $D_i g^{\frac{q(i)}{t_i}}$ for the $i$-th attribute based on the master secret key. Hence without direct access to $\mk$, it is not possible for an external adversary to forge a new secret key with desired attributes without knowing the corresponding secret values $t_i$. The secret values cannot be retrieved from the public parameters $g^{t_i}$ due to the fundamental hardness of the discrete log, which the DBDH builds upon. Moreover, an existing unauthorized user cannot modify their secret key to selectively add attributes to their set $\omega$ to satisfy $|\omega\cap\omega'|\geq d$, and decrypt a ciphertext.

Traditional ABE schemes provide access control to encrypted messages. Homomorphic ABE (ABHE) schemes can further support evaluations on these ciphertexts, which adds a fifth function, $\eval$ to the ABE functions. Gentry~et~al.~\cite{gentry2013homomorphic} proposed a leveled homomorphic ABE based on LWE. The scheme is based on an earlier work by Gorbunov~et~al.~\cite{gorbunov2015attribute} which uses a variation of garbled circuits as the access policy structure, but it is the first ABE scheme that also supports computation on the encrypted data. Mainly, let $\ell$ be the number of defined attributes, and $\mathcal{C}$ be a garbled circuit with $\ell$ input wires and one output wire, each input wire corresponds to the encoded input of two independent public/secret key pair.
The scheme uses an encoding mechanism~\cite{gorbunov2015attribute} with a random seed to encode each input wire and use the encoded output of the circuit as a masking value when encrypting the message. Each authorized user receives a secret key which embeds a set of recoding keys associated with their attributes and corresponding to those used in the encryption. These recoding keys are similar to a translation table for a garbled circuit, such that the decryption is only possible when a user is able to recreates the encoded value which was used to mask the message.

Any ABE scheme has to be collision-resistant to be ensure security of the data. In other words, users cannot combine their decryption keys to gain access to encrypted data when none of them is initially authorized. This can be ensured by choosing an independent polynomial embedding unique random elements for each user's share (secret keys) as in~\cite{sahai2005fuzzy, goyal2006attribute}, or by using a masking technique which uniquely randomizes the user's secret key as in~\cite{bethencourt2007ciphertext}. In the homomorphic ABE~\cite{gorbunov2015attribute}, the collision is not possible since each user's secret key hides a different ``translation table" and hence even if two users collude, they will not be able to decrypt if none of them can independently satisfy the circuit predicate.

Revocation of user's secret key is another important requirement for ABE schemes. One naive revoking approach is to generate new public key and master key and update all users' secret keys except for the revoked user's. However this approach is infeasible in practice since it requires invoking the setup and key generation functions, and re-encrypting the existing ciphertexts with the new public key. Alternatively, the scheme can include time stamps~\cite{pirretti2010secure} as additional attributes which state an expiry date $TS$ for each generated secret key and an encryption date $TS'$ for each ciphertext. Hence, a user can decrypt only if their secret key's expiry date is beyond the ciphertext's last decryption date (or $TS \geq TS'$).   

Homomorphic ABE/IBE schemes provide means for data owners to control the access to their encrypted messages. Homomorphic evaluations can be performed but only on ciphertexts encrypted under the same identity or attributes. Several works~\cite{clear2015multi} in the literature proposed multi-identity IBE schemes, which support computation on ciphertexts under different identities. However, the users with those identities are required to cooperate in a distributed decryption. The design of these schemes is similar in the design to multi-key HE schemes, discussed in the subsequent section, which support homomorphic evaluations on ciphertexts encrypted under different keys.

% \begin{table}[t]
% \caption{Overview of different ABE schemes features}
% % \resizebox{.9\textwidth}{!}{
% \begin{tabular}{|l|c|c|c|}
% \hline
% \multicolumn{1}{|c|}{Scheme}  & Security assumption & Policy structure & Homomorphic? \\ \hline
% Sahai and Waters~\cite{sahai2005fuzzy} & DBDH & LSSS &  \\ \hline
% Goyal~et~al.~\cite{goyal2006attribute} & DBDH & Access Tree &  \\ \hline
% Bethencourt~et~al.~\cite{bethencourt2007ciphertext} & DBDH & Access Tree &  \\ \hline
% Gorbunov~et~al.~\cite{gorbunov2015attribute} & LWE & Circuit &  \\ \hline
% Gentry~et~al.~\cite{gentry2013homomorphic} & LWE & Circuit & \checkmark \\ \hline
% Boneh~et~al.~\cite{boneh2013attribute} & LWE & - & \checkmark \\ \hline
% \end{tabular}
% % }
% \end{table}

\section{Multi-key approaches}
\label{sec:multi}
Many outsourced computations require homomorphic evaluations on data provided by different owners and encrypted using their own keys. For example in Fig.~\ref{fig:outsourced}, the function, or trained model, $f$ takes two inputs $[x]_A$ and $[y]_B$ encrypted under Alice's and Bob's public keys, respectively. Single-key approaches, discussed in previous section, enable re-encryption or decryption of ciphertexts with a different key, but computations \emph{must} be performed under the same key. Multi-key approaches extend many base HE schemes to support homomorphic evaluation with data encrypted by multiple keys. 

\begin{figure*}[t]
\centering
  \subfloat[Threshold HE]
 {\includegraphics[width=.48\textwidth]{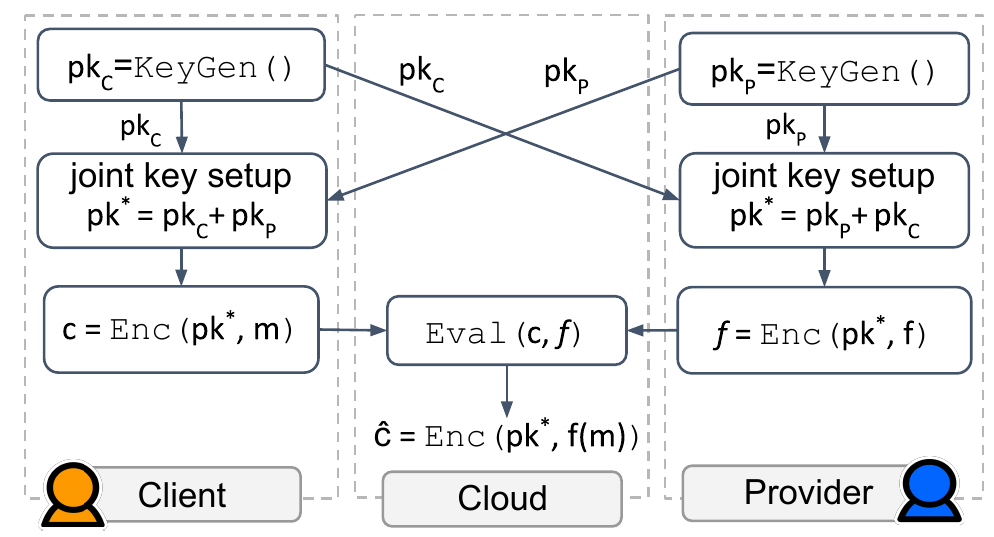} \label{fig:the}}\hfill
  \subfloat[Multi-key HE]
 {\includegraphics[width=.5\textwidth]{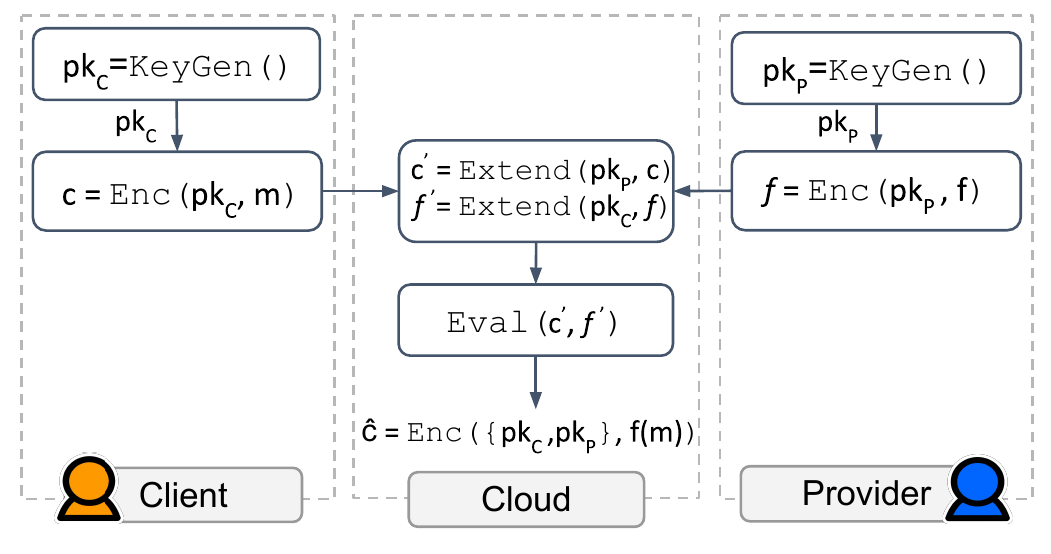} \label{fig:mkhe}}
\caption{Encryption step of schemes supporting multiple keys.}
\label{fig:multi-key}
\end{figure*}

In general, HE schemes can be extended to \emph{threshold} or \emph{multi-key} settings, or a \emph{hybrid} of both. In threshold HE schemes (ThHE), participants generate a joint public key $pk^*$ in advance from their individual public keys, e.g., a linear combination of their public keys as shown in Fig.~\ref{fig:the}. Inputs are homomorphically encrypted and evaluated under this joint key $pk^*$. In contrast, multi-key HE schemes (MKHE) support ``on-the-fly'' evaluation under different keys without a prior key setup as shown in Fig.~\ref{fig:mkhe}. A hybrid approach utilizing both ThHE and MKHE techniques can be used to improve practicality in special scenarios, where a group of system users always participate in evaluation. 

In all three approaches, schemes are often often designed in the common reference string (CRS) model as discussed in Sec.~\ref{subsec:notation}. This means the participants have access to a public parameter in the form of a ring element and use it to generate their individual keys. This produces individual keys that are related to each other and ensures correct computation with multiple keys. Moreover, in all three approaches, decryption is commonly done in a distributed manner among participants because no single user has the corresponding decryption key (i.e., the aggregation and concatenation of individual secret keys for ThHE and MKHE, respectively). 

We review in this section the fundamentals and challenges of the state-of-the-art approaches and discuss their security and efficiency. For simplicity, we describe the notations use throughout this section in Table~\ref{tab:MKnotations}.

\begin{table}[t]
\centering
\caption{Notations specific to multi-key approaches.}
\label{tab:MKnotations}
\resizebox{\textwidth}{!}{
\begin{tabular}{|c|c|l|}
\hline
Category & Notation & \multicolumn{1}{c|}{Description}  \\ \hline
\multirow{5}{*}{ThHE-specific} & $N$      & Total number of users in the system. \\ \cline{2-3} 
& $T$      & Required number (threshold) of secret shares to decrypt. \\ \cline{2-3} 
& $\pk^*$  & A joint public key generated from individual public keys $\{\pk_1,\dots, \pk_N\}$ \\ \cline{2-3} 
& $\sk^*$  & A joint secret key, secretly shared among $N$ users. \\ \cline{2-3} 
& $\ek^*$  & A joint evaluation key generated corresponding to $\pk^*$.            \\ \hline
\multirow{6}{*}{MKHE-specific} & $\mathcal{K}$ & A bound on number of keys we can extend a ciphertext under. \\ \cline{2-3} 
& $K$      & Total number of keys, such that $K \leq \mathcal{K}$  \\ \cline{2-3} 
& \parboxc{b}{2.5ex}{$\Bar{\pk}$}  & An extended public key concatenated as $(\pk_1,\dots,\pk_K)$ \\ \cline{2-3} 
& \parboxc{b}{2.5ex}{$\Bar{\sk}$}  & An extended secret key concatenated as $(\sk_1,\dots,\sk_K)$ \\ \cline{2-3}
&  \parboxc{b}{2.5ex}{$\Bar{\ek}$}  & An extended evaluation key generated corresponding to $\Bar{\pk}$. \\  \cline{2-3} 
& $\Bar{c}$ & An extended ciphertext encrypted under the concatenated public keys. \\ \hline
Distributed decryption & $\rho_i$ & The decryption component constructed by the $i$-th user in distributed decryption. \\ \hline
\end{tabular}
}
\end{table}

\subsection{Threshold Homomorphic Encryption}
\label{subsec:THE}
Supporting homomorphic computations on inputs from different data owners without compromising data privacy requires data to be encrypted under individual's keys. Threshold encryption is a common approach for fulfilling this requirement, with the generation of a joint key in a key setup before any computation. This joint key can be changed or revoked at any point but with the cost of rerunning the key setup and re-encrypting all input data. 

In a threshold encryption scheme~\cite{desmedt1989threshold, boneh2006chosen, delerablee2008dynamic, asharov2012multiparty}, a set of users can encrypt their data under a joint key but have to cooperate for decryption. Shamir's secret sharing~\cite{shamir1979share} forms the base of threshold schemes. Mainly, a secret $s$ is divided into shares $s_i$ and distributed among $N$ users, such that no single user knows the secret $s$. The secret is reconstructed if and only if a predefined number of shares (say a threshold $T$) are provided. This threshold encryption scheme follows the design of Shamir's $(T,N)$ threshold scheme. 

The threshold value $T$ is chosen according to the security model and requirements. For example, if all $N$ users in the system are required to participate in decryption, the threshold $T$ is set to $T = N$. This setting follows a \emph{dishonest-majority} assumption, where $T \geq N/2$ system users may be corrupted. In this case, even with $N-1$ corrupted users, it is not possible to decrypt. This assumption can be relaxed to a \emph{honest-majority} assumption, which implies that the number of dishonest users is less than $N/2$ and a subset of users can decrypt. On the other hand, if every user in the system has the right to independently decrypt, the threshold is set to $T = 1$ which corresponds to threshold \emph{broadcast} encryption~\cite{ghodosi1996dynamic, canetti1999efficient, daza2007cca2}. 

A ThHE scheme combines threshold functionality with the ability to compute on encrypted data under a joint key. There are three main components in ThHE schemes: a distributed key setup protocol, a homomorphic encryption function, and a distributed decryption protocol~\cite{Schoenmakers2011}. Formally, we can define a general ThHE scheme as tuple of \textsf{PPT} algorithms $\mathsf{ThHE} = (\setup, \jkgen, \enc, \eval,\dec)$.
\begin{itemize}
\item[-] $\mathsf{ThHE}.\setup(1^\lambda)\rightarrow((\pk_1,\sk_1)\dots,(\pk_N,\sk_N))$: Given a security parameter $\lambda$, the setup algorithm outputs a set of $N$ key pairs $(\pk_i, \sk_i)$.
\item[-] $\mathsf{ThHE}.\jkgen(\pk_1,\dots,\pk_N)\rightarrow \pk^*, \ek^*$: Given the input of $N$ public keys $(\pk_1,\dots,\pk_N)$, the interactive algorithm outputs a joint public key $\pk^*$ and the evaluation key $\ek^*$, if required.
\item[-] $\mathsf{ThHE}.\enc(\pk^*, m)\rightarrow c$: Given a joint public key $\pk^*$ and a message $m$, the encryption algorithm outputs a ciphertext $c$.
\item[-] $\mathsf{ThHE}.\eval(\pk^*, f, c, c')\rightarrow c_{\mathsf{eval}}$: Given a joint public key $\pk^*$ and two ciphertexts $c, c'$, the evaluation algorithm outputs the evaluated ciphertext $c_{\mathsf{eval}} = f(c, c')$.
\item[-] $\mathsf{ThHE}.\pdec(\sk_i, c)\rightarrow \rho_i$ Given a ciphertext $c$ encrypted under $\pk^*$ and a secret share of the key $\sk_i$, perform the partial decryption algorithm and output a partially decrypted message $\rho_i$.
\item[-] $\mathsf{ThHE}.\comb(\rho_1,\dots,\rho_T)\rightarrow m$: Given a set of partial decryptions $(\rho_1,\dots,\rho_T)$ for some threshold set, combine decryption components to perform the final decryption and output the message $m$.
\end{itemize}

In earlier threshold schemes in the literature~\cite{cramer1997secure, desmedt1989threshold}, the joint key pair may be generated first by a trusted dealer, then the public key and shares of the secret key are distributed among users. Alternatively, the public key can be generated from users' individual public keys, as implied in $\mathsf{ThHE}.\jkgen$, and they need to collaborate to decrypt with their individual secret keys at the end of computation. The two algorithms $\mathsf{ThHE}.\pdec$ and $\mathsf{ThHE}.\comb$ are the routines performed in distributed decryption, which we discussed as a common technique in Sec.~\ref{subsec:distdec}. Another way to decrypt may utilize the key switching technique (Sec.~\ref{subsec:keyswitching}) to re-encrypt the ciphertext result from an encryption under the joint key to one under the intended user's key. This method avoids performing distributed decryption and enable the user to directly decrypt with their own secret key. But, this method can lead to security problem as discussed in Sec.\ref{subsubsec:HEDiscussion}. Mouchet~et~al.~\cite{cryptoeprint:2020:304} proposed two versions of key switching, depending on whether parties have access to the shares of the new secret or public key, based on their construction of multiparty HE cryptosystem. In the rest of this section, we survey ThHE schemes based on two defined security assumptions, dishonest-majority, and honest-majority.

\subsubsection{Dishonest-majority threshold HE} Many HE schemes can be extended to a threshold setting by leveraging the additive homomorphism of the key space. This property enables the establishment of a joint key from individual keys that are generated and owned by individual users without a trusted party. By direct aggregation of the users' public keys, we can effortlessly set up a ($N$-out-of-$N$) threshold encryption scheme following the dishonest-majority model. In other words, threshold HE transforms the setting from computing with different keys to computing with a single joint key. 

Both ElGamal and Paillier schemes can be extended to a threshold version~\cite{desmedt1989threshold, pedersen1991threshold, cramer1997secure}. The aggregation of public keys in the former is less complex than the latter. Specifically in the multiplicatively homomorphic ElGamal scheme, assume we have a set of $N$ users where each user $i$ has an independently generated key pair $(\pk_i = g^{s_i}, \sk_i = s_i)$, where $g$ is a group generator. Then, the joint key is computed as $\pk^* = \prod_{i=1}^{N} \pk_i = g^{\sum^{N}_{i=1}s_i} = g^{s^*}$. A user can encrypt data under the generated joint public key $\pk^*$ with a uniform random $r$ for semantic security as $\boldsymbol{c} = (c_0, c_1) = (g^r, m\cdot (g^{s^*})^r)$. Homomorphic computations can be done on the ciphertext using the homomorphic primitives of the scheme. Decryption of ciphertexts must be performed with the help of all $N$ users to remove the element $g^{rs^*}$ from $c_1$. In particular, each user $i$ uses their own secret key $s_i$ to construct a component $\rho_i = (c_0)^{-s_i} = (g^r)^{-s_i} = g^{-rs_i}$ such that combining these components yields $\sum^{N}_{i=1}\rho_i = g^{-rs^*}$. Then, each user $i$ takes turn and uses their component $\rho_i$ to partially remove their secret key $s_i$ by computing $\hat{c}_1\cdot\rho_i = m\cdot g^{rs^*}\cdot g^{-rs_i}$. After partial decryption by all parties, we yield a plaintext message $m$.

In a similar manner, (R)LWE-based HE schemes can also be extended to a threshold setting. Asharov~et~al.~\cite{asharov2012multiparty} proposed a threshold variant of the BGV scheme~\cite{brakerski2012leveled} that we present in Scheme~\ref{scheme:ThBGV}. The scheme is designed as a 3-round MPC protocol, where the joint public key $\pk^*$ and the joint evaluation key $\ek^*$ are generated in the first two rounds, and the distributed decryption is performed in the third round. 
Formally, given a set of $N$ public keys $\pk_i= (b_i, a) = (as_i+t e_i, a)$, where the element $a\in R_q$ is a shared element in the CRS model and $s_i$ is the $i$-th user's secret key. We generate a joint public key $\pk^*= (\sum^{N}_{i=1}b_i, a)$ such that we obtain $(a\sum^{N}_{i=1}{s_i}+t\sum^{N}_{i=1}{e_i}, a) = (as^*+te^*, a)$. Note that only the first component $b$ of the individual public keys is aggregated such that the underlying secret keys $s_i$ are homomorphically added under the RLWE assumption. If we add both components of all the public keys, the joint key will be $\pk^*=(as^*+te^*, Na)$. Subsequently, a ciphertext encrypted under the joint key will be $\boldsymbol{c} = (c_0, c_1) = (ras^*+te^*+ m, Nra)$. Obviously, the decryption $c_0-c_1s^*$ will fail because $(Nras^* \neq ras^*)$.

\begin{pabox}[label={scheme:ThBGV}]{A Threshold-variant of BGV scheme~\cite{asharov2012multiparty}}
\begin{itemize}[leftmargin=*]
\item[-] {$\mathsf{ThBGV.}\setup(1^\lambda,1^L)$}: Given the security parameter $\lambda$ and a multiplicative depth $L$, run $\mathsf{BGV}.\setup$ and output $\param = (d, \chi, \psi, q, t, \boldsymbol{a})$ as the public parameters of scheme. Let $\boldsymbol{a} \sample R^{\ell}_q$ be the shared CRS vector.

\item[-] {$\mathsf{ThBGV}.\kgen(\param)$}: Given the public parameters $\param$, choose $a\sample \boldsymbol{a}$ and uniformly sample a set of $N$ secrets $s_i \sample \psi$ and errors $e_i \sample \chi$, where $i\in \{1,\dots,N\}$. Output the set of key pairs $((\pk_1,\sk_1)\dots,(\pk_n,\sk_N))$ such as $\pk_i=(as_i+te_i, a)$ and $\sk_i = s_i$.
%\rightarrow((\pk_1,\sk_1)\dots,(\pk_n,\sk_N))

\item[-] {$\mathsf{ThBGV}.\jkgen(\pk_1,\dots,\pk_N)$}: Given the input of $N$ public keys $(\pk_1,\dots,\pk_N)$, aggregate the first component of $\pk_i$ to generate the joint public key as $\pk^*= (\sum(as_i+te_i), a)$. 

\item[-]$\mathsf{ThBGV}.\evalkgen(\sk_1,\dots,\sk_N)$: Given a set of $N$ secret keys $\sk_i= s_i$, jointly compute the evaluation key as $\ek^* = \sum_{1\le i,j \le N} (\{\enc(\pk^*, \sum s_i[\alpha]s_j[\beta])\})$ where $0\le \alpha,\beta<d$ are the indices of coefficients of the secret keys.
%\rightarrow\pk^*,\ek^*

\item[-] {$\mathsf{ThBGV}.\enc(\pk^*,m)$}: Given a joint public key $\pk^*$ and a message $m$, the encryption algorithm outputs a ciphertext $\boldsymbol{c} = (c_0, c_1)$ where $c_0 = r(as^* + te^*) + m$ and $c_1 = ra$. %\rightarrow \boldsymbol{c}

\item[-] {$\mathsf{ThBGV}.\eval(\pk^*, f, \boldsymbol{c}, \boldsymbol{c}')$}: Given a joint public key $\pk^*$ and two ciphertexts $\boldsymbol{c}, \boldsymbol{c}'$, the evaluation algorithm outputs the evaluated ciphertext $\boldsymbol{c}_{\mathsf{eval}} = f(\boldsymbol{c}, \boldsymbol{c}')$. Similar to the base BGV scheme, the evaluation function $f$ can be either homomorphic addition or multiplication.

\item[-]$\mathsf{ThBGV}.\relin(\ek^*,\tilde{\boldsymbol{c}}_{\mult})$: Given the joint evaluation key $\ek^*$ and the long ciphertext $\tilde{\boldsymbol{c}}_{\mult} = (\tilde{c_0}, \tilde{c_1}, \tilde{c_2})$, perform relinearization similar to $\mathsf{BGV}.\relin$ to output $\boldsymbol{c}_{\mult} = (c_0, c_1)$. %\rightarrow \boldsymbol{c}_{\mathsf{eval}}

\item[-] {$\mathsf{ThBGV}.\pdec(\sk_i, \boldsymbol{c})$}: Given a ciphertext $\boldsymbol{c}$ encrypted under $\pk^*$ and a secret share of the key $\sk_i$, generate a decryption component as $\rho_i = (c_1s_i + te'_i)$. %\rightarrow \rho_i

\item[-] {$\mathsf{ThBGV}.\comb(\rho_1,\dots,\rho_N, \boldsymbol{c})$}: Given decryption components $(\rho_1,\dots,\rho_N)$, decrypt the ciphertext as $c_0-\sum_i{\rho_i} \pmod t$.
\end{itemize} %\rightarrow m
\end{pabox} 

As mentioned, homomorphic multiplication requires additional evaluation keys to perform relinearization that brings the quadratic ciphertext in $(s^*)^2$ back to be linear in $s^*$. Recall that the evaluation key encrypts the powers of base of the secret key $(s^*)^2 = (s_1+\dots+s_N)^2$. Since the secret key is shared among $N$ users, they cannot simply calculate the sum of the square of their individually generated secret keys $s_i$, because $(s_1^2+\dots+s_N^2) \neq (s_1+\dots+s_N)^2$. The generation of this evaluation key $ek^*$ is trickier than the public key due to its fundamentally complex structure. It requires all $N$ parties to cooperate in a 2-round setup phase to compute the joint evaluation key from their secret shares. Let $0\le \alpha,\beta<d$ be the indexes of coefficients of the secret $s^*$, which is a polynomial with degree $d-1$, and let $\ell \in \{0, \lfloor\log{q}\rfloor\}$. In the first round, every party will broadcast $\{\enc(\pk^*, t^\ell s_i[\alpha])\}$, which is a set of encryption of all the powers of base $t$ of the $i$-th secret share's coefficients. Note if we aggregate the shares from all parties, we will obtain $\{\enc(\pk^*, t^\ell\cdot s^*[\alpha])\}$. In the second round, each party performs a pairwise multiplication between the coefficients of the aggregated encrypted secret key $s^*$ and the coefficients of its secret shares $s_i$, yielding $ek^*=\{\enc(\pk^*,t^\ell \cdot s^*[\alpha]\cdot s_i[\beta])\}$. Combining the shares from all parties in the second round outputs the joint evaluation key $\{ek^*\}$, such that for $a\sample R_{q}$, the evaluation key is $ek^* = (as^* + te + t^\ell\cdot s^*\cdot s^*, a)$. To avoid making circular security assumption, the secret $(s^*)^2$ should be encrypted under a different secret $s^*$, hence we can generate $L$ joint keys, one for each level $l \in L$, and encrypt  $(s_{l-1}^*)^2$ corresponding to level $l-1$ under $\pk^*_{l}$ as discussed in Sec.~\ref{subsubsec:BGV}. 

To decrypt, each user contributes their partial key $s_i$ by computing the component $\rho_i = c_1s_i+te_i$. Then, all users collaboratively produce a component that contains the sum of all secret keys shares, that is $\sum^{N}_{i=1}\rho_i =(c_1\sum^{N}_{i=1}{s_i}+t\sum^{N}_{i=1}{e_i}) = (c_1s^*+te^*) = \rho^*$. Refer to the decryption process $\dec(\sk,\boldsymbol{c})=c_0-\rho^*$, the message can be decrypted correctly when computing $c_0-(c_1s^*+te^*)$.

These $(N,N)$ threshold schemes enable the group of $N$ users to homomorphically compute on their data that is encrypted under a joint key from their individual keys. Decryption however is impossible unless all of them participate with their secret shares. In practice, this may become a single point of failure if a user becomes uncooperative or was offline at the time for decryption. Hence, designing a more flexible threshold scheme is necessary where only a subset of secret shares is sufficient to decrypt.

\subsubsection{Honest-majority threshold HE}
Threshold HE scheme proposed by Asharov~et~al~\cite{asharov2012multiparty} targets a dishonest-majority security assumption. However, some applications may need to relax this assumption to an honest-majority one to avoid single point of failure by enabling only a subset of keys to decrypt. Desmedt and Frankel~\cite{desmedt1989threshold} proposed a threshold version of ElGamal scheme~\cite{elgamal1985public} where only $T$ users out of $N$ are needed to decrypt. The scheme leverages Shamir's linear secret sharing ~\cite{shamir1979share}. Given a threshold $T$ and a large prime $p$ as parameters, a \emph{trusted} dealer performs a key setup as follows. The dealer uniformly sample a secret key $s^*~\mod p$ and set it as the constant term of a random polynomial $f(x)~\mod p$, such that $f(0) = s^*$. The degree of this polynomial is determined as $T-1$ if we want $T$ users to reconstruct the secret. The dealer then uses this polynomial to generate $N$ secret shares $s_i = f(i)~\mod~p$ as unique points for each user $i\in\{1,N\}$. Finally,broadcast $\pk^* = g^{s^*}$ as the joint public key and keep both $s^*$ and $f(x)$ secret from the users. The intuition is that Lagrange interpolation can be used to reconstruct the secret $s^*$ if at least $T$ points (shares) were provided.

Each user can encrypt a message $m_i$ as $(a_i, b_i) = (g^{r_i}, m_ig^{r_is^*})$ using the public key $g^{s^*}$ and a random value $r_i$. However, at least $T$ users are required to use their secret shares to decrypt. This construction of threshold ElGamal requires a trusted dealer to generate and distribute the secret in the key setup. Pedersen~\cite{pedersen1991threshold} improved the scheme later by removing the need for this trusted dealer at key setup and enabled validation of the secret shares for robustness. However, the decryption still requires a trusted key authority to generate the unique polynomial and decrypt. 

Threshold schemes proposed after that were based on different encryption schemes~\cite{cramer2001multiparty, shoup2002securing}, not necessarily homomorphic, and removed the need of this trusted dealer by relying on Diffie-Hellman key generation instead, such as in threshold RSA~\cite{rabin1998simplified, damgaard2001practical}. 

Boneh~et~al~\cite{boneh2018threshold} proposed the first ($T$-out-of-$N$) ThHE scheme based on LWE problem. In this scheme, the decryption key is split into $N$ shares among the participants. Shamir's linear secret sharing scheme (LSSS) technique is also leveraged here to allow $T$ key shares to sufficiently decrypt according to a defined threshold access structure. Moreover, a \emph{universal thresholdizer} was proposed to extend general HE schemes to threshold HE schemes. Specifically, a universal thresholdizer split a given cryptographic key into $N$ valid shares that can be used as individual public keys. 

\subsubsection{Discussion} 
Key management in ThHE schemes is essential to system security. In the key setup, the joint key is often generated based on participants' individual keys before any computation. Each user encrypts their inputs under this one joint key to produce compact ciphertexts, which means the ciphertext size is \emph{independent} of the number of users $N$. The security depends on the fact that no single user has the decryption key, which can be only constructed if a threshold set of users cooperated. The joint key can be revoked when a user is added or removed at any point, but this requires running the threshold key setup again and re-encrypting all inputs with the new joint key. It may be a better choice to leverage ThHE schemes in outsourced computations if participants do not change often. If they do, the efficiency quickly degrades because the joint key needs to be updated accordingly. In this case, a more flexible approach is needed to allow computation with multiple keys ``on-the-fly" when required.

Potentially, we investigate if the key homomorphism can be used to directly transform ciphertexts from encryptions under individual keys to ones encrypted under a joint key without prior key setup. For example in the threshold BGV scheme~\cite{asharov2012multiparty}, let $[m]_A = (c_0, c_1) = (r a(s_A)+ te_A + m, ra)$ be a ciphertext encrypting a message $m$ under Alice's public key $\pk_A$. A cloud evaluator may perform joint homomorphic evaluation on this ciphertext for Alice and Bob such that the result is encrypted under their joint key $\pk_{AB}$. Using Bob's public key $\pk_B = (as_B + te_B, a)$, the cloud can construct a component $\xi_B = (ras_B + te_B)$ which is the encryption of zero if the randomness $r$ is known. Then, the cloud can add this component to the ciphertext as $[m]_{AB} = (c_0+ \xi_B, c_1) = (ra(s_A + s_B) + t(e_A + e_B) + m, ra)$, which is an encryption under $\pk_{AB}$. 

Unfortunately, this approach has two major flaws which breaches the security of the scheme. First, the encryption randomness $r$ may need to be \emph{fixed} and shared with the cloud similar to the CRS element $a$. Recall that the BGV scheme encryption algorithm depends on $r$ to randomize the ciphertexts. Fixing this element renders deterministic encryption scheme which loses its semantic security property. i.e., encrypting the same message at different times will yield the same ciphertexts. The second flaw is more vital since this technique relies on \emph{disclosing} the randomness $r$ to the cloud evaluator. Obviously, this break the security of the scheme. Specifically, given a ciphertext $[m]_A$, the cloud evaluator can easily decrypt it by constructing a component $(ras_A)$ using the public key $\pk_A$ and compute $\tilde m =c_0 - ras_A = m+te_A \pmod q$, which removes the large element $ras_A$ and yield $m = \tilde{m} \pmod t$.

Due to these security issues, ThHE cannot be simply modified to support on-the-fly computation with different keys. Alternatively, a multi-key technique targets this problem and extends HE schemes to dynamic transformation of ciphertexts by concatenating the different public keys and treating them as one key. We review this technique in depth in the following section.

\subsection{Multi-key Homomorphic Encryption}
\label{subsec:MKHE}
The first notion of multi-key HE (MKHE) schemes was introduced by L\'{o}pez-Alt~et~al.~\cite{lopez2012fly} to support the homomorphic evaluation on ciphertexts encrypted under different keys. 
% What is the main advantage? no prior key setup -> dynamic computation
Compared to ThHE schemes, this type of HE schemes remove the need of a key setup phase to generate a joint key from individual keys prior to any computation. Instead, a cloud evaluator can dynamically extend ciphertexts from encryption under individual keys to ones under the concatenation of individual users' keys $(\pk_1,\dots,\pk_K)$. However, both ThHE and MKHE schemes are similar in requiring users to cooperate and run a distributed decryption protocol when retrieving the evaluated result. % \fix{need to elaborate on how this concatenated key looks like}

% What are the main types? single-hop, multi-hop define them
We categorize MKHE scheme as \textit{single-hop} if the extended ciphertext cannot be further extended to additional keys after being homomorphically evaluated. Otherwise, we categorize them as \textit{multi-hop} MKHE schemes. In general, an MKHE scheme is a tuple of \textsf{PPT} algorithms $\mathsf{MKHE} =(\setup, \kgen, \enc, \extend,$ $\evalkgen, \eval, \dec)$. We formally define each algorithm as follow.
\begin{itemize}
\item[-] $\mathsf{MKHE}.\setup(1^\lambda,1^\mathcal{K})\rightarrow\param$: Given a security parameter $\lambda$ and a bound $\mathcal{K}$ on the number of keys, the setup algorithm outputs the public parameters $\param$.
\item[-] $\mathsf{MKHE}.\kgen(\param)\rightarrow (\pk, \sk)$: Given the public parameters $\param$, the key generation algorithm outputs a public key $\pk$, a private key $\sk$.
\item[-] $\mathsf{MKHE}.\enc(\pk,m)\rightarrow c$: Given a public key $\pk$ and a message $m$, the encryption algorithm outputs a ciphertext $c$. 
\item[-] $\mathsf{MKHE}.\extend(\{\pk_1,\cdots,\pk_K\}, c)\rightarrow \Bar{c}$: Given a set of $K$ public keys $\pk_1,\cdots,\pk_K$ where $ K \le \mathcal{K}$, and a ciphertext $c$, the output is the extended ciphertext $\Bar{c}$ under the concatenated public key $\Bar{\pk}$.
\item[-] $\mathsf{MKHE}.\evalkgen(\Bar{\pk})\rightarrow \Bar{\ek}$: Given a concatenated public key $\Bar{\pk}$ generate the corresponding evaluation (linearization) key $\Bar{\ek}$.
\item[-] $\mathsf{MKHE}.\eval(\Bar{pk}, f, \Bar{c}, \Bar{c}')\rightarrow c_{\mathsf{eval}}$: Given two extended ciphertexts $\Bar{c}, \Bar{c}'$ under the same concatenated public key $\Bar{\pk}$, the evaluation algorithm outputs the evaluated ciphertext $\Bar{c}_{\mathsf{eval}} = f(\Bar{c}, \Bar{c}')$.
\item[-] $\mathsf{MKHE}.\dec((\sk_1,\dots,\sk_K), \Bar{c})\rightarrow m$: Given a set of concatenated secret shares $(\sk_1,\dots,\sk_K)$ and an extended ciphertext $\Bar{c}$, the interactive decryption algorithm performs decryption and outputs the message $m$ 
\end{itemize}

% What are the challenges? the evaluation key, sharing the randomness without disclosing them
To design an MKHE scheme, two main points have to be addressed: the generation of the evaluation key and maintaining the encryption randomness. Homomorphic multiplication on the extended ciphertexts requires providing an evaluation key corresponding to the concatenated key to perform dimension reduction (i.e., relinearization). Also, decryption in its essence depends on canceling out the large randomized elements in the ciphertext to retrieve the message. Hence, we must ensure that decrypting with different secret keys is done correctly in MKHE schemes, i.e., it does not yield remaining randomized elements.  

\subsubsection{MKHE via onion encryption} In theory, any standard HE scheme can be extended to mutli-key for a \emph{constant} number of keys through an \emph{onion} encryption and decryption technique~\cite{lopez2012fly}. This technique is similar to onion routing~\cite{goldschlag1999onion} used in anonymous communication in Tor~\cite{dingledine2004tor}.  L\'{o}pez-Alt~et~al.~\cite{lopez2012fly} shows an HE construction for encrypting a message $m$ under multiple keys. 
% Given an HE scheme which encrypts a message $m \in \{0,1\}$ into a ciphertext $c \in \{0,1\}^\mu$ where $\mu = \mathsf{poly}(\lambda)$ for some polynomial $\mathsf{poly}(\cdot)$. Let $m_j$ be the $j$-th bit of a message such that $m = (m_1, \dots, m_l)$ for a bit length $l$. Define the bit-wise encryption of a message $m$ and the decryption of ciphertext $c$ as following. 
Given an HE scheme which encrypts a message $m \in \ZZ_t$, say $t = 2$, into a ciphertext $c \in \ZZ_q$. In this scenario, let $m$ be bit-decomposed such that $m = (m_1, \dots, m_\ell) \in \{0,1\}^\ell$ for a bit length $\ell$. Define the bit-wise encryption of a message $m$ and the decryption of ciphertext $c$ as following. 
\begin{align*}
\overline{\enc}(\pk,m) &= (\enc(\pk, m_1), \dots, \enc(\pk, m_\ell)) \\
\overline{\dec}(\sk,c) &= (\dec(\sk, c_1), \dots, \dec(\sk, c_\ell))
\end{align*}
Now, define the onion encryption and decryption for a message $m$ under a set of $K\in\NN$ keys as following. 
\begin{align*}
\enc^*(\pk, m) &= \overline{\enc}(\pk, m) \\
\enc^*(\pk_1, \dots, \pk_K, m) &= \enc^*(\pk_K, \pk_{K-1}, \dots, \overline{\enc}(\pk_1, m)) \\
&= \overline{\enc}(\pk_K,\overline{\enc}(\pk_{K-1},\dots,\overline{\enc}(\pk_1,m)))\\
\\
\dec^*(\sk, c) &= \overline{\dec}(\sk, c) \\
\dec^*(\sk_1, \dots, \sk_K, c) &= \dec^*(\sk_{K-1}\dots, \sk_{1}, \overline{\dec}(\sk_K, c)) \\
&= \overline{\dec}(\sk_1,\overline{\dec}(\sk_{2},\dots,\overline{\dec}(\sk_K,c)))
\end{align*} 

\begin{figure*}
\centering
\includegraphics[width=.65\textwidth]{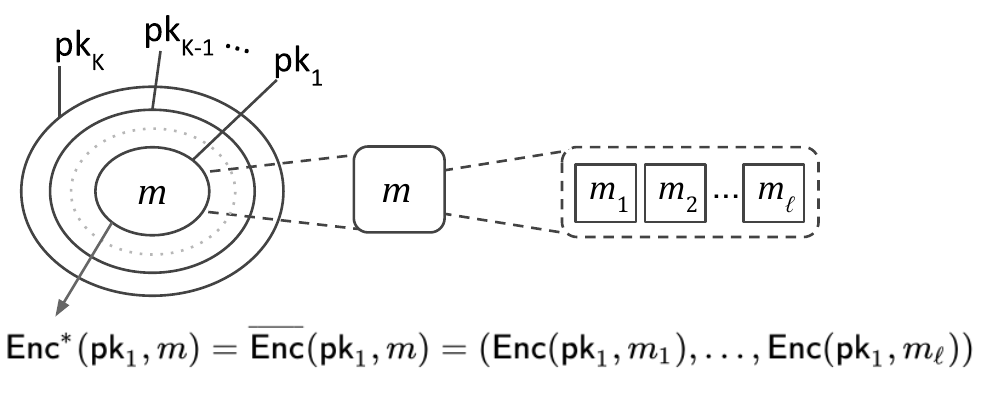}
\caption{Achieving MKHE via onion encryption.}
\label{fig:onionEnc}
\end{figure*}

Figure ~\ref{fig:onionEnc} shows an example of this HE scheme which adds encryption layers repeatedly to extend ciphertexts to additional keys. Retrieving the result is done by applying decryption in reverse order of encryption using the set of corresponding secret keys. 

Before adding a new encryption layer, the ciphertext is represented in binary as $c \in \{0,1\}^\mu$, where $\mu = \mathsf{poly}(\lambda)$ for some polynomial $\mathsf{poly}(\cdot)$. Each layer of encryption expands the ciphertext size by a factor of $\mu$ because the encryption function $\overline{\enc}(\pk, m)$ outputs a bit-wise encrypted ciphertexts. Hence, the size of a ciphertext grows exponentially with each additional encryption layer. Specifically, in the case where a message space $\{0,1\}$ and a ciphertext space $\{0,1\}^\mu$, the size of a ciphertext encrypted under $K$ keys is approximately $\mu^K$. Due to this large ciphertext expansion after each layer, onion encryption method is feasible only for a very small number of keys, which must be chosen before computation.

Given $K$ ciphertexts $c_i = \overline{\enc}(\pk_i, m_i)$, we can extend them to be encryptions under $K$ key via onion encryption described above. Homomorphic operations are performed on ciphertexts encrypted under the same set of keys but must be \textit{in order}. The result is decrypted with the corresponding ordered secret keys such that $\dec^*(\sk_1, \dots, \sk_K, f(m_1,\dots,m_K)) = f(m_1,\dots,m_K)$.  

% Discuss limitations. Defer BGV extension to when you talk about BGV
% A ciphertext $\enc(\pk_1, m)$ can be extended to a second key by adding a second layer of encryptions such as $\enc(\pk_2,\enc(\pk_1, m))$. 
As mentioned before, this technique is generic to extend any standard HE scheme to the multi-key setting. It proposes the notion of on-the-fly computation, but a minimum setup is required before computation; i.e., choosing the number of keys $K$ as a parameter. Onion encryption shares the setup requirement with ThHE schemes. A key setup phase must be done in the ThHE to generate the joint key from participants' keys, and in onion encryption the number of keys is selected. However, onion encryption has a restriction on the order of evaluations and decryptions as they must be done in the order of the keys. In contrast, evaluations in ThHE are performed on data encrypted under the joint key, and decryption is done in a distributed manner that does not depend on the order. 

\begin{pabox}[label={scheme:LTV}]{The L\'{o}pez-Alt-Tromer-Vaikuntanathan (LTV) scheme~\cite{lopez2012fly}}
\begin{itemize}[leftmargin=*]
\item[-]{$\mathsf{LTV}.\setup(1^\lambda)$}: Given a security parameter $\lambda$, choose a cyclotomic polynomial $\Phi(x) = x^d + 1$ and define $R = \ZZ[x]/(\Phi(x))$ as a polynomial ring of degree $d$ with integer coefficients. Let $R_q$ be the quotient ring $R/qR$ where $q$ is a modulus and $\chi$ be an error distribution. Output the set of parameters $\param = (d, q, \chi)$.

\item[-]{$\mathsf{LTV}.\kgen(\param)$}: Given a security parameter $\lambda$, sample $f',g\sample\chi$. Compute $f=2f'+1$. If $f$ is not invertible in $R_q$, re-sample. Otherwise, find $f^{-1}$ and set $h = 2gf^{-1} \in R_q$. Output the key pair as $(\pk, \sk) = (h, f)$. 

\item[-]$\mathsf{LTV}.\evalkgen(f)$: Given the secret key $f$, output the evaluation key $\ek = h\cdot \boldsymbol{u}' + 2\boldsymbol{e}' + f\cdot \bg \in R_q $, where $\boldsymbol{u}', \boldsymbol{e}'\sample\chi^\ell$.

\item[-]{$\mathsf{LTV}.\enc(\pk,m)$}: Given a public key $\pk = h$ and a message $m \in \{0,1\}$. Sample $u, e\sample\chi$ and output the ciphertext as $c = hu + 2e + m \in R_q$.

\item[-]{$\mathsf{LTV}.\eval(\Bar{\pk}, f, c_1, \dots, c_k)$}: Given a set of public keys $\Bar{\pk} = (\pk_1,\dots,\pk_K)$ and $k$ ciphertexts, perform the evaluation $f$ and output $\Bar{c}_{\mathsf{eval}}$. Evaluation can be addition or multiplication. For addition, output $\Bar{c}_\add = \Bar{c}_1+\dots+\Bar{c}_k \in R_q$. For multiplication, perform $\Bar{\tilde{c}}_\mult = \Bar{c}_1\cdot\ldots\cdot \Bar{c}_k \in R_q$. % and use evaluation keys $\{\ek_1,\dots,\ek_K\}$ for relinearization.

\item[-]$\mathsf{LTV}.\relin(\{\ek_1,\dots,\ek_K\},\Bar{\tilde{c}}_\mult)$: Given a ciphertext $\Bar{\tilde{c}}_\mult = \Bar{c}\cdot \Bar{c'}$, output $\Bar{c}_\mult = \Bar{\tilde{c}}_\mult$ if $\Bar{c},\Bar{c'}$ are encrypted under the same extended key. Otherwise, for the set of $k$ additional keys in $\Bar{\pk}\cap\Bar{\pk'}$, let $c_0 = \Bar{\tilde{c}}_\mult$ and compute $ c_j = \langle \bg^{-1}(c_{j-1}), \ek_j \rangle \pmod q$. Output $\Bar{c}_\mult = c_k$. 

\item[-]{$\mathsf{LTV}.\dec((\sk_1,\dots,\sk_K),\Bar{c})$}: Given a set of secret keys $(\sk_1,\dots,\sk_K) = \{f_1,\dots,f_K\}$ and an extended ciphertext $\Bar{c}$, perform the decryption as $z = f_1,\dots,f_n\Bar{c} \in R_q$. Output the result as $m = z\pmod 2$. 
\end{itemize}
\end{pabox}

% add limitation of this technique, what can be done beyond onion encryption
L\'{o}pez-Alt~et~al.~\cite{lopez2012fly} proposed an MKHE based on NTRU HE scheme~\cite{stehle2010faster}, which we describe in Scheme~\ref{scheme:LTV}. Let $c_1 = h_1u_1 + 2e_1 + m_1$ and $c_2 = h_2u_2 + 2e_2 + m_2$ be two ciphertexts encrypted under two public keys $h_1, h_2$. We can add them and correctly decrypt to the sum using the joint secret keys of $f_1f_2$ as long as the noise $e_{\add}$ is not too large. Also, note that $f_1 \equiv f_2 \equiv 1 \pmod 2$.
\begin{align*}
    f_1f_2(c_1 + c_2) &= f_1f_2h_1u_1 + 2f_1f_2e_1 + f_1f_2m_1 + f_1f_2h_2u_2 + 2f_1f_2e_2 + f_1f_2m_2 \\
    &= f_1f_2(h_1u_1 + h_2u_2) + 2f_1f_2(e_1 + e_2) + f_1f_2(m_1 + m_2) \\
    &= 2(f_2g_1u_1 + f_1g_2u_2 + f_1f_2e_1 + f_1f_2e_2) + f_1f_2(m_1 + m_2) \\
    &= 2e_{\add}  + f_1f_2(m_1 + m_2) \\ 
    &= m_1 + m_2 \pmod 2
\end{align*}

Similarly, we can decrypt to the product using the joint secret keys $f_1f_2$ as following.
\begin{align*}
    f_1f_2(c_1c_2) &= f_1f_2(h_1u_1 + 2e_1 + m_1)  (h_2u_2 + 2e_2 + m_2) \\
    &= f_1f_2h_1u_1h_2u_2 + 2f_1f_2h_1u_1e_2 + f_1f_2h_1u_1m_1 + 2f_1f_2h_2u_2e_1 + 4f_1f_2e_1e_2 \\ &\;+ 2f_1f_2e_1m_2 + f_1f_2m_1h_2u_2 + 2f_1f_2m_1e_2 + f_1f_2m_1m_2 \\ 
    &= 4g_1u_1g_2u_2 + 4f_2g_1u_1e_2 + 2f_2g_1u_1m_1 + 4f_1g_2u_2e_1 + 4f_1f_2e_1e_2 \\ &\;+ 2f_1f_2e_1m_2 + 2f_1m_1g_2u_2 + 2f_1f_2m_1e_2 + f_1f_2(m_1m_2) \\
    &= 2e_{\mult}  + f_1f_2(m_1m_2) = m_1m_2 \pmod 2
\end{align*}

A naive decryption in LTV requires keys to be constructed based on the evaluated circuit. For example, the ciphertext $c_1^2c_2$ is decrypted by multiplying it with the keys $f_1^2f_2$, and $c_1c_2 + c_2c_3$ is decrypted with the keys $f_1f_2^2f_3$. In other words, the power of the secret key $f_i$ corresponds to the times the ciphertext $c_i$ is used in the evaluated circuit. This impacts the efficiency of the scheme since the size of the decryption key grows based on both the number of involved keys and the depth of the circuit. Instead, they apply key switching on evaluated ciphertexts to make keys independent of the circuit. Specifically, the technique makes a ciphertext decryptable with $f_i$ instead of $f^2_i$, e.g., the ciphertext $c_1c_2 + c_2c_3$ is decrypted with $f_1f_2f_3$ instead of $f_1f_2^2f_3$. 

% \fix{need to elaborate further}

%~\cite{clear2015multi, mukherjee2016two, brakerski2016lattice, peikert2016multi, dodis2016spooky} 
After the LTV scheme, many works propose MKHE constructions with different capabilities and security assumptions. We provide a  comparison of these different schemes in Table~\ref{tab:mkhe}. Observed from the table, we can improve the ciphertext size from growing quadratically in early works~\cite{lopez2012fly, mukherjee2016two, peikert2016multi, dodis2016spooky} to growing linearly with the number of keys in the recent works~\cite{brakerski2016lattice, chen2017batched, li2019efficient, chen2019MKSEAL}. The majority of schemes are multi-hop, which means an extended ciphertext can be further extended to additional keys after homomorphic evaluations. Most schemes~\cite{mukherjee2016two,peikert2016multi,chen2017batched,chen2019MKSEAL} also require key generation to use a public parameter in the CRS model. Moreover, the RLWE-based MKHE schemes~\cite{chen2017batched,aloufi2019collaborative, chen2019MKSEAL} support packing multiple messages in one ciphertext which enables SIMD evaluations. 

\begin{table*}[t]
% \centering{
\begin{center}
\caption{Comparison between proprieties of existing different MKHE schemes.}
\label{tab:mkhe}
\resizebox{\columnwidth}{!}{%
\begin{tabular}{|c|c|c|c|c|c|c|c|c|}
\hline
Scheme & Security & \begin{tabular}[c]{@{}c@{}}Ctxt size\\  growth\end{tabular} & \begin{tabular}[c]{@{}c@{}}Protocol\\ rounds\end{tabular} & Multi-hop & \begin{tabular}[c]{@{}c@{}}Bounded \#\\ of keys\end{tabular} & CRS & Packing & \begin{tabular}[c]{@{}c@{}}Need\\ bootstrapping\end{tabular}\\ \hline
% Scheme                            & Security              & C size    & Rnd & 1hop & 2keys & 3param & 4Pack & 5bootstrap
LTV12~\cite{lopez2012fly}           & NTRU                  & Quadratic & - & \checkmark & \checkmark &   &   & \\ \hline
CM15~\cite{clear2015multi}          & \multirow{4}{*}{LWE}  & Quadratic & - &  &  & \checkmark &  &  \\ \cline{1-1} \cline{3-9}
MW16~\cite{mukherjee2016two}        &                       & Quadratic & 2 &  &  & \checkmark &  & \\ 
\cline{1-1} \cline{3-9}
PS16~\cite{peikert2016multi}        &                       & Quadratic & 2 & \checkmark & \checkmark  & \checkmark &  & \\ 
\cline{1-1} \cline{3-9}
BP16~\cite{brakerski2016lattice}    &                       & Linear    & 2 & \checkmark &  & \checkmark &  & \checkmark \\ 
\hline
DHRW16~\cite{dodis2016spooky}       & piO                   & Quadratic & 2 & \checkmark &  &  &  & \checkmark\\ \hline
CZW17~\cite{chen2017batched}        & \multirow{5}{*}{RLWE}  & Linear    & 2 & \checkmark & \checkmark & \checkmark & \checkmark & \\ \cline{1-1} \cline{3-9}
YKHK18~\cite{yasuda2018multi}       &                       & Linear    & 2 & \checkmark & \checkmark & \checkmark & \checkmark& \\ \cline{1-1} \cline{3-9}
LZYH+19~\cite{li2019efficient}      &                       & Linear    & 2 & \checkmark & \checkmark & \checkmark & \checkmark & \\  \cline{1-1} \cline{3-9}
CDKS19~\cite{chen2019MKSEAL}        &                       & Linear    & 2 & \checkmark & \checkmark & \checkmark & \checkmark & \\ 
\cline{1-1} \cline{3-9}
AH19~\cite{aloufi2019collaborative} &                       & Constant  & 4~$^*$ & \checkmark & \checkmark & \checkmark & \checkmark & \\
\hline
CCS19~\cite{chen2019MKTFHE}         & TLWE                 & Linear    & 2 & \checkmark & \checkmark & \checkmark &  & \checkmark \\ \hline
\end{tabular}
}
\end{center}
\begin{flushleft}
\footnotesize{$^*$ Two extra rounds are for the ThHE setup phase to generate the joint key.} 
\end{flushleft}
\end{table*}

\subsubsection{Single-hop MKHE}
\label{subsec:single-hop}
Clear and McGoldrick~\cite{clear2015multi} proposed a compiler for a multi-identity homomorphic IBE scheme based on the GSW scheme~\cite{gentry2013homomorphic}. As discussed in Sec.~\ref{subsec:ABE}, this compiler produces ciphertexts that are encrypted under different identities, instead of only supporting single-identity in the original GSW scheme. Identities can be viewed as an analog to keys in HE schemes; hence, we can also obtain a multi-key HE scheme in this way. Mukherjee and Wichs~\cite{mukherjee2016two} improved this IBE-based MKHE and built a general two-round MPC protocol on top of it, including the first introduction of a detailed distributed decryption protocol. We describe the latter multi-key variant of the GSW in Scheme~\ref{scheme:MW}. 

\begin{pabox}[label={scheme:MW}]{The Mukherjee-Wichs MKHE scheme~\cite{mukherjee2016two}}
\begin{itemize}[leftmargin=*]
\item[-]\highlight{$\mathsf{MW}.\setup(1^\lambda, 1^L)$:} Given a security parameter $\lambda$ and a circuit depth $L$, run the $\mathsf{GSW}.\setup$ and choose $\boldsymbol{B} \sample \ZZ^{(n-1)\times n\ell}_q$ as a common shared matrix among users. Output the set of public parameters $\param = (q,n,\psi, \chi,\boldsymbol{B})$.

\item[-]{$\mathsf{MW}.\kgen(\param)$}: Given the public parameters $\param$, run the key generation algorithm $\mathsf{GSW}.\setup$ for each user and output the individual key pair $(\pk,\sk) = (\boldsymbol{A}, \boldsymbol{s})$ where $\boldsymbol{A} = (\boldsymbol{b}=\boldsymbol{\grave{s}}\boldsymbol{B}+\boldsymbol{e}, \boldsymbol{B})$ and $\boldsymbol{s}=(1, -\boldsymbol{\grave{s}})$ where $\grave{s}\sample \psi$.

\item[-]{$\mathsf{MW}.\enc(\pk,m)$}: To encrypt a message bit $m$ with the public key $\pk = \boldsymbol{A}$ and a randomness $\boldsymbol{R}$, run $\mathsf{GSW}.\enc$ to obtain the GSW ciphertext $\boldsymbol{C} = \boldsymbol{A}\boldsymbol{R} + m \boldsymbol{G}$. Additionally, encrypt each entry of the randomness $\boldsymbol{R}$ such that $\boldsymbol{U} = \{\boldsymbol{U}_{\alpha, \beta}\}_{\alpha, \beta = 1,\dots,n\ell}$ where  $\boldsymbol{U}_{\alpha, \beta} = \enc(\pk, r_{\alpha, \beta})$. Output the ciphertext as Output the ciphertext as the pair $\hat{\boldsymbol{C}} = (\boldsymbol{U}, \boldsymbol{C})$.

\item[-]{$\mathsf{MW}.\extend((\pk_1,\dots,\pk_K), \hat{\boldsymbol{C}})$}: Given a list of $K$ public keys $(\pk_1,\dots,\pk_K)$ and a ciphertext $\hat{\boldsymbol{C}} = (\boldsymbol{U},\boldsymbol{C})$ encrypting a message $m$ under $\pk_i$, compute the auxiliary matrix $\boldsymbol{X}_j = \boldsymbol{b}_j - \boldsymbol{b}_i$ for each $\pk_{j\neq i}$. Construct a matrix $\Bar{\boldsymbol{C}} \in \ZZ^{nK\times n\ell K}_q$ such that each sub-matrix $\Bar{\boldsymbol{C}}_{a, b} \in \ZZ^{n\times n\ell}$ for $a,b \in 1,\dots,K$ is defined as follows.
\[\Bar{\boldsymbol{C}}_{a, b} =     
\begin{cases}
      \boldsymbol{C}, & \text{if}\ a=b \\
      \boldsymbol{X}_j, & \text{if}\ a=i\neq j\ \text{and}\ b = j\\
      \boldsymbol{0}_{n\times n\ell}, & \text{otherwise}
    \end{cases}\]
% \[\Bar{\boldsymbol{C}} = \begin{pmatrix} \boldsymbol{C} & 0 & \cdots & 0  & 0\\
% 0 & \ddots & \cdots & \cdots & \vdots\\
% \vdots & 0 & 0 & \cdots & 0\\
% \boldsymbol{X}_1 & \cdots & \boldsymbol{C} & \cdots & \boldsymbol{X}_K \\
% 0 & \cdots & \cdots & \cdots & 0\\
% \vdots & \cdots& \ddots & \cdots & 0\\
% 0 & 0 & \cdots & 0 &\boldsymbol{C} \end{pmatrix}\]
Output $\Bar{\boldsymbol{C}}$ as the extended ciphertext under the concatenated key $\Bar{\pk} = (\pk_1,\dots,\pk_K)$. For the corresponding secret key $\Bar{\sk} = (\sk_1,\dots,\sk_K)$, the property $(\boldsymbol{s}_1,\dots,\boldsymbol{s}_K)\boldsymbol{C} \approx m (\boldsymbol{s}_1,\dots,\boldsymbol{s}_K)\boldsymbol{\bar{G}}$ holds, where $\boldsymbol{\bar{G}} = \boldsymbol{I}_{nK}\otimes\boldsymbol{g}$. 

\item[-]{$\mathsf{MW}.\eval(\Bar{\pk}, f, \Bar{\boldsymbol{C}}_1, \dots, \Bar{\boldsymbol{C}}_u)$}: Given a set of $u$ extended ciphertexts $(\Bar{\boldsymbol{C}}_1, \dots, \Bar{\boldsymbol{C}}_u)$, directly perform the original GSW evaluation algorithms $\mathsf{GSW}.\evaladd$ and $\mathsf{GSW}.\evalmult$ but in the extended dimensions $nK\times n\ell K$.    

\item[-]{$\mathsf{MW}.\dec(\Bar{\sk},\Bar{\boldsymbol{C}})$}:
Let $\Bar{\boldsymbol{C}}$ be an extended ciphertext $\Bar{\boldsymbol{C}}$ encrypted under $K$ concatenated keys $\Bar{\sk}$ and consisting of $K$ sub-matrices, such that $\boldsymbol{V}_i$ includes the ciphertext $\boldsymbol{C}$ and the auxiliary matrix $\boldsymbol{X}_i$. Each party $i$ performs locally the partial decryption using their secret $\sk_i$ and $\boldsymbol{V}_i$ and output $\rho_i = \boldsymbol{s}_i\Bar{\boldsymbol{C}}_i\Bar{\boldsymbol{G}}^{-1}(\Bar{\boldsymbol{w}}^T) + e_i$, where $\bar{\boldsymbol{w}}=(0,\dots,0,\ceil{q/2})$ and $e_i$ is a smudging noise. The message is retrieved as $m = \abs{\near{\frac{\sum^K_{i=1}(\rho_i)}{q/2}}}$.
\end{itemize}
\end{pabox}

Compared to the LTV scheme~\cite{lopez2012fly}, the new GSW-based scheme \textit{does not} limit the maximum number of keys the ciphertext can be extended to. But, the ciphertext \emph{cannot} be further extended after performing a homomorphic evaluation.

To understand how the scheme works, let us consider the following toy example. Let $\boldsymbol{C}_1 = \boldsymbol{A}_1\boldsymbol{R}_1 + m_1\boldsymbol{G}$ be a fresh GSW ciphertext encrypted with the public key $\pk_1 = \boldsymbol{A}_1 = (\boldsymbol{b}_1, \boldsymbol{B})$, where $\boldsymbol{B}$ is a shared matrix in the CRS model. For the corresponding secret key $\sk_1 = \boldsymbol{s}_1$, the property $\boldsymbol{s}_1\boldsymbol{C}_1 \approx m_1\boldsymbol{s}_1\boldsymbol{G}$ holds as mentioned in Sec.~\ref{subsubsec:GSW}. 

To involve this ciphertext in an homomorphic evaluation with another ciphertext under a different key, we need to extend it first to the new key. Suppose $\pk_2 = (\boldsymbol{b}_2, \boldsymbol{B})$ is another public key which encrypts a ciphertext $\boldsymbol{C}_2$. Before performing evaluation between $\boldsymbol{C}_1$ and $\boldsymbol{C}_2$, we need to extend these ciphertexts to the other public keys. We extend $\boldsymbol{C}_1$ to $\pk_2$ as $\bar{\boldsymbol{C}}_1 = \begin{pmatrix}\boldsymbol{C}_1 & 0 \\ 0 & \boldsymbol{C}_1 \end{pmatrix} \in \ZZ^{2n\times 2n\ell}$, which is encrypted under the concatenated key $(\pk_1,\pk_2)$. The same process applies to extending $\boldsymbol{C}_2$ to $\pk_1$, but we will focus our discussion on $\boldsymbol{C}_1$. Note that, the property $(\boldsymbol{s}_1,\boldsymbol{s}_2)\bar{\boldsymbol{C}}_1 \approx  m_1(\boldsymbol{s}_1,\boldsymbol{s}_2)\boldsymbol{\bar{G}}$, where $\boldsymbol{\bar{G}} = \boldsymbol{I}_{2n}\otimes\boldsymbol{g}$ and $\boldsymbol{I}_{2n}$ is the identity matrix of dimension $2n$, no longer holds. 

Observe that $\begin{pmatrix}\boldsymbol{s}_1 & \boldsymbol{s}_2 \end{pmatrix} \begin{pmatrix}\boldsymbol{C}_1 & 0 \\ 0 & \boldsymbol{C}_1 \end{pmatrix} = \begin{pmatrix}\boldsymbol{s}_1\boldsymbol{C}_1 & \boldsymbol{s}_2\boldsymbol{C}_1 \end{pmatrix}$. Since $\boldsymbol{s}_1$ corresponds to the public key $\pk_1 = \boldsymbol{A}_1$, we correctly get $\boldsymbol{C}_1\boldsymbol{s}_1 \approx m_1\boldsymbol{s}_1\boldsymbol{G}$ for the first element. Note that a correct decryption should obtain $m_1(\boldsymbol{s}_1,\boldsymbol{s}_1)\boldsymbol{G}$. However, decrypting with $\boldsymbol{s}_2$ does not retrieve the message $m_1\boldsymbol{s}_2\boldsymbol{G}$ alone because $\boldsymbol{s}_2$ does not cancel $\pk_1$ as $\boldsymbol{s}_2\boldsymbol{A}_1 \approx 0$. Hence, it yields a ``lingering'' element $(\boldsymbol{b}_1 - \boldsymbol{b}_2)\boldsymbol{R}_1$ according to the decryption of $\boldsymbol{s}_2\boldsymbol{C}_1$ shown below. %Then, we obtain the message $m$ by summing $m_1\boldsymbol{s}_1\boldsymbol{G}+m_1\boldsymbol{s}_2\boldsymbol{G}$, as described in the decryption step later.
\begin{align*}
\boldsymbol{s}_2(\boldsymbol{A}_1\boldsymbol{R}_1 + m_1\boldsymbol{G}) &= \boldsymbol{s}_2\boldsymbol{A}_1\boldsymbol{R}_1 + m_1\boldsymbol{s}_2\boldsymbol{G}
\\ &= (1, -\boldsymbol{\grave{s}}_2)(\boldsymbol{b}_1, \boldsymbol{B})\boldsymbol{R}_1 + m_1\boldsymbol{s}_2\boldsymbol{G}
\\ &= (1, -\boldsymbol{\grave{s}}_2)(\boldsymbol{\grave{s}}_1\boldsymbol{B}\boldsymbol{R}_1 + \boldsymbol{e}, \boldsymbol{B}) + m_1\boldsymbol{s}_2\boldsymbol{G}
\\ &= (\boldsymbol{\grave{s}}_1\boldsymbol{B} - \boldsymbol{\grave{s}}_2\boldsymbol{B})\boldsymbol{R}_1 + m_1\boldsymbol{s}_2\boldsymbol{G}
\\ &= (\boldsymbol{b}_1 - \boldsymbol{b}_2)\boldsymbol{R}_1 + m_1\boldsymbol{s}_2\boldsymbol{G}
\end{align*}
To solve this issue, we have to provide an auxiliary matrix $\boldsymbol{X}_2$ such that $ \boldsymbol{s}_1\boldsymbol{X}_2 = (\boldsymbol{b}_2 - \boldsymbol{b}_1)\boldsymbol{R}_1$, which is needed to eliminate this lingering element. To create this matrix, we need to provide the randomness $\boldsymbol{R}_1$ that is used to encrypt the message $m_1$ for semantic security but \emph{without} disclosing it to avoid security breach as we discussed in Sec.~\ref{subsec:THE}. We encrypt each element $r_{\alpha,\beta}$ of the randomness matrix $\boldsymbol{R}_1$ under $\pk_1$ as $\boldsymbol{U}_{\alpha, \beta} = \boldsymbol{R'}_{\alpha, \beta}\boldsymbol{A}_1 + r_{\alpha, \beta}\boldsymbol{G}$, where $\alpha, \beta$ are in the indexes and $\boldsymbol{R'}$ is another randomness. We provide this encrypted randomness $\boldsymbol{R_1}$ with the ciphertext $\boldsymbol{C}_1$ as the tuple $(\boldsymbol{U}, \boldsymbol{C}_1)$. 

At the ciphertext extension step, we define a matrix $\boldsymbol{Z}_{\alpha, \beta} = (\boldsymbol{0}_{n\times n\ell-1}, \boldsymbol{b}_2 - \boldsymbol{b}_1) \in \ZZ_q^{n\times n\ell}$, which means all entries are $0$ except the last column is the vector $\boldsymbol{b}_2 - \boldsymbol{b}_1$. Note that $\boldsymbol{b}_1, \boldsymbol{b}_2$ are  components of the public keys $\pk_1, \pk_2$, respectively. Then we compute $\boldsymbol{X}_2 = \sum_{\alpha, \beta} \boldsymbol{U}_{\alpha, \beta}\cdot \boldsymbol{G}^{-1}(\boldsymbol{Z}_{\alpha, \beta})$, such that the property $\boldsymbol{s_1}\boldsymbol{X}_{2,\alpha, \beta} \approx r_{\alpha, \beta}(\boldsymbol{b}_2 - \boldsymbol{b}_1)$ holds. Finally, output the extended ciphertext as $\bar{\boldsymbol{C}}_1 = \begin{pmatrix}\boldsymbol{C}_1 & \boldsymbol{X}_2 \\ {\bf{0}} & \boldsymbol{C}_1 \end{pmatrix}$. 

This process can be generalized to extending $\boldsymbol{C}_1$ to $K$ different keys. For each additional key $\pk_j$, we compute the matrix $\boldsymbol{X}_j$ such that $\boldsymbol{s}_1\boldsymbol{X}_j \approx (\boldsymbol{b}_j - \boldsymbol{b}_1)\boldsymbol{R}_1$. The new structure of the extended ciphertext is illustrated in Fig.~\ref{fig:single-hop} as follows.
\[\Bar{\boldsymbol{C}}_1 = \begin{pmatrix} \boldsymbol{C}_1 & \boldsymbol{X}_2 & \cdots & \boldsymbol{X}_K\\
{\bf{0}} & \boldsymbol{C}_1 & \cdots & {\bf{0}} \\
\vdots & \vdots & \ddots & \vdots \\
{\bf{0}} & {\bf{0}} & \cdots & \boldsymbol{C}_1 \end{pmatrix} \in \ZZ^{nK\times n\ell K}\]
As observed, the dimension of the extended ciphertext matrix can increase quadratically with the number of involved keys because we have to add the auxiliary matrix and an additional copy of the ciphertext. With each extension, the dimension of the gadget matrix $\boldsymbol{G}$ is also changed in correspondence with the size of the extended key. Namely, we create a $(nK\times n\ell K)$ matrix where the diagonal is the vector $\boldsymbol{g} = (2^0,\dots,2^{\ell-1})$.

Homomorphic evaluation on extended ciphertexts performs the usual GSW addition and multiplication. The only difference is performing on ciphertexts with higher dimensions based on the $K$ involved keys.

Let $\bar{\boldsymbol{C}}_{\Eval}$ be the result of some evaluations on $\bar{\boldsymbol{C}}_1$ with respect to the concatenated key $(\pk_1, \pk_2)$. Suppose that we want to extend the result to an additional key $\pk_3$. In this case, we must provide an encryption of the randomness $\boldsymbol{R}_\Eval$, associated with the ciphertext $\boldsymbol{\bar{C}}_\Eval$, under the corresponding set of keys so it can be used to generate $\boldsymbol{X}_3$. However, it is not clear how to obtain such an encryption for $\boldsymbol{R}_\Eval$. Note the randomness is affected by the homomorphic evaluation. For example, the product of $\boldsymbol{\bar{C}}_1$ with another ciphertext $\boldsymbol{\bar{C}}_2$ under the same keys $(\pk_1, \pk_2)$ may contain randomness $\boldsymbol{R}_\Eval \approx \boldsymbol{R}_1 \cdot \boldsymbol{R}_2$, where $\boldsymbol{R}_2$ is a randomness associated with $\boldsymbol{\bar{C}}'_2$. Constructing such term from the encryptions is not trivial. The scheme does not describe either a method to dynamically extend the matrix $\boldsymbol{X}_1$ to other keys. Hence, this renders a single-hop scheme. 

Given a ciphertext $\boldsymbol{\bar{C}}$ under $(\pk_1,\dots,\pk_K)$, previous MKHE schemes~\cite{lopez2012fly,clear2015multi} assumed the presence of a \textit{trusted} party who holds all the corresponding secret keys $(\sk_1,\dots, \sk_K)$ and can directly decrypt. To achieve a stronger security and avoid disclosing individual secret keys to anyone, this scheme performs a one-round distributed decryption protocol (similar to the ThHE technique). Let $\bar{\boldsymbol{C}}$ be a ciphertext extended under $K$ keys and composed of $K$ sub-matrices such that $\Bar{\boldsymbol{C}} = (\Bar{\boldsymbol{C}}_1, \dots, \Bar{\boldsymbol{C}}_K)$, $\Bar{\boldsymbol{C}}_i \in \ZZ^{n\times n\ell K}_q$. All $K$ participants will agree on a vector $\Bar{\boldsymbol{w}} = (0,\dots,0,\ceil{q/2}) \in \ZZ^{nK}$. In the one round of the protocol, each participant $i$ partially decrypts $\Bar{\boldsymbol{C}}$ by locally computing a masked decryption component \highlight{$\rho_i = \boldsymbol{s}_i\Bar{\boldsymbol{C}}_i\Bar{\boldsymbol{G}}^{-1}(\Bar{\boldsymbol{w}}^T) + e_i$}, where $e_i$ is a smudging noise used to protect $\boldsymbol{s}_i$ under LWE according to the lemma in Sec.~\ref{subsec:distdec}. After receiving the broadcast components $(\rho_1,\dots, \rho_K)$, the intended participant performs the final decryption by aggregating components as $\rho = \sum^K_{i=1}(\rho_i)$ and computing $m = \abs{\near{\frac{\rho}{q/2}}}$ to retrieve the message. %\YS{please check this}

Based on this scheme, Mukherjee and Wichs~\cite{mukherjee2016two} constructed an MPC protocol with two rounds to enable secure computation with multiple keys. In the first round, users encrypt their inputs with their individual keys and output the ciphertexts and encryption of the randomness. Then, the evaluator can dynamically extend a ciphertext to $K$ different keys without knowing any secret components, such as the encrypted randomness, and perform the homomorphic evaluations. In the second round, each user helps in a distributed decryption protocol by partially decrypting a part of the extended ciphertext such that the message is retrieved by combining these partial decryptions. 

As mentioned, there is no limit on the number of involved keys $K$. Yet, the scheme is still not robust because no additional keys, or even further homomorphic evaluations, can be supported on evaluated ciphertexts. Moreover, the ciphertext size expansion is significant, which affects the overall practicality of the scheme. We discuss techniques for realizing multi-hop MKHE schemes in the following section.

\subsubsection{Multi-hop MKHE} 
\label{subsec:multi-hop}
Recent MKHE schemes adopt new design that overcomes the limitations of earlier MKHE schemes and allows ciphertext extension after homomorphic evaluation. We review the state-of-the-art MKHE schemes and show how these schemes adopt the new design to support multi-hop.

\subsubsection*{Multi-hop MK-variant of GSW} 
Building on the single-hop Mukhenrjee-Wichs scheme~\cite{mukherjee2016two}, two concurrent works~\cite{peikert2016multi, brakerski2016lattice} proposed techniques to extend it to the multi-hop setting. 

Peikert and Shiehian~\cite{peikert2016multi} proposed a \textit{leveled} scheme that is based on the Mukhenrjee-Wichs scheme but has a new ciphertext structure and extension function. In the setup step, a public parameter in the form of uniformly LWE matrix $\boldsymbol{B}\sample \ZZ^{n\times 2n\ell}_q$ is chosen. The ciphertext in the new scheme consists of three components $(\boldsymbol{C}, \boldsymbol{F}, \boldsymbol{D})$ compared to the former ciphertext tuple $(\boldsymbol{C}, \boldsymbol{U})$. The component $\boldsymbol{C} \in \ZZ^{n\times n\ell}_q$ remains the same as an original GSW ciphertext of the message $m$. The new component $\boldsymbol{F} = \boldsymbol{BR}+m\boldsymbol{G} \in \ZZ^{n\times n\ell}_q$ is the commitment of the message $m$ under the randomized public parameter $\boldsymbol{B}$. Note the message is protected since the randomness $\boldsymbol{R}$ is not disclosed. The randomness is encrypted and provided as a ciphertext $\boldsymbol{D} \in \ZZ^{2n^2\ell^2\times n\ell}_q$ such that $\boldsymbol{s}\boldsymbol{D} \approx \boldsymbol{R}$ holds. 

Given a ciphertext $(\boldsymbol{C}, \boldsymbol{F}, \boldsymbol{D})$ under $\boldsymbol{s}_1$, we can extend it to a second key $\boldsymbol{s}_2$ as $(\bar{\boldsymbol{C}}, \bar{\boldsymbol{F}}, \bar{\boldsymbol{D}})$, where $\bar{\boldsymbol{F}}=\boldsymbol{F}$ and $\bar{\boldsymbol{D}}$ is extended by padding it with rows and columns of zeros such that the new padded component is $\bar{\boldsymbol{D}} = \boldsymbol{I}_{2n\ell} \otimes \begin{pmatrix}\boldsymbol{I}_{n}\\ \boldsymbol{0}_{n\times n}\end{pmatrix} \cdot \boldsymbol{D} \in \ZZ^{4n^2\ell^2\times n\ell}_q$. Note that we denote all extended elements with a bar on the top, such as $\bar{\boldsymbol{D}}$.
The new structure of the extended ciphertext is $\bar{\boldsymbol{C}} = \begin{pmatrix}\boldsymbol{C} & \boldsymbol{X} \\ 0 & \boldsymbol{F} \end{pmatrix}$. Note that decrypting $\boldsymbol{s}_2\boldsymbol{F} = \boldsymbol{s}_2(\boldsymbol{BR}+m\boldsymbol{G}) = \boldsymbol{b}_2\boldsymbol{R}+m\boldsymbol{s}_2\boldsymbol{G}$
where $\boldsymbol{b}_2\boldsymbol{R}$ is a lingering term.
The component $\boldsymbol{X}$ encrypts information of the inverse of this lingering term such that $\boldsymbol{s}_1\boldsymbol{X} \approx -\boldsymbol{b}_2\boldsymbol{R}$ which cancels the resulted lingering term. 

\begin{figure}[t]
\centering
  \subfloat[Single-hop MK-GSW~\cite{mukherjee2016two}]
 {\includegraphics[width=.47\textwidth]{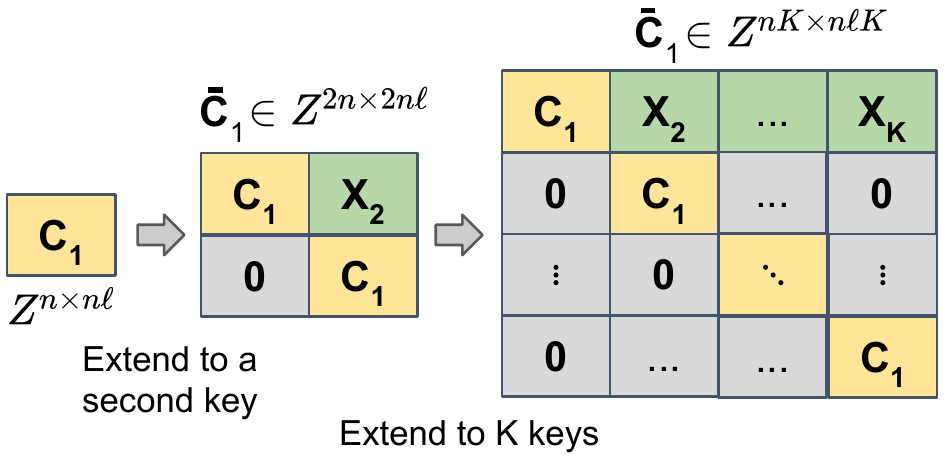} \label{fig:single-hop}}
 \hspace{.5cm}
 \subfloat[Multi-hop MK-GSW~\cite{peikert2016multi}]
 {\includegraphics[width=.47\textwidth]{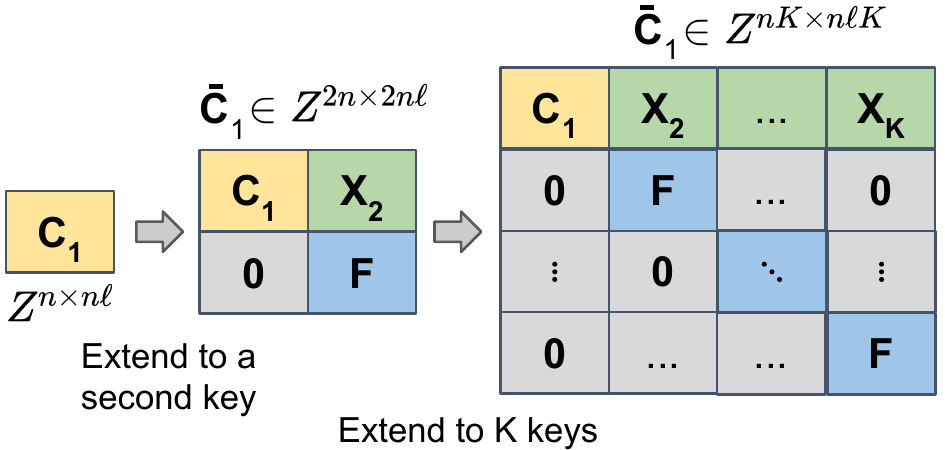} \label{fig:multi-hop}}
\caption{A comparison between single-hop and multi-hop MK-variants of the GSW scheme}
\label{fig:mkgsw}
\end{figure}

With this new design, as illustrated in Fig.~\ref{fig:multi-hop}, it is possible to further extend the ciphertext $\bar{\boldsymbol{C}}$ to $K$ additional keys. Note in the Fig.~\ref{fig:mkgsw}, some elements in the extended ciphertexts have been substituted from $\boldsymbol{C}_1$ to $\boldsymbol{F}$. The new scheme is multi-hop because the randomness used in $\boldsymbol{F},\boldsymbol{D}$ is tied to the same public parameter rather than a new key at every key extension process. Hence, there is no need to provide an encryption of some randomness under a specific key to construct the cancelling term, as discussed in Sec.~\ref{subsec:single-hop}.

% With this new design, it is possible to further extend the ciphertext $\bar{\boldsymbol{C}}$ to additional keys since the two components $\boldsymbol{F},\boldsymbol{D}$ are tied to the public parameter rather than on a specific key. Figure~\ref{} Hence, this solves the issue discussed in Sec.~\ref{subsec:single-hop}, which requires providing the encryption of the randomness corresponding to the evaluated ciphertext. 

The size of the ciphertext grows \textit{quadratically} with the number of involved keys. Hence, the scheme is limited by a bounded number of keys. An alternative scheme is introduced by authors to obtain smaller ciphertexts. Specifically, the extended ciphertext remains as a GSW encryption but the additional extension information is embedded in its corresponding public key instead. This scheme yields smaller ciphertexts but larger public keys, which reduces the space overhead since the number of evaluated ciphertexts is likely to be more than the number of extended public keys. Both schemes are leveled, which support homomorphic evaluations up to a predefined level, but they can be made fully through the bootstrapping technique.

Brakerski and Perlman~\cite{brakerski2016lattice} proposed a \textit{fully} MK-variant of GSW scheme to overcome the single-hop limitation in \cite{mukherjee2016two}. Ciphertexts in the single-hop scheme cannot be further extended without being decrypted and encrypted again. This process can be done homomorphically via bootstrapping, as discussed in Sec.~\ref{subsec:bootstrapping}. Hence, the Brakerski-Perlman method depends on the use of bootstrapable NAND gates for evaluation. Note that a NAND gate has a function completeness property, which means we can support any other operation by combining a set of NAND gates. Let $\boldsymbol{C}_1$ and $\boldsymbol{C}_2$ be two ciphertexts encrypted under two different keys $\pk_1$ and $\pk_2$. Evaluating a bootstrapable NAND on the two ciphertexts homomorphically performs $\mathsf{NAND}(\dec(\sk_1, \boldsymbol{C}_1),\dec(\sk_2, \boldsymbol{C}_2))$ and yields the extended ciphertext $\bar{\boldsymbol{C}}$ that is encrypted under the concatenated keys $(\pk_1,\pk_2)$. As observed, we must provide as inputs the secret keys $\sk_1, \sk_2$ encrypted under their corresponding public keys to enable homomorphic decryption. Therefore, a circular-security (see Sec.~\ref{subsec:assumptions}) assumption is required. Moreover, the size of the ciphertexts can be significantly reduced in this scheme. As discussed in Sec.~\ref{subsubsec:GSW}, the decryption process can be sufficiently done with one single column vector of the GSW ciphertext. Since we need to perform decryption first in the evaluation, the scheme proposes to keep the last column of the ciphertext and discard the rest. Hence, the size of the ciphertexts grows linearly, instead of quadratically, with the number of involved keys. There is no bound on the number of because ciphertexts are refreshed after each evaluation which keeps the noise level reduced. However, the bootstrapping can be slow and the efficiency of the scheme may degrade as the number of bootstrapable gates increases.

\subsubsection*{MK-variant of BGV}
\label{subsec:MKBGV}
The BGV scheme~\cite{brakerski2012leveled} is one among the first RLWE-based schemes to enable ciphertext packing, where multiple messages can be packed in one ciphertext to support efficient SIMD computation. The first attempt of extending the BGV scheme~\cite{brakerski2012leveled} to multi-key setting was by L\'{o}pez-Alt~et~al.~\cite{lopez2012fly}. Unfortunately, this initial MKHE scheme did not support key switching technique (see Sec.~\ref{subsec:keyswitching}) or bootstrapping which prevented extending it to a leveled or FHE scheme. 

Later, Chen~et~al.~\cite{chen2017batched} proposed a multi-key variant of the BGV scheme that is multi-hop and supports extending a ciphertext to a bounded number of keys. We describe this new multi-key BGV (MKBGV) scheme in Scheme~\ref{scheme:MKBGV}. Similar to other LWE-based MKHE schemes, the MKBGV scheme is designed in the CRS model. Participants use a pre-shared vector $\boldsymbol{a}\sample R_q^\ell$ to generate their key pairs. Messages are initially encrypted under a single key following the base BGV encryption function discussed in Sec.~\ref{subsubsec:BGV}. 

\begin{pabox}[label={scheme:MKBGV}]{An MK-variant of BGV scheme~\cite{chen2017batched}}
\begin{itemize}[leftmargin=*]
\item[-] {$\mathsf{MKBGV}.\setup(1^\lambda, 1^L, 1^\mathcal{K})$}: Given the security parameter $\lambda$, a multiplicative depth $L$, and a bound $\mathcal{K}$ on the number of keys, run the $\mathsf{BGV}.\setup$ function. Output the public parameters $\param = (d, q, t, \chi, \psi, \boldsymbol{a})$ where $\boldsymbol{a}\sample R^{\ell}_{q}$ is the CRS vector.

\item[-]{$\mathsf{MKBGV}.\kgen(\param)$}: Given the public parameters $\param$, run $\mathsf{BGV}.\kgen$ to generate a key pair $(\sk, \pk)$ as $\sk = s \sample \psi$, and $\pk = (b=-as+te, a)$ where $a\sample R_q$ and $e\sample \chi$.

\item[-]$\mathsf{MKBGV}.\evalkgen(\sk)$: Given the secret $\sk=s$, generate a helper component $\ek'$ as a set of ciphertexts encrypting each bit of secret key $s$ and their encryption randomnesses under a ring variant of GSW scheme, where messages are ring elements instead of bits. For $0\le i<\ell$, output the RGSW ciphertexts
$\Theta_i = \mathsf{RGSW}.\enc(\pk,g_i\cdot s)$, $\Psi_i = \mathsf{RGSW}.\enc(\pk,\bg^{-1}(s)[i])$. Similarly, provide RGSW encryptions of the randomnesses $r_i, r'_i$ used in $\Psi_i$ and $\Theta_i$ such that  $F_i = \mathsf{RGSW}.\enc(\pk, r_i)$ and $F'_i = \mathsf{RGSW}.\enc(\pk, r'_i)$. Output the helper element as $\ek'=\{(\Theta_{i},F_{i}),(\Psi_{i},F'_{i})\}_{i\in\{0,\dots,\ell-1\}}$.

\item[-]{$\mathsf{MKBGV}.\enc(\pk,m)$}: Given a message $m\in R_t$ and a public key $\pk$, perform $\mathsf{BGV}.\enc$ to obtain the ciphertext as $\boldsymbol{c} = (c_0, c_1)\in R^{2}_{q}$ where $c_0 = rb+m+te_0$ and $c_1 = ra+te_1$. Assume each participant has a unique index $i$ and initiate an ordered set $\mathcal{S} = (i)$ for the ciphertext. Output the fresh BGV ciphertext as the tuple $\hat{\boldsymbol{c}} = (\boldsymbol{c}, \mathcal{S}, L)$, where $L$ is the initial level. 

\item[-]{$\mathsf{MKBGV}.\extend((\pk_1,\dots,\pk_K), \hat{\boldsymbol{c}})$}: To extend a BGV ciphertext to one encrypted under a set of $K$ users' keys, simply set the ciphertext $\boldsymbol{c}$ as a concatenated $K$ sub-vectors $\Bar{\boldsymbol{c}} = (\boldsymbol{c}'_1,\dots,\boldsymbol{c}'_K)\in R^{2N}_{q_l}$, such that $\boldsymbol{c}'_i = \boldsymbol{c}_i$ if the index $i\in \mathcal{S}$, and $\boldsymbol{c}'_i = 0$ otherwise. Update the index set $\mathcal{S}$ and the level correspondingly.    

\item[-]{$\mathsf{MKBGV}.\eval(\Bar{\pk}, f, \Bar{\boldsymbol{c}}, \Bar{\boldsymbol{c}}')$}: Given two extended ciphertexts $\Bar{\boldsymbol{c}}, \Bar{\boldsymbol{c}}' \in R_{q_l}^{2K}$ encrypted under the same concatenated key $\Bar{\pk}$, perform homomorphic addition as $\Bar{\boldsymbol{c}}_{\mathtt{add}} = \Bar{\boldsymbol{c}} + \Bar{\boldsymbol{c}}'\pmod q$ and homomorphic multiplication as  $\Bar{\boldsymbol{c}}_{\mathtt{mult}} = \Bar{\boldsymbol{c}} \otimes \Bar{\boldsymbol{c}}'\pmod q$. %Then perform the $\mathsf{KeySwitch}$ technique using the generated evaluation key $\Bar{\ek}$ to reduce the ciphertext dimension and the $\mathsf{ModulusSwitch}$ technique to decrease the resultant noise by switching to a smaller ciphertext modulus. 

\item[-]$\mathsf{MKBGV}.\relin(\{\ek'_1,\dots,\ek'_K\}, \tilde{\boldsymbol{c}}_\mult)$: Given a set of helper elements, generate an evaluation key $\Bar{\ek}$ for the extended initial ciphertext product $\Bar{\boldsymbol{c}}_\mult$ as follows.
For each user $1\le k\le K$, extend each RGSW encryption in $\ek'_k$ to the other set of keys $\{\pk_{j\neq k}\}_{1\le j\le K}$ to obtain $\Bar{\ek}'$. Then, homomorphically compute $\Bar{\ek} = \Bar{\ek}' \otimes \Bar{\ek}'$, which encrypts $\Bar{\sk}^2$. Use this extended evaluation key to perform relinearization as $\Bar{\boldsymbol{c}}_\mult = \mathsf{BGV}.\relin(\Bar{\ek}, \Bar{\boldsymbol{c}}_\mult)$. 

\item[-]{$\mathsf{MKBGV}.\dec((\sk_1,\dots,\sk_K),\Bar{\boldsymbol{c}})$}: Given an extended ciphertext $\Bar{\boldsymbol{c}}$, and the secret key $\Bar{s}$ concatenating all the secret keys of users in the set $\mathcal{S}$. Decrypt as $(\langle\Bar{\boldsymbol{c}},\Bar{s}\rangle \pmod q) \pmod t = (\sum^{K}_{i=1}\langle \boldsymbol{c}'_i, s_{i}\rangle \pmod q) \pmod t$. 
\end{itemize}
\end{pabox}

\begin{figure*}[t]
\centering
  \subfloat[Ciphertext extension]
 {\includegraphics[width=.4\textwidth]{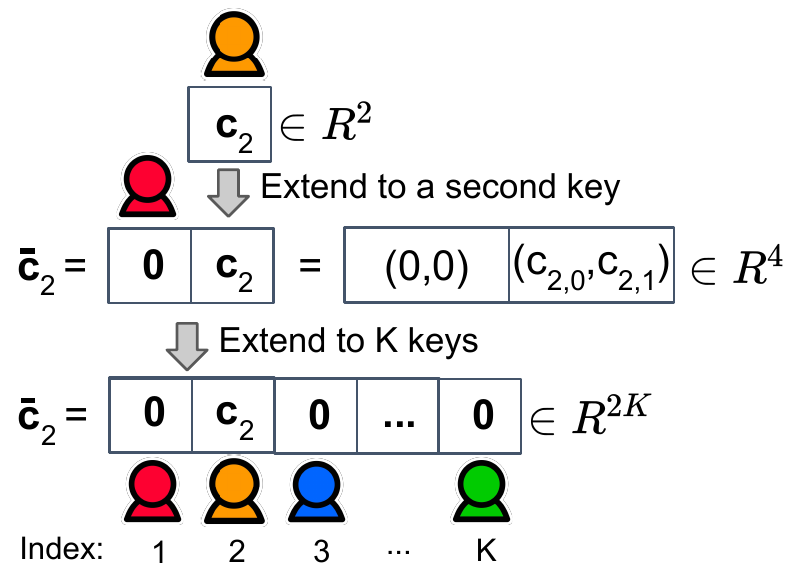} \label{fig:bgvextend}}%\hfill
 \hspace{.5cm}
 \subfloat[Homomorphic evaluation]
 {\includegraphics[width=.5\textwidth]{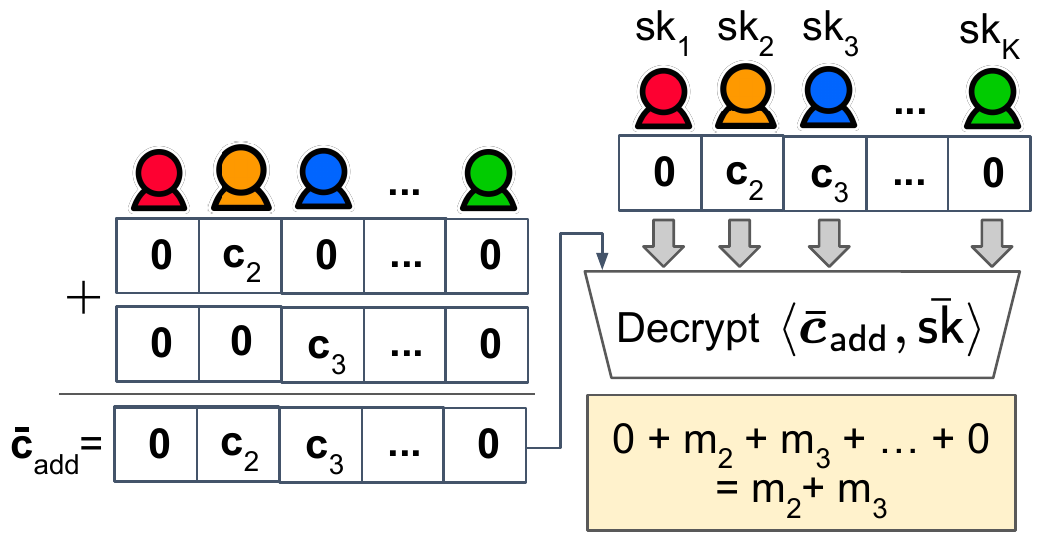} \label{fig:bgvdec}}%\hfill
\caption{MK-variant of the BGV scheme}
\label{fig:mkbgv}
\end{figure*}

The scheme supports extending the ciphertexts to at most $\mathcal{K}$ keys. Let $K \leq \mathcal{K}$ be the number of participants who wish to compute with their keys. Each participant is assigned a unique index $i \in \{1,\dots,K\}$ such their index is recorded in a special ordered set $\mathcal{S}$ for each ciphertext under the corresponding key. Let $\boldsymbol{c}_1\in R^2_q$ be a BGV ciphertext encrypting $m_1$ under the public key $\pk_1$. The set for the ciphertext $\boldsymbol{c}_1$ is populated with the index $1$ as $\mathcal{S} = (1)$ since the encryption is under the first participant's key $\pk_1$. Extending the ciphertext to a second key $\pk_2$ is done by constructing a new ciphertext with two slots as $\bar{\boldsymbol{c}}_1 = (\boldsymbol{c}'_1, \boldsymbol{c}'_2) \in R^{4}_q$. Each slot holds a sub-ciphertext corresponding to the index set. In particular, we define $\boldsymbol{c}'_1 = \boldsymbol{c}_1\in R^{2}_q$ as the original ciphertext since the index $1 \in \mathcal{S}$. The second sub-ciphertext is initiated as zero $\boldsymbol{c}'_2 = (0,0)\in R^{2}_q$ because the index $2 \notin \mathcal{S}$. The index set is updated after the extension to include the new index $\mathcal{S}=(1,2)$. In a similar manner, we can further extend this ciphertext to additional $K$ keys as illustrated in  Fig.~\ref{fig:bgvextend}. The size of the ciphertext increases \textit{linearly} with the number of keys such that for a ciphertext encrypted under $K$ keys, we have $\bar{\boldsymbol{c}} \in R^{2K}_q$. 

Homomorphic evaluations can be performed on the extended ciphertexts. Given two ciphertexts $\bar{\boldsymbol{c}}_1, \bar{\boldsymbol{c}}_2 \in R^{4}_q$ encrypted under the same set of keys, their sum is obtained through slot-wise homomorphic addition $\bar{\boldsymbol{c}}_{\add} = \bar{\boldsymbol{c}}_1 + \bar{\boldsymbol{c}}_2 \in R^{4}_q$. Note that sub-ciphertexts in different slots do not interact directly with each other. Rather, we perform slot-wise operations and the final result can be aggregated at the decryption step as shown in the example in Fig.~\ref{fig:bgvdec}. On the other hand, homomorphic multiplication is done by performing tensor-product of the two ciphertexts. The initial product is a long ciphertext $\bar{\boldsymbol{c}}_{\mult} = \bar{\boldsymbol{c}}_1 \otimes \bar{\boldsymbol{c}}_2 \in R^{16}_q$ that is encrypted under $\bar{\boldsymbol{s}}\otimes \bar{\boldsymbol{s}}$ instead of $\bar{\boldsymbol{s}}$. We need to apply key switching to convert $\bar{\boldsymbol{c}}_{\mult}$ to an encryption under $\bar{\boldsymbol{s}}$ and reduce its dimension to $R^{4}_q$. As mentioned in Sec.~\ref{subsec:keyswitching}, this process requires providing an evaluation key $\ek$, which encrypts the long secret key $\bar{\boldsymbol{s}}\otimes \bar{\boldsymbol{s}}$. To facilitate the generation of the evaluation key corresponding to the extended secret key, the scheme proposes a ring-GSW scheme, which allows encrypting ring elements. Specifically, each participant provides a GSW encryption of their secret key $\boldsymbol{s}_i$ such that each row can be viewed as a BGV ciphertext. Then, these ciphertexts can be extended to their concatenated keys. Lastly, one homomorphic multiplication is performed to obtain the encryption of $\bar{\boldsymbol{s}}\otimes \bar{\boldsymbol{s}}$ under the extended key that is used in the key switching procedure.  

The decryption process can be done with a two-round distributed decryption protocol similar to previous schemes~\cite{mukherjee2016two,peikert2016multi,brakerski2016lattice}. It requires all participants, whose indexes are in $\mathcal{S}$, to collaborate in a distributed decryption protocol. Each participant uses their secret key $\sk_i$ to decrypt the corresponding sub-ciphertext $\boldsymbol{c}'_i$. Observe that decryptions of zero sub-ciphertexts are cancelled and the final result can be obtained through the aggregation of the decrypted BGV sub-ciphertexts.
% The decryption is correct since each secret key $s_{i}$ decrypts the corresponding sub-vector ciphertext $\boldsymbol{c}'_i$, and only one of these sub-vectors is an encryption of the message $m$ while others are all zero and canceled out. 

\begin{figure*}[t]
\centering
  \subfloat[Nested ciphertext extension]
 {\includegraphics[width=.4\textwidth]{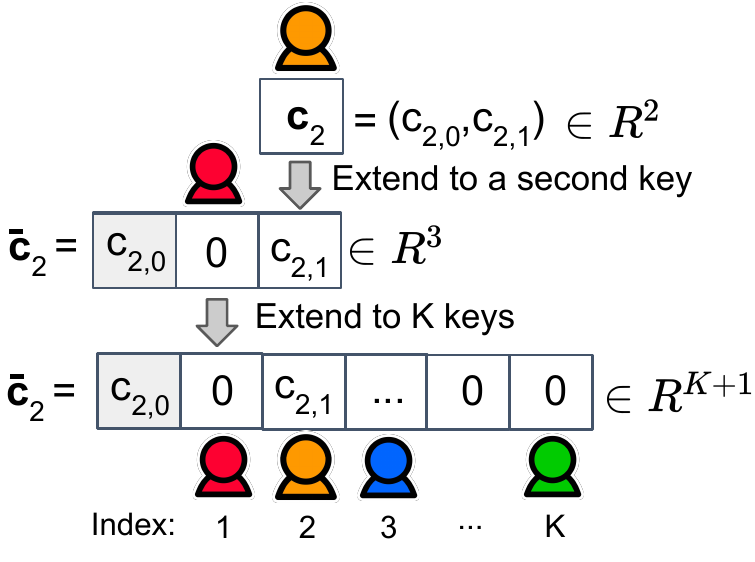} \label{fig:bgvk1extend}}
 \hspace{.3cm}
 \subfloat[Nested ciphertext evaluation]
 {\includegraphics[width=.56\textwidth]{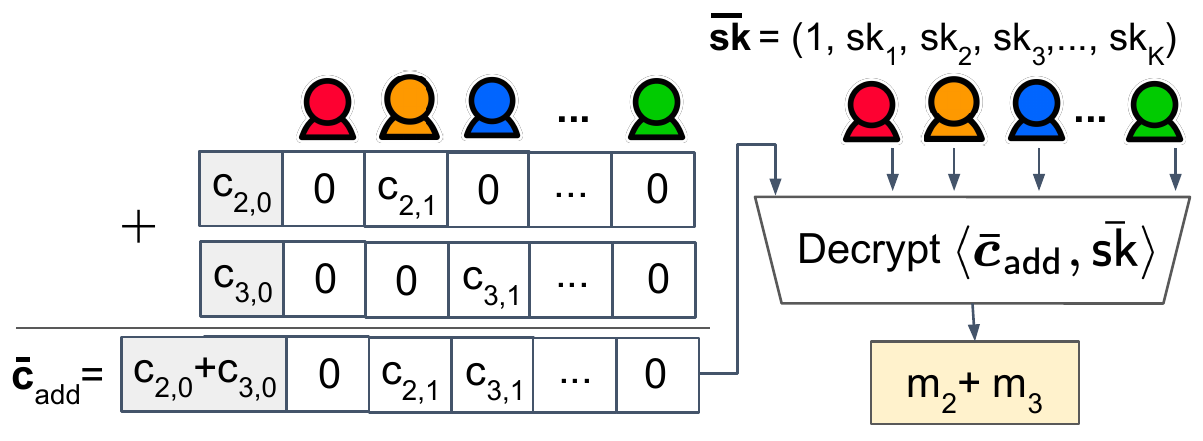} \label{fig:bgvk1dec}}
\caption{Improved MK-variant of the BGV scheme with reduced ciphertexts}
\label{fig:mkbgvk1}
\end{figure*}

Li~et~al.~\cite{li2019efficient} proposed an alternative design for MKBGV scheme that reduced the size of extended ciphertexts roughly by half. In particular, the dimension (number of ring elements) of ciphertext extended under $K$ keys will be $K+1$ instead of $2K$. Recall that a BGV ciphertext has two components $\boldsymbol{c} = (c_0, c_1)\in R^2_q$, where $c_0$ includes the message and $c_1$ encodes the randomness used in encryption. In the initial MKBGV scheme, the extended ciphertext under $K$ keys has each slot holding two ring elements as a full BGV ciphertext or padded with zeros $(0,0)$. Sub-ciphertexts still need to be combined at decryption step. However, the new scheme follows a nested ciphertext design which allocates the first slot of the extended ciphertext to the first ciphertext component $c_0$ which holds the message. Subsequent $K$ slots are allocated to the second ciphertext component which holds the randomness or padded with zero $0$ as illustrated in Fig.~\ref{fig:bgvk1extend}. For example, we can extend a ciphertext $\boldsymbol{c}_1$, with a corresponding index set $\mathcal{S}=(1)$, to a second key $\pk_2$ as $\bar{\boldsymbol{c}}_1 = (c'_0, c'_1, c'_2)\in R^3_q$ such that $c'_0 = c_0$ and $c'_i= c_i$ if $i \in \mathcal{S}$ or $c'_i=0$ otherwise.

Homomorphic operations in this scheme are performed similar to the initial MKBGV scheme. An example of homomorphic addition is shown in Fig.~\ref{fig:bgvk1dec}. Homomorphic multiplication is performed via tensor product. The generation of evaluation key is modified by combining operations between GSW and BGV ciphertexts instead of between pair of GSW ciphertexts to reduce the space overhead. The scheme also suggested a \textit{directed} distributed decryption protocol to prevent all participants from learning the decrypted result. In a nutshell, each participant sends their partial decryption to the designated participant, who is intended to receive the result. This participant does not share their partial decryption but collects all partial decryptions and combine them locally to retrieve the message. This design enforces the dishonest-majority assumption since none of the participants is able retrieve the message without having the partial decryption of the designated participant. 

\subsubsection*{MK-variant of TFHE} The MKHE primitive remained abstract constructions that are not practical for years. Chen, Chillotti, and Song~\cite{chen2019MKTFHE} designed an MK variant of the TFHE scheme~\cite{chillotti2016faster} and provided the first proof-of-concept implementation of MKHE.
% Based on the TFHE scheme which operates based on the torus. 

TFHE is an FHE over the torus $\mathbb T=\mathbb R \pmod 1$ which supports the evaluation of a binary gate followed by a fast bootstrapping procedure, which takes about 13ms.
Its multi-key variant is multi-hop and has the same functionality as the single-key TFHE.
We provide a description in Scheme~\ref{scheme:mktfhe}.
For the ring dimension $d$, we denote by $R=\mathbb Z[x]/(x^d+1)$ and $T=\mathbb T[x]/(x^d+1)$.
The gadget toolkit is defined in an approximate manner, i.e., for a gadget vector $\bg\in \mathbb Z^\ell$, the gadget decomposition is a function $\bg:T\rightarrow R^\ell$ such that $\langle{\bg^{-1}(a),\bg}\rangle\approx a \pmod 1$ for all $a\in T$.

\begin{pabox}[label={scheme:mktfhe}]{An MK-variant of gate bootstrapping~\cite{chen2019MKTFHE}}
\begin{itemize}[leftmargin=*]
\item[-]{$\mathsf{MKTFHE.}\setup(1^\lambda)$}: Given a security parameter $\lambda$, choose the RLWE dimension $N$ and the LWE dimension $n$. Set the LWE and RLWE error distributions $\chi'$ over $\mathbb T$ and $\chi$ over $T$, respectively. Sample an element $\ba$ uniformly at random from $T^\ell$. Output the public parameters as $\param = (n, N, \chi', \chi, \boldsymbol{a})$.
Set the RLWE and LWE gadget toolkits $\bg\in\mathbb Z^\ell$, $\bg^{-1}:T\rightarrow R^\ell$ and $\bg'\in\mathbb Z^{\ell'}$, $\bg'^{-1}:\mathbb T\rightarrow \mathbb Z^{\ell'}$ respectively.

\item[-]{$\mathsf{MKTFHE.}\kgen(\param)$}: Sample the LWE secret $\boldsymbol{t}=(t_j)_{0\le j< n}\sample\{0,1\}^n$. Sample $\boldsymbol{s}=(s_j)_{0\le j<N}\sample\{0,1\}^N$ and set the RLWE secret as $s=\sum_j s_j X^j$.
Sample $\boldsymbol{e}\sample \chi^\ell$ and set the public key as $\boldsymbol{b}=-s\cdot \boldsymbol{a}+\boldsymbol{e} \pmod 1$.
For $1\le j\le n$, generate $\mathsf{BK}_j\leftarrow \unienc(\ba,t_j,s)$ and set the bootstrapping key as $\mathsf{BK}=\{\mathsf{BK}_j\}_{1\le j\le n}$. Generate a key-switching key $\mathsf{KS}$ from $\boldsymbol{s}$ to $\boldsymbol{t}$.

\item[-]{$\mathsf{MKTFHE.}\HomNAND(\Bar{\boldsymbol{c}}_1, \Bar{\boldsymbol{c}}_2, \{\bb_i, \mathsf{BK}_i, \mathsf{KS}_i\}_{1\le i\le K})$}: For given two extended ciphertexts $\Bar{\boldsymbol{c}}_i\in \mathbb T^{nK+1}$ under the same extended secret key $\Bar{\sk} = (s_1,\dots,s_K)$,
\begin{enumerate}
    \item $\Bar\bc \leftarrow (5/8,{\bf 0})-\Bar\bc_1-\Bar\bc_2 \pmod 1$.
    \item Write $\Bar\bc=(b,\ba_1,\dots,\ba_K)$. Compute $\tilde b=\near{2N\cdot b}$ and $\tilde a_{i,j}=\near{2N\cdot \ba_i[j]}$ for $1\le i\le K$ and $1\le j\le n$.
    \item Let $\mathsf{ACC}\leftarrow \left(\frac{1}{8}(1+X+\dots+X^{N-1})\cdot X^{N/2+\tilde b}, {\bf 0}\right)\in T^{K+1}$
    \item For $1\le i\le K$ and $1\le j\le n$, do $\mathsf{ACC}\leftarrow \cmux(\mathsf{ACC}, x^{\tilde a_{i,j}}\cdot \mathsf{ACC}, \mathsf{BK}_{i,j})$.
    \item Write $\mathsf{ACC}$ as $(\sum_{0\le i<N} b_i'\cdot X^i, \sum_{0\le j<N} a_{1,j}' \cdot X^j, \dots, \sum_{0\le j<N} a_{N,j}'\cdot X^j)$, and let $$\bc'=\left(\frac{1}{8}+b_0',a_{1,0}',-a_{1,N-1}',\dots,-a_{1,1}',\dots,a_{K,0}',-a_{K,1}',\dots,-a_{K,N-1}'\right)\in \mathbb T^{NK+1}.$$
\item Perform the key-switching from $(\boldsymbol s_1,\dots, \boldsymbol s_K)$ to $(\boldsymbol t_1,\dots, \boldsymbol t_K)$ using $\{\mathsf{KS}_i\}_{1\le i\le K}$, and return the output ciphertext in $\mathbb T^{nK+1}$.
\end{enumerate}
\end{itemize}
\end{pabox}

In the original TFHE, the bootstrapping procedure relied on the \emph{external product} which takes a pair of ring-GSW and RLWE ciphertexts as input and returns an RLWE ciphertext.
Two MK variants of the external product are proposed in \cite{chen2019MKTFHE}: the first approach is an improvement of the previous ciphertext extension technique of \cite{mukherjee2016two} and the other is a new solution which does not require the precomputation of extended ciphertexts.
Instead, the involved parties' evaluation keys directly act on the corresponding elements of extended ciphertext during bootstrapping.
In the following, we describe the second method which has advantages in terms of noise growth and complexity.
Each party generates a public key $\bb_i\approx -s\cdot \ba \pmod 1$ which is taken as an input of the external product.
We write $\bb_0=-\ba$ for the simplicity.

\begin{itemize}
    \item[-]$\unienc(\ba,\mu,s_i)$: The CRS $\ba\in R_q^\ell$, message $\mu$ and the $i$-th secret $s_i$ are given as input. Sample $r_i\sample \psi$, $\boldsymbol{f}_{i,1}$ uniformly at random from $R_q^\ell$, and $\be_{i,1}, \be_{i,2}\leftarrow\chi^\ell$.
    Return the uni-encryption $C_i=(\bd_i,\boldsymbol{f}_{i,0},\boldsymbol{f}_{i,1})$ 
    where $\boldsymbol{d}_i=r_i\cdot \ba+\be_{i,1}+\mu\cdot \bg \pmod 1$ and $\boldsymbol{f}_{i,0}=-s_i\cdot \boldsymbol{f}_{i,1}+\boldsymbol{e}_{i,2}+r_i\cdot \bg \pmod 1$. 
    \item[-]$\extprod(\bar \bc, C_i, \{\bb_i\}_{0\le i\le K})$: Given an uni-encryption $C_i=(\bd_i,\boldsymbol{f}_{i,0},\boldsymbol{f}_{i,1})$ and an extended RLWE ciphertext $\bar \bc=(c_{i})_{0\le i\le K}$ and the set of public keys $\{\bb_i\}_{0\le i\le K}$, compute and return the ciphertext $\bar\bc'=(c_0',\dots,c_K')\in R_q^{K+1}$ as follows:
    \begin{enumerate}
        \item Compute $u_j=\langle{\bg^{-1}(c_j),\bd_j}\rangle$,  $v_j=\langle{\bg^{-1}(c_j),\bb_j}\rangle$, $w_{j,0}=\langle{\bg^{-1}(v_j),\boldsymbol{f}_{i,0}}\rangle$, and $\langle{\bg^{-1}(v_j), \boldsymbol{f}_{i,1}}\rangle$ for all $0\le j\le K$.
        \item Compute $c_0'=u_0+\sum_{0\le j\le K}w_{j,0} \pmod 1$, $c_i'=u_i+\sum_{0\le j\le K}w_{j,1} \pmod 1$ and $c_j'=u_j$ for $j\neq i$.
    \end{enumerate}
\end{itemize}
We will denote the external product by $\extprod(\bar \bc, C_i, \{\bb_i\}_{0\le i\le K})=\bar\bc\boxdot C_i$. 
Similar to the TFHE scheme, we can define a multi-key variant of the CMUX gate using the MK external product above.
For given two MK RLWE ciphertexts $\bar\bc_0$, $\bar\bc_1$ and an uni-encryption $C_i$ of $\mu\in\{0,1\}$ from the $i$-th party, it returns the RLWE ciphertext $\cmux(\bar\bc_0,\bar\bc_1, C_i)=\bar\bc_0 +(\bar \bc_1-\bar \bc_0) \boxdot C_i$ encrypting the same message as $\bar \bc_\mu$.

A MK-variant of the gate bootstrapping was constructed by replacing the external product and cmux gate of TFHE by their MK variants described above.
The final multi-key-switching step is done by repeating the single key-switching procedure on individual ciphertext entries.

%This scheme implements a Ring-variant of the GSW scheme and operates over torus $T$. 
%It is more efficient than other LWE-based schemes. The bootstrapping is fast (less than 0.1 second).
%Computations are done by evaluating binary \textsf{NAND} gate followed by a bootstrap.

% \begin{pabox}[label={scheme:mktfhe}]{MK-variant of the TFHE scheme~\cite{chen2019MKTFHE}}
% \begin{itemize}[leftmargin=*]
% \item[-]{$\mathsf{MK-TFHE.}\setup(1^\lambda)$}: Given a security parameter $\lambda$, uniformly choose a random element $a \sample T$, where $T$ is the 
% \item[-]{$\mathsf{MK-TFHE.}\kgen(1^\lambda)$}: 
% \item[-]{$\mathsf{MK-TFHE.}\eval(\Bar{\pk}, f, c_1, \dots, c_\ell)$}: 
% \item[-]{$\mathsf{MK-TFHE.}\dec(\{\sk_1,\dots,\sk_n\},\Bar{c})$}: 
% \end{itemize}
% \end{pabox}

\subsubsection*{MK-variant of BFV/CKKS}
Chen~et~al.~\cite{chen2019MKSEAL} introduced MK variants of the BFV and CKKS schemes.
A ciphertext corresponding to the tuple of $K$ secrets $\bar{\boldsymbol s}=s_1,\dots, s_K$ is the form of $\bar\bc=(c_0,\dots,c_K)\in R_{q}^{K+1}$ where $c_0+\sum_i c_i\cdot s_i$ is a randomized encoding of the underlying plaintext.

Different from the MKBGV scheme~\cite{chen2017batched}, the authors proposed a new relinearization algorithm which does not require any precomputation on the evaluation keys.
To be precise, homomorphic multiplication between $\bar\bc_1, \bar\bc_2\in R_q^{K+1}$ first computes $\hat \bc=\bar\bc_1\otimes \bar \bc_2 \in R_q^{(K+1)^2}$, and then perform the relinearization procedure on the entries of $\hat\bc=(\hat c_{i,j})_{0\le i,j\le K}$.
Each relinearization procedure on $\hat c_{i,j}$ requires only two evaluation keys from the parties $i$ and $j$.
As a result, the complexity grows quadratically with the number of involved parties $K$.

In the following, we first describe the common relinearization method and then provide the descriptions of MKBFV and MKCKKS in Schemes~\ref{scheme:mkbfv} and \ref{scheme:mkckks}, respectively.
The public key of the $i$-th party is denoted by $\pk=\bb_i \in R_q^\ell$.

\begin{itemize}
    \item[-]$\rlkgen(\ba,s_i)$: For the CRS $\ba\in R_q^\ell$ and the secret $s_i$, sample $r_i\sample \psi$. Sample $\boldsymbol{d}_{i,1}$ uniformly at random from $R_q^\ell$ and errors $\be_{i,1}, \be_{i,2}$ from $\chi^\ell$.
    Return the relinearization key $\rlk_i=(\bd_{i,0},\bd_{i,1},\bd_{i,2})$ 
    where $\boldsymbol{d}_{i,0}=-s_i\cdot \boldsymbol{d}_{i,1}+\boldsymbol{e}_{i,1}+r_i\cdot \bg \pmod q$ and $\boldsymbol{d}_{i,2}=r_i\cdot \ba+\be_{i,2}+s_i\cdot \bg \pmod q$.
    \item[-]$\relin(\{\pk_i,\rlk_i\}_{1\le i\le K}, \hat \bc)$: Given a ciphertext $\hat \bc=(c_{i,j})_{0\le i,j\le K}$ and the set of public and relinearization keys $\pk_i=\bb_i$, $\rlk_i=(\bd_{i,0},\bd_{i,1},\bd_{i,2})$, compute and return the ciphertext $\bc'=(c_0',\dots,c_K')\in R_q^{K+1}$ as follows:
    \begin{enumerate}
        \item Initialize the ciphertext $\bc'$ as $c_0'\leftarrow c_{0,0}$ and $c_i\leftarrow c_{i,0}+c_{0,i} \pmod q$ for $1\le i\le K$.
        \item For $1\le i,j\le K$, do $c_{i,j}'\leftarrow \langle\bg^{-1}(c_{i,j}),\bb_j\rangle \pmod q$, $(c_0',c_i')\leftarrow (c_0',c_i')+\bg^{-1}(c_{i,j}')\cdot [\bd_{i,0}|\bd_{i,1}] \pmod q$, and $c_j'\leftarrow c_j'+\langle{\bg^{-1}(c_{i,j}),\bd_{i,2}}\rangle \pmod q$.
    \end{enumerate}
\end{itemize}

It is also presented in \cite{chen2019MKSEAL} how to perform the rotation operation on the plaintext slots using an MK variant of the key-switching technique. Moreover, the authors extended the bootstrapping methods of BFV and CKKS into the MK case. We refer the reader to the original manuscript for more details and experimental results.

\begin{pabox}[label={scheme:mkbfv}]{ An MK-variant of BFV scheme~\cite{chen2019MKSEAL}}
\begin{itemize}[leftmargin=*]
\item[-]{$\mathsf{MKBFV.}\setup(1^\lambda)$}: Given a security parameter $\lambda$, choose the RLWE dimension $n$, the ciphertext module $q$, and the plaintext module $t$. Set the error distribution $\chi$ and the key distribution $\psi$ over $R$. Sample a random vector $\boldsymbol{a} \sample R^\ell_q$. Output the public parameters $\param = (n, q, t, \chi, \psi, \boldsymbol{a})$. 

\item[-]{$\mathsf{MKBFV.}\kgen(\param)$}: Given the public parameters, sample $s \sample \psi$ and $\boldsymbol{e}\sample R_q^\ell$, and let $\boldsymbol{b}=-\ba\cdot s+\be \pmod q$. Output the key pair as $(\pk=\bb, \sk=s)$. Run $\rlk\sample\rlkgen(\ba,s)$ to generate a relinearization key.

\item[-]{$\mathsf{MKBFV.}\enc(\pk, m)$}: Given a public key $\pk=\bb$, let $b=\bb[0]$ and $a=\ba[0]$. Sample $r \sample \psi$ and $e_0, e_1 \sample \chi$. Compute the ciphertext as $\boldsymbol{c} =r \cdot(b,a)+(\Delta m+e_0, e_1) \in R_q$, where $\Delta = \near{q/t}$. 
% \{EvalKeyGeneration}
% \item[-]{$\mathsf{MKBFV.}\evalkgen$}:

\item[-]{$\mathsf{MKBFV.}\evaladd(\Bar{\boldsymbol{c}}_1, \Bar{\boldsymbol{c}}_2)$}: For given two extended ciphertexts under the same extended secret key $\Bar{\sk} = (s_1,\dots,s_K)$, output $\Bar{\boldsymbol{c}}_\add = \Bar{\boldsymbol{c}}_1 + \Bar{\boldsymbol{c}}_2 \in R^{K+1}_q$. 

\item[-]{$\mathsf{MKBFV.}\evalmult(\{\pk_i,\rlk_i\}_{1\le i\le K}, \Bar{\boldsymbol{c}}_1, \Bar{\boldsymbol{c}}_2)$}: Compute $\hat \bc_\mult = \near{t/q} (\Bar{\boldsymbol{c}}_1 \otimes \Bar{\boldsymbol{c}}_2) \in R^{(K+1)^2}_q$ and return the ciphertext $\Bar{\bc}_\mult=\relin(\{\pk_i,\rlk_i\}_{1\le i\le K},\hat \bc_\mult)\in R_q^{K+1}$.

\item[-]{$\mathsf{MKBFV.}\dec((\sk_1,\dots,\sk_K),\Bar{\boldsymbol{c}})$}: Given a set of secret keys and an extended ciphertext $\Bar{\boldsymbol{c}} = (c_0,c_1,\dots,c_K)$, set the extended secret key as $\Bar{\sk} = (1, \sk_1,\dots,\sk_K)$ and output $\near{\near{t/q}\langle\Bar{\boldsymbol{c}}, \Bar{\sk}\rangle}\pmod t$.
\end{itemize}
\end{pabox}

% \subsubsection*{MK-variant of CKKS scheme}
\begin{pabox}[label={scheme:mkckks}]{An MK-variant of CKKS scheme~\cite{chen2019MKSEAL}}
\begin{itemize}[leftmargin=*]
\item[-]{$\mathsf{MKCKKS.}\setup(1^\lambda)$}: Given a security parameter $\lambda$, choose the RLWE dimension $n$ and a chain of ciphertext moduli $q_0<q_1<\dots<q_L$. Assume $q = \prod_{i=0}^{L}p_i$ for some integers $p_i$, such that $q_\ell = \prod_{i=0}^{\ell}p_i$. Set the error distribution $\chi$ and the key distribution $\psi$ over $R$. Uniformly choose a random vector $\boldsymbol{a} \sample R^\ell_q$. Output the public parameters as $\param = (n, q, \chi, \psi, \boldsymbol{a})$. 

\item[-]{$\mathsf{MKCKKS.}\kgen(\param)$}: Given the public parameters, sample $s \sample \psi$ and $\be\sample \chi^\ell$, and let $\bb=-\ba\cdot s+\be \pmod q$. Output the key pair as $(\pk=\bb, \sk=s)$. 

\item[-]{$\mathsf{MKCKKS.}\enc(\pk, m)$}: Given a public key $\pk = \bb$, let $b=\bb[0]$ and $a=\ba[0]$. Sample $r \sample \chi_k$ and $e_0, e_1 \sample \chi$, and return the ciphertext $\boldsymbol{c} =  r\cdot (b,a)+(m + e_0,e_1) \in R^2_q$.

\item[-]{$\mathsf{MKCKKS.}\eval(\Bar{\pk}, f, \Bar{\boldsymbol{c}}_1, \Bar{\boldsymbol{c}}_2)$}: Given two extended ciphertexts under the same extended secret key $\Bar{\sk} = (s_1,\dots,s_K)$ at level $\ell$ for $K$ the number of involved parties, perform homomorphic evaluation as follows. For addition, $\Bar{\boldsymbol{c}}_{add} = \Bar{\boldsymbol{c}}_1 + \Bar{\boldsymbol{c}}_2 \in R^{K+1}_{q_\ell}$.
For multiplication, compute $\hat\bc_\mult = \Bar{\boldsymbol{c}}_1 \otimes \Bar{\boldsymbol{c}}_2 \in R^{(K+1)^2}_{q_\ell}$. Perform the relinearization and return the ciphertext $\Bar{\bc}_\mult=\relin(\{\pk_i,\rlk_i\}_{1\le i\le K},\hat \bc_\mult)\in R_q^{K+1}$.

\item[-]{$\mathsf{MKCKKS.}\rescale(\Bar{\boldsymbol{c}})$}: Given a ciphertext $\Bar{\boldsymbol{c}} = (c_0, \dots, c_K) \in R^{K+1}_{q_\ell}$ at level $\ell$, compute $c'_i = \near{p_\ell^{-1}c_i}$ for $0\le i \le K$ and return the re-scaled ciphertext $\Bar{\boldsymbol{c}'} = (c'_0, \dots, c'_K) \in R^{K+1}_{q_{\ell -1}}$

\item[-]{$\mathsf{MKCKKS.}\dec((\sk_1,\dots,\sk_K),\Bar{\boldsymbol{c}})$}: Given a set of secret keys and an extended ciphertext $\Bar{\boldsymbol{c}} = (c_0,c_1,\dots,c_K) \in R_{q_\ell}$, set the extended secret key as $\Bar{\sk} = (1, \sk_1,\dots,\sk_K)$ and output $\langle\Bar{\boldsymbol{c}}, \Bar{\sk}\rangle\pmod {q_\ell}$.
\end{itemize}
\end{pabox}

\subsection{Hybrid Multi-key Homomorphic Encryption}
\label{subsec:MKHE+}
The proportional increase in ciphertext size remains a bottleneck for efficient MKHE schemes. The size of a ciphertext grows correspondingly to the number of involved participants' keys. This may pose an efficiency limitation in many outsourced computation scenarios with a large number of participants. Suppose we have a system with $N$ model owners and $P$ clients. The model owners want collaborate and jointly compute on their trained models or functions and  evaluate their client's requests. The model owners want to keep their sensitive models private, so they encrypt them under their individual keys and delegate them to a cloud. The cloud evaluates these encrypted models but should not learn anything from them. In this system, both model owners and clients want to protect their inputs under their keys. 

Leveraging MKHE schemes for this scenario will require each ciphertext to be extended to $N+1$ different keys (i.e., the model owners keys plus the key of the requesting client) before any computation. The efficiency of the system is affected, especially if the number of model owners is large because it proportionally increases the ciphertext size. On the other hand, one may leverage ThHE schemes which compute under a joint key generated from participants keys. However, this approach is also not practical because we need to generate a joint key for every client and the group of $N$ model owners; that is, we will need to produce and maintain $P$ joint keys. This requirement also means that each model owner has to provide $P$ copies of the model, each encrypted under one of the $P$ joint keys, to the Cloud. Moreover, if there is any change (adding or removing) in the participants requires generating a new joint key and re-encrypting all ciphertexts with this new joint key, which is inefficient. 

To address this issue, Aloufi~and~Hu~\cite{aloufi2019collaborative} proposed a hybrid approach (MKHE$+$) based on the BGV scheme~\cite{brakerski2012leveled}. The scheme supports homomorphic computation over ciphertexts encrypted under multi-key and produces small ciphertexts. The approach combines the advantages of both ThHE and MKHE techniques to reduce computation complexity and ciphertext size. We present the MKHE$+$ scheme in Scheme~\ref{scheme:mkhe+}.

\begin{pabox}[label={scheme:AH}]{\highlight{A Threshold MKHE scheme}~\cite{aloufi2019collaborative}}\label{scheme:mkhe+}
\begin{itemize}[leftmargin=*]%[noitemsep,topsep=10pt]

\item[-] {$\mathsf{MKHE+}.\setup(1^\lambda, 1^L, 1^\mathcal{K})$}: Given the security parameter $\lambda$, a multiplicative depth $L$, and a bound $\mathcal{K}$ on the number of keys, run the $\mathsf{BGV}.\setup$ and output $\param = (R, \chi, q, \boldsymbol{a}, t)$ where $\boldsymbol{a}\sample R^{\ell}_{q}$ is the CRS vector.

\item[-]{$\mathsf{MKHE+.}\kgen(\param$}: Given the public parameters $\param$, choose $a \sample \boldsymbol{a}$ and generate a key pair $(\pk_{M_i}, \sk_{M_i})$ for each model owner $\mathcal{M}_i$ where $i\in\{1,\dots, N\}$ and $(\pk_{\mathcal{C}_i}, \sk_{\mathcal{C}_i})$ for each client $\mathcal{C}$. Then for the set of model owners, generate a joint key $\pk_\mathcal{M}$ as in $\mathsf{ThBGV}.\jkgen$.

\item[-]{$\mathsf{MKHE+.}\enc(\pk, m)$}: Each model owner encrypts their model under the joint key $\pk_\mathcal{M}$ and delegate them to the evaluator (if in outsourced system model, see Fig.~\ref{fig:outsourced}). The client encrypts their input under their public key $\pk_\mathcal{C}$. 

\item[-]{$\mathsf{MKHE+.}\extend((\pk_\mathcal{M},\pk_\mathcal{C}), \boldsymbol{c})$}: When evaluation is requested, the evaluator first extends each ciphertext under the set of the two keys $\Bar{\pk} = (\pk_\mathcal{M},\pk_\mathcal{C})$ using the algorithm $\mathsf{MKBGV.}\extend((\pk_1,\dots,\pk_N), \boldsymbol{c})$ described in the MKBGV scheme.

\item[-]\highlight{$\mathsf{MKHE+.}\evalkgen(\sk_{\mathcal{M}_i}, \sk_\mathcal{C})$}: Given a set of model owners' secret keys and the client secret key, first generate a joint helper element $\ek'_{\mathcal{M}}$ (in a threshold manner) and a helper element $\ek'_{\mathcal{C}}$ as in $\mathsf{MKBGV}.\evalkgen$. 

\item[-]{$\mathsf{MKHE+.}\eval(\Bar{\boldsymbol{c}}_1, \Bar{\boldsymbol{c}}_2)$}: Computations now can be done on extended ciphertexts $\Bar{\boldsymbol{c}}, \Bar{\boldsymbol{c}}_2$ that are encrypted under the same extended key $\Bar{\pk}$. The homomorphic addition of two extended ciphertexts is performed as element-wise addition, and the homomorphic multiplication is performed as the tensor product. 

\item[-]\highlight{$\mathsf{MKHE+.}\relin(\ek'_{\mathcal{M}}, \ek'_{\mathcal{C}}, \Bar{\boldsymbol{c}}_\mult)$}: Given a two helper elements, generate the extended evaluation key $\bar{\ek}$ for the extended initial ciphertext product and apply the key switching technique as described in $\mathsf{MKBGV}.\relin$ to reduce the dimension and allow further homomorphic operations.

\item[-] {$\mathsf{MKHE+.}\dec((\sk_\mathcal{M},\sk_\mathcal{C}), \Bar{\boldsymbol{c}})$}: Given an extended ciphertext that is encrypted under the extended key $\Bar{\pk} = (\pk_\mathcal{M}, \pk_\mathcal{C})$, invoke the distributed decryption protocol between the model owners and the client to perform the decryption as the following algorithm, $\dec(\Bar{\boldsymbol{s}},\Bar{\boldsymbol{c}}) = (\langle \boldsymbol{c}_\mathcal{C},\boldsymbol{s}_\mathcal{C}\rangle + \sum^{N}_{i=1}\langle \boldsymbol{c}_\mathcal{M}, \boldsymbol{s}_{M_i}\rangle \pmod q) \pmod t$.
\end{itemize}
\end{pabox}

The fundamental observation in MKHE$+$ is that the group of model owners do not often change and is likely to remain the same during the evaluation for each client. Hence, it is better to generate one joint key $\pk_{\mathcal{M}}$, using the ThHE key setup, for the model owners group to encrypt their models under this joint key. On the other hand, client's request can be dynamic as clients can come and go; therefore, a client can encrypt their data under their own key $\pk_{\mathcal{C}}$. This way the number of keys we are computing with becomes two keys -- the model owner's joint key $\pk_{\mathcal{M}}$ and the client's key $\pk_{\mathcal{C}}$.

At evaluation, the Cloud dynamically extends the ciphertexts to the concatenated key $\bar{\pk} = (\pk_{\mathcal{M}}, \pk_{\mathcal{C}})$. More specifically, each encrypted model is transformed from a ciphertext under one joint key into one encrypted under two keys, i.e., the joint key and the client's key. Similarly, the encrypted client's request is extended to the model owners' joint key. Homomorphic evaluations are done in a similar manner as described in Sec.~\ref{subsec:MKBGV}. One core difference is that model owners also need to collaborate to jointly generate the evaluation key corresponding to the joint key $\pk_{\mathcal{M}}$ to be used in the homomorphic multiplication. This can be done in an additional round as proposed by Asharov~et~al.~\cite{asharov2012multiparty} which we discuss in Sec.~\ref{subsec:THE}. Similar to ThHE and MKHE schemes, the hybrid scheme requires invoking a distributed decryption protocol so participants can jointly decrypt the encrypted result. 

This hybrid approach combines ThHE and MKHE techniques to remove the requirement of generating a joint key for every client with the model owners. It also produces short extended ciphertexts since the number of involved keys is two instead of $K+1$, which is adequate in scenarios where $K > 2$, the number of model owners is large.
\section{Open Research Problems}
\label{sec:lessons}

In this survey, we reviewed different techniques that support homomorphic computation on data contributed by different owners and encrypted using different encryption keys. We categorized these techniques into the single-key and multi-key approaches, as illustrated in Fig.~\ref{fig:taxonomy}.

Selecting the right scheme for an application depends on the system design and requirements. For example, if the application goal is to share encrypted input between a set of authorized users, it is better to leverage PRE schemes, or IBE/ABE schemes if fine-grained access control is required. For outsourced computations, such as training an ML model based on data contributed by multiple parties without leaking them, multi-key approaches is more suitable. It may be a better choice to leverage ThHE schemes in this case \textit{if} participants do not change often. If they do, the efficiency quickly degrades because the joint key needs to be updated accordingly. In this case, MKHE schemes is more flexible to enable \emph{on-the-fly} computations. Based on our observations and lesson-learned, we share open research problems and our view on methods that may overcome these problems.

\subsection{Distributed decryption}
In multi-key approaches discussed in Sec.~\ref{sec:multi}, ciphertexts are encrypted under multiple keys -- a joint key generated from multiple keys in ThHE or a concatenation of those keys in MKHE. Both ThHE and MKHE schemes share a similar requirement where all involved parties must collaborate for decryption. A trusted party can perform centralized decryption on behalf of involved parties, but it must have access to all secret keys. This setting may not secure, especially with the increase of security breaches. Alternatively, a distributed decryption based on MPC protocol (see Sec.~\ref{subsec:distdec}) can be invoked such that each participant partially decrypts the ciphertext with their own secret key. All participants then combine the partial decryptions to retrieve the message. In this setting, all participants learn the final decrypted result. To add an extra layer of security, the distributed decryption protocol can be made directed. This means only one \textit{intended} participant can learn the result by keeping their partial decryption private and be the last one to perform the decryption step. One way to render a non-interactive decryption protocol is to key switch the ciphertext's key to the intended participant's key before returning the result. But, the key switching step is interactive, as further elaborated in Sec.~\ref{subsec:CombiningMultikeyTechniques}.

\begin{figure*}[t]
\centering
  \subfloat[Distributed decryption]
 {\includegraphics[width=.4\textwidth]{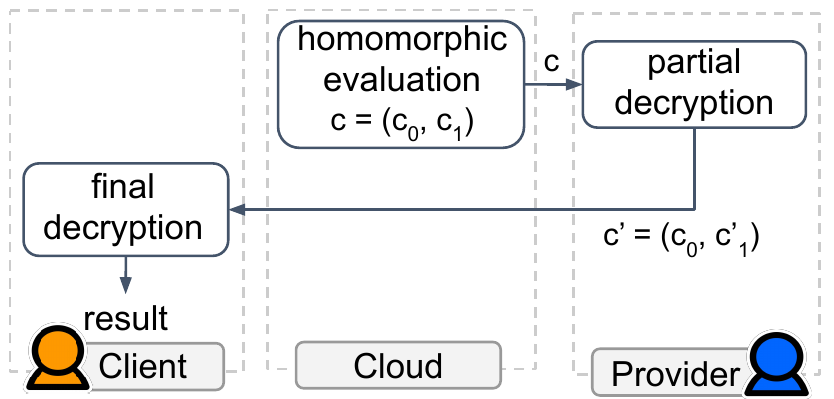} \label{fig:partial}}\hfill
  \subfloat[Optimized distributed decryption]
 {\includegraphics[width=.58\textwidth]{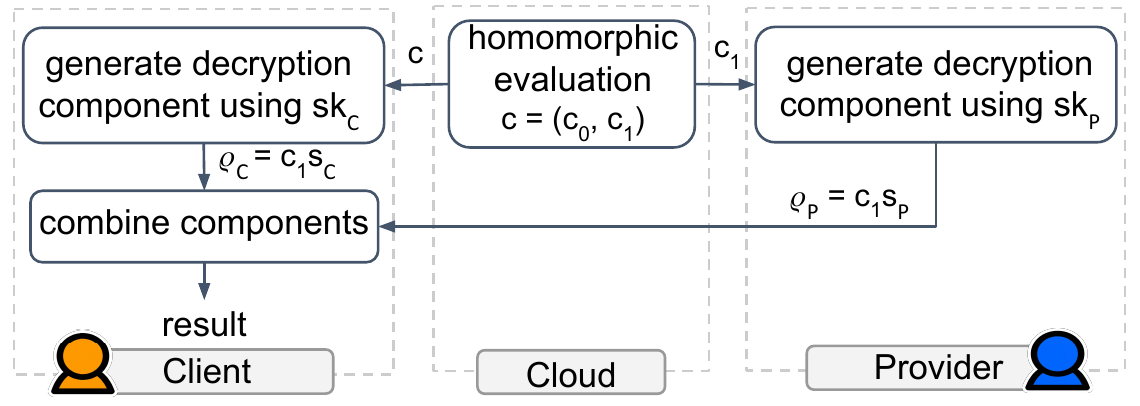} \label{fig:component}}
\caption{Overview of different approaches for distributed decryption protocol}
\label{fig:decryption}
\end{figure*}

In the outsourced system model (in Fig.~\ref{fig:outsourced}), the Cloud evaluator may return the resulting ciphertext to the client, who then invokes the distributed decryption protocol with other participants. This setting does not involve the Cloud in the decryption step. On the other hand, the Cloud can help and invoke the distributed decryption by sending the ciphertext directly to the other participants to perform partial decryption and return the partially decrypted ciphertext to the client, as illustrated in Fig.~\ref{fig:partial}. The client then completes the decryption with their own secret key to retrieve the result. This design involves moving a large ciphertext, containing two large components, among participants. The design can be improved, as illustrated in Fig.~\ref{fig:component}. The Cloud returns the full ciphertext to the clients and sends only the randomness component to the participants who use their secret keys to generate decryption components and send them to the client who completes the decryption step by combining the received components. Note the improved method may have additional round of communication, but it suits scenarios with a large number of participants because it avoids forwarding large partially decrypted ciphertexts between participants. 

\subsection{A combination of multi-key techniques}
\label{subsec:CombiningMultikeyTechniques}
Most of our surveyed multi-key approaches are designed by adding extra protocols on top of some base HE schemes. These extra protocols include PRE which enables switching the key of a given ciphertext to another key without revealing the plaintext, IBE/ABE which provides access control for the homomorphic computation. ThHE and MKHE support computation with ciphertexts encrypted, respectively, under joint and concatenated keys. An interesting observation is that some of these features, such as PRE and ThHE, can be combined to obtain an encryption scheme with a more efficient design.
% based on the desired system requirements to achieve a more efficient design. 

As mentioned earlier, PRE can be integrated with multi-key schemes such as ThHE and MKHE to avoid distributed decryption. The idea is to key switch the ciphertext from a join or concatenated key to the intended client's key. Although this method remove the interactiveness within the decryption step, it still requires all participants to collaboratively generate a re-encryption key in advance either based on their secret or public key. Mouchet et al.~\cite{cryptoeprint:2020:304} proposed methods for performing a distributed re-encryption key generation and distributed key-switching techniques.

Alternatively, we can develop hybrid MKHE approach~\cite{aloufi2019collaborative} based on ThHE and MKHE techniques to obtain a more practical design with shorter extended ciphertexts. More specifically, this approach is more suitable for computation outsourcing scenarios where a fixed group of data owners wanting to offer data-driven services to customers. In this system model, the group membership of data owners does not change often hence ThHE can used to create a joint key, whereas MKHE can support dynamic extension of this joint key to a customer's key, as we discussed in Sec.~\ref{subsec:MKHE+}. Unfortunately, the research work on this approach is still scarce. More work is needed to address other system models with one group of users. Another open research question is how to achieve more generalized and dynamic hybrid schemes, specifically in cases where participants often change. One may investigate the generation of joint keys for subsets of participants and allow the cloud to use PRE and switch \textit{on-the-fly} to a less number of concatenated keys to obtain shorter extended ciphertexts.
% Another open research question is whether it is possible to include the PRE technique to dynamically switch when appropriate from multiple concatenated keys to a less number concatenated keys (which include joint keys). \fix{what do you mean? why? please elaborate}  

\subsection{Public parameters}
To facilitate computing with multiple keys in LWE-based schemes, it is common to design them in the CRS model. That is, participants use a publicly shared parameter in the form of a lattice (or ring in RLWE) element and use it in generating their public keys. This is to ensure that all keys are related, which is a requirement for correct elimination of large lattice elements during decryption. One disadvantage of this model is that it often relies on a trusted party to uniformly choose the shared element before any computation is done. Even with removing this trusted party and generating independent public keys~\cite{kim2018towards}, we still need to perform a \textit{linking algorithm} on the keys to make them related. This linking algorithm is held at run time and is interactive, which increases both computation and communication overheads. Hence, a research question still stands as if we can compute efficiently with multiple keys without relying on public parameters.

Another form of public parameters that must be provided is encrypted randomness in GSW-based MKHE schemes and evaluation keys in BV-based MKHE schemes. In the MK-variants of GSW scheme, there is no need to provide evaluation keys. However, decrypting an extended ciphertext with concatenated keys fails if no information about the encryption randomness was provided. Hence, it is crucial to include additional encrypted randomness components within the extended ciphertext. Because of this requirement, the size of the ciphertexts increases quadratically at worst case~\cite{mukherjee2016two,peikert2016multi}, which leads to compute on significantly high-dimensional matrices. Even with proposed optimizations using bootstrapping~\cite{brakerski2016lattice} to obtain linearly increased size, we need to bootstrap at each evaluation and the initial evaluated ciphertexts have quadratic growth. 

On the other hand, MK-variants of BV schemes do not require encrypting the randomness, but they need providing evaluation keys to apply relinearization after each homomorphic multiplication as discussed in Sec.~\ref{subsec:keyswitching}. Generating the evaluation keys corresponding to the concatenated key is not a trivial task. In earlier works proposed by Chen~et~al.~\cite{chen2017batched} and Li~et~al.~\cite{li2019efficient}, this task requires combining different schemes, namely BGV and ring-GSW. However, Chen~et~al.~\cite{chen2019MKSEAL} proposed that relinearization can be applied directly on extended ciphertexts with individually generated evaluation keys.
\section{Applications}
\label{sec:applications}
Many outsourced computations involve sensitive data contributed by individual data owners. HE protects this sensitive data in-use without requiring decryption first. Multi-key HE extends the protection of private data to individual owners, with each possesses a private key for decrypting his/her inputs and results but nothing else from others. There are many applications that can benefit from multi-key HE. We classify these applications based on the two system models described in Section~\ref{subsec:systemAndSecurityModel}.

\subsection{Computation outsourced to a cloud evaluator}
Nowadays, large-scale data analysis and machine learning tasks are outsourced to the Cloud for its compelling economic efficiency and wide accessibility. Outsourcing these computational-intensive tasks to the Cloud where the homomorphically encrypted data resides can reduce the overheads of moving large ciphertext around. Many existing work proposed homomorphic algorithms based on this system model.

CryptoNets~\cite{gilad2016cryptonets} and CryptoDL~\cite{hesamifard2017cryptodl} are earlier works on supporting secure inference of encrypted neural networks that are outsourced to the Cloud. These works assumed a system model in which both network parameters and feature inputs are contributed by the same owner who just wants to offload data storage and computations to the cloud, hence these initial works assumed private data is encrypted under a single key. However, this setup is not suitable in a collaborative setting where each data owner wants to protect their private data using their own keys, as illustrated in Fig.~\ref{fig:outsourced}.

Wang et~al.~\cite{wang2013computing} proposed a framework with two non-colluding servers $\mathcal{A}, \mathcal{B}$ to support homomorphic computation on ciphertexts encrypted under different keys. The core idea is based on proxy re-encryption (PRE) which has been discussed in Section~\ref{subsec:PRE}. The framework leverages one of two proposed ElGamal variant schemes, Vitamin$^+$ and Vitamin$^*$, to support partial homomorphic evaluation (addition or multiplication) on ciphertexts. In the Vitamin$^+$ scheme, server $\mathcal{A}$ can independently perform addition on ciphertexts but needs to interact with server $\mathcal{B}$ to perform multiplication on the blinded messages. In particular, users encrypt their messages with their keys and send ciphertexts to the server $\mathcal{A}$. The server $\mathcal{A}$ converts ciphertexts from under users' public keys to ones encrypted under the public key of server $\mathcal{B}$ and locally perform homomorphic addition on ciphertexts under the same key. To evaluate multiplication, ciphertexts are first blinded and sent to the server $\mathcal{B}$ who decrypts and perform multiplication of blinded messages in the plain. Then, the server $\mathcal{B}$ encrypts the product with its key and returns it. The server $\mathcal{A}$ can then remove the blinding factor and convert ciphertext results back to ones encrypted under users' keys. The same process follows in the Vitamin$^*$ scheme but with changing the operations. The re-encryption and blinding operations are required due to the use of ElGamal variant schemes which cannot support homomorphic addition and multiplication on ciphertexts. These operations add significant overhead impacting the practicality of this framework. Also, many existing works explore the use of partial HE schemes to support homomorphic computation in a two-party setting (Fig.\ref{fig:2pc}). These solutions typically have limitations making them not straightforward to be generalized.

Nevertheless, the idea of using PRE to transform ciphertexts encrypted under one key to another has been adopted in later works~\cite{yasuda2018multi,raisaro2018m} to avoid the need of a trusted crypto server for decryption or invoking a distributed decryption. For example, the work of \cite{yasuda2018multi} leverages an MK-variant of BGV scheme to extend ciphertexts to multiple keys before evaluation. The ciphertext result, which is encrypted under concatenated keys, can be then converted via key switching to the receiver's public key so it is decryptable directly with the corresponding secret key.

Hu et~al.~\cite{hu2017geosocial} proposed homomorphic algorithms for evaluating geosocial query with user-controlled privacy on the Cloud. The authors proposed a system model that allows users to submit periodically encrypted geo-location data to a cloud evaluator (i.e., service provider). The cloud evaluator performs homomorphic evaluation of location queries upon receiving users' requests. In this case, the data privacy of individual users needs to be protected with multi-key approaches. The authors leveraged the idea of ThHE (Sec.~\ref{subsec:THE}) to support homomorphic evaluation with multiple keys and proposed the use of a semi-trusted crypto server to create joint public key using individual public key from each pair of users. Users encrypt their location data using this joint key. The homomorphically evaluated results are then sent to the crypto server for decryption, or passed around all involved parties for a distributed decryption. This work inherits the limitation of ThHE, and it is not salable to a large number of users.

Aloufi~et~al.~\cite{aloufi2019blindfolded} proposed homomorphic protocols for blindfolded evaluation of random forests that are assembled from collaborating model owners. Each model owner encrypts their decision trees under their keys and outsource them to the Cloud evaluator. The authors proposed the use of the hybrid MKHE (Sec.~\ref{subsec:MKHE+}) to reduce the number of joint keys needed for encrypting the models while preserving the ability to dynamically extend the encrypted models to a user's public key just before evaluation. Similar design has been proposed by Chen~et~al.~\cite{chen2019MKSEAL} for oblivious neural network inference based on encrypted inputs under multiple owners' keys. Both work require the semi-honest cloud to work with individual model owners to perform a MPC-based distributed decryption protocol.

%  \highlight{perform the extension function} 
\subsection{Computation among multi-party}
Earlier schemes, namely LTV~\cite{lopez2012fly} and Clear-McGoldrick~\cite{clear2015multi}, extended HE schemes to the multi-key setting. However, they assumed the existence of a trusted party that has access to all secret keys and can decrypt the evaluated result at the end. This setting is not feasible when multiple participants do not want to disclose their secret keys. Mukherjee and Wichs~\cite{mukherjee2016two} laid the first construction of a two-round MPC protocol that leverages MKHE evaluations and enable distributed decryption. In the first round, each participant encrypts their inputs under their individually generated keys and broadcasts the ciphertexts. In the second round, each participant locally extends the ciphertexts to the additional keys (Sec.~\ref{subsec:MKHE}) and homomorphically evaluates the algorithm. Then, each participant uses their secret key to decrypt the locally evaluated result and broadcasts this partially decrypted result to other participants. At the end, all participants aggregate all partial decrypted result and obtain the final result. The protocol avoids delegating secret keys to a trusted party, but it requires all the parties to have access to the evaluated algorithm in order to concurrently compute the same algorithm on ciphertexts. Building on top of this construction, Gavin and Bonnevay~\cite{gavin2019securely} proposed a system to securely aggregate testimonies from different parties in which some parties may be colluded to perturb the results. Another secure data aggregation protocol were proposed for Vehicular ad-hoc networks (VANETs)~\cite{mi2019secure}. 

Technically, it is possible to extended most (R)LWE based HE schemes to a multiparty setting using a \emph{universal thresholdizer}~\cite{boneh2018threshold}. The construction is based on the linear secret sharing of key as in ThHE (Sec.~\ref{subsec:THE}). For example, Mouchet et al.~\cite{cryptoeprint:2020:304} proposed the construction of the multiparty variant of recent (R)LWE-based schemes such as BFV~\cite{brakerski2012fully,Fan:2012aa} and CKKS~\cite{cheon2018full}.
Fundamentally, this work shares many similarities with other ThHE scheme. Hence, it requires the collaboration of all parties to decrypt the evaluated results or performs a collaborative key switching protocol to transform the evaluation results encrypted under a joint key to a ciphertext that is decryptable by the user's secret key.

Clinical and Genomic data is typically locked up on secure servers of individual data owners such as hospitals, biobanks. Due to privacy regulations (e.g. HIPAA, GDPR), these data owners are reluctant to outsource any private data to a third-party. Medco~\cite{raisaro2018m} combined HE with MPC protocol to facilitate federated processing of homomorphic encrypted data that is on-premises of individual data owners. Medco adopted ThHE to support multi-key HE, assuming the membership of these data owners does not change often.
Hence, they can setup a joint public key for clients to encrypt their inputs. Note, the client inputs are protected since no single data owner can decrypt without the joint secret key which are generated using partial secret keys owned by each data owner. The encrypted client queries are sent to each data owner for homomorphic evaluation. The evaluated results are verified and passed through a distributed decryption protocol which converts the final results to be encrypted under only the client's key.

\section{Conclusion}
\label{sec:conclusion}

Many secure computation applications requires stronger security models where data are protected under different keys. In this survey, we reviewed existing techniques that support this model of computation. We categorized the techniques, based on the number of keys involved in the encryption, to single-key and multi-key approaches. Single-key approaches encrypt and process data then enable an authorized user to independently decrypt with their key. On the other hand, multi-key approaches enable a set of users to construct one joint key based on their individual keys as in ThHE schemes, or dynamically extends their individually encrypted data to multiple keys as in MKHE schemes. More recent techniques support combining different techniques for specific system models for efficiency. We also analyzed the security and complexity of different existing schemes and discussed open problems in this emerging research area.

\bibliographystyle{ACM-Reference-Format}
\bibliography{ref}

\end{document}